%Paper: hep-th/9310198
%From: denjoe@stp.dias.ie
%Date: Sun, 31 Oct 93 17:33:32 GMT

%Environmentally Friendly Renormalization
%%%%%%%%%%%%%%%%%%%%%%%%%%%%%%%%%%%%%%%%%%%%%%%%%%%%%%%%%%%%%%%%%%%%
%part of jnl
\font\twelverm=cmr10 scaled 1200    \font\twelvei=cmmi10 scaled 1200
\font\twelvesy=cmsy10 scaled 1200   \font\twelveex=cmex10 scaled 1200
\font\twelvebf=cmbx10 scaled 1200   \font\twelvesl=cmsl10 scaled 1200
\font\twelvett=cmtt10 scaled 1200   \font\twelveit=cmti10 scaled 1200

\skewchar\twelvei='177   \skewchar\twelvesy='60

%  Define \...point macros to change fonts and spacings consistently

\def\twelvepoint{\normalbaselineskip=12.4pt
  \abovedisplayskip 12.4pt plus 3pt minus 9pt
  \belowdisplayskip 12.4pt plus 3pt minus 9pt
  \abovedisplayshortskip 0pt plus 3pt
  \belowdisplayshortskip 7.2pt plus 3pt minus 4pt
  \smallskipamount=3.6pt plus1.2pt minus1.2pt
  \medskipamount=7.2pt plus2.4pt minus2.4pt
  \bigskipamount=14.4pt plus4.8pt minus4.8pt
  \def\rm{\fam0\twelverm}          \def\it{\fam\itfam\twelveit}%
  \def\sl{\fam\slfam\twelvesl}     \def\bf{\fam\bffam\twelvebf}%
  \def\mit{\fam 1}                 \def\cal{\fam 2}%
  \def\tt{\twelvett}
  \textfont0=\twelverm   \scriptfont0=\tenrm   \scriptscriptfont0=\sevenrm
  \textfont1=\twelvei    \scriptfont1=\teni    \scriptscriptfont1=\seveni
  \textfont2=\twelvesy   \scriptfont2=\tensy   \scriptscriptfont2=\sevensy
  \textfont3=\twelveex   \scriptfont3=\twelveex  \scriptscriptfont3=\twelveex
  \textfont\itfam=\twelveit
  \textfont\slfam=\twelvesl
  \textfont\bffam=\twelvebf \scriptfont\bffam=\tenbf
  \scriptscriptfont\bffam=\sevenbf
  \normalbaselines\rm}

%       tenpoint

%%
%%      Various internal macros
%%

\def\beginparmode{\endmode
  \begingroup \def\endmode{\par\endgroup}}
\let\endmode=\par
{\obeylines\gdef\
{}}
\def\singlespace{\baselineskip=\normalbaselineskip}

\def\doublespace{\baselineskip=\normalbaselineskip \multiply\baselineskip by 2}

\newcount\firstpageno
\firstpageno=2
\footline={\ifnum\pageno<\firstpageno{\hfil}\else{\hfil\twelverm\folio\hfil}\fi}
\let\rawfootnote=\footnote              % We must set the footnote style
\def\footnote#1#2{{\rm\singlespace\parindent=0pt\rawfootnote{#1}{#2}}}
\def\raggedcenter{\leftskip=4em plus 12em \rightskip=\leftskip
  \parindent=0pt \parfillskip=0pt \spaceskip=.3333em \xspaceskip=.5em
  \pretolerance=9999 \tolerance=9999
  \hyphenpenalty=9999 \exhyphenpenalty=9999 }
\def\dateline{\rightline{\ifcase\month\or
  January\or February\or March\or April\or May\or June\or
  July\or August\or September\or October\or November\or December\fi
  \space\number\year}}
\def\received{\vskip 3pt plus 0.2fill
 \centerline{\sl (Received\space\ifcase\month\or
  January\or February\or March\or April\or May\or June\or
  July\or August\or September\or October\or November\or December\fi
  \qquad, \number\year)}}

%%
%%      Page layout, margins, font and spacing (feel free to change)
%%

\hsize=6.5truein
\hoffset=.1truein
\vsize=8.9truein
\voffset=.05truein
\parskip=\medskipamount
\twelvepoint            % selects twelvepoint fonts (cf. \tenpoint)
\doublespace            % selects double spacing for main part of paper (cf.
                        %       \singlespace, \oneandahalfspace)
\overfullrule=0pt       % delete the nasty little black boxes for overfull box

\def\preprintno#1{
 \rightline{\rm #1}}    % Preprint number at upper right of title page

\def\head#1{                    % Head;  NOTE enclose the text in {}
  \filbreak\vskip 0.5truein     %  e.g., \head{I. Introduction}
  {\immediate\write16{#1}
   \raggedcenter \uppercase{#1}\par}
   \nobreak\vskip 0.25truein\nobreak}

\def\references                 % Begin references -- basic format is Phys Rev
  {\head{References}            % I.e., volume, page, year (space after
%%commas).
   \beginparmode
   \frenchspacing \parindent=0pt \leftskip=1truecm
   \parskip=8pt plus 3pt \everypar{\hangindent=\parindent}}

\def\frac#1#2{{\textstyle{#1 \over #2}}}
\def\square{\kern1pt\vbox{\hrule height 1.2pt\hbox{\vrule width 1.2pt\hskip 3pt
   \vbox{\vskip 6pt}\hskip 3pt\vrule width 0.6pt}\hrule height 0.6pt}\kern1pt}
%%%%%%%%%%%%%%%%%%%%%%%%%%%%%%%%%%%%%%%%%%%%%%%%%%%%%%%%%%%%%%%%%%%%%
%nashmac
\newcount\pagenumber
\newcount\questionnumber
\newcount\sectionnumber
\newcount\appendixnumber
\newcount\equationnumber
\newcount\referencenumber

\def\ifundefined#1{\expandafter\ifx\csname#1\endcsname\relax}
\def\docref#1{\ifundefined{#1} {\bf ?.?}\message{#1 not yet defined,}
\else \csname#1\endcsname \fi}

\newread\bib
\newcount\linecount
\newcount\citecount
\newcount\localauthorcount
\def\article{\def\eqlabel##1{\edef##1{\sectionlabel.\the\equationnumber}}
\def\seclabel##1{\edef##1{\sectionlabel}}
\def\feqlabel##1{\ifnum\passcount=1
\immediate\write\crossrefsout{\relax}  %this has to be written out purely to
% allow one to open the .ref file to see if it has zero bytes or not
\immediate\write\crossrefsout{\def\string##1{\sectionlabel.
\the\equationnumber}}\else \fi }
\def\fseclabel##1{\ifnum\passcount=1
\immediate\write\crossrefsout{\relax}   %this has to be written out purely to
% allow one to open the .ref file to see if it has zero bytes or not
\immediate\write\crossrefsout{\def\string##1{\sectionlabel}}\else\fi}
\def\cite##1{\immediate\openin\bib=bib.tex\global\citecount=##1
\global\linecount=0{\loop\ifnum\linecount<\citecount \read\bib
to\temp \global\advance\linecount by 1\repeat\temp}\immediate\closein\bib}
\def\docite##1 auth ##2 title ##3 jour ##4 vol ##5 pages ##6 year ##7{
\par\noindent\item{\bf\the\referencenumber .}
 ##2, ##3, ##4, {\bf ##5}, ##6,
(##7).\par\vskip-0.8\baselineskip\noindent{
\global\advance\referencenumber by1}}
\def\dobkcite##1 auth ##2 title ##3 publisher ##4 year ##5{
\par\noindent\item{\bf\the\referencenumber .}
 ##2, {\it ##3}, ##4, (##5).
\par\vskip-0.8\baselineskip\noindent{\global\advance\referencenumber by1}}
\def\doconfcite##1 auth ##2 title ##3 conftitle ##4 editor ##5 publisher ##6
year ##7{
\par\noindent\item{\bf\the\referencenumber .}
##2, {\it ##3}, ##4,  {edited by: ##5}, ##6, (##7).
\par\vskip-0.8\baselineskip\noindent{\global\advance\referencenumber by1}}}

\def\appendixlabel{\ifcase\appendixnumber\or A\or B\or C\or D\or E\or
F\or G\or H\or I\or J\or K\or L\or M\or N\or O\or P\or Q\or R\or S\or
T\or U\or V\or W\or X\or Y\or Z\fi}

\def\sectionlabel{\ifnum\appendixnumber>0 \appendixlabel
\else\the\sectionnumber\fi}

\def\beginsection #1
 {{\global\subsecnumber=1\global\appendixnumber=0\global\advance\sectionnumber
by1}\equationnumber=1
\par\vskip 0.8\baselineskip plus 0.8\baselineskip
 minus 0.8\baselineskip
\noindent$\S$ {\bf \sectionlabel. #1}
\par\penalty 10000\vskip 0.6\baselineskip plus 0.8\baselineskip
minus 0.6\baselineskip \noindent}

\newcount\subsecnumber
\global\subsecnumber=1

\def\subsec #1 {\bf\par\vskip8truept  minus 8truept
\noindent \ifnum\appendixnumber=0 $\S\S\;$\else\fi
$\bf\sectionlabel.\the\subsecnumber$ #1
\global\advance\subsecnumber by1
\rm\par\penalty 10000\vskip6truept  minus 6truept\noindent}

\def\beginappendix #1
{{\global\subsecnumber=1\global\advance\appendixnumber
by1}\equationnumber=1\par
\vskip 0.8\baselineskip plus 0.8\baselineskip
 minus 0.8\baselineskip
\noindent
{\bf Appendix \appendixlabel . #1}
\par\vskip 0.8\baselineskip plus 0.8\baselineskip
 minus 0.8\baselineskip
\noindent}

\def\no{\eqno({\rm\sectionlabel}
.\the\equationnumber){\global\advance\equationnumber by1}}

\def\beginref #1 {\par\vskip 2.4 pt\noindent\item{\bf\the\referencenumber .}
\noindent #1\par\vskip 2.4 pt\noindent{\global\advance\referencenumber by1}}

\def\ref #1{{\bf [#1]}}

\def\specialbar#1{\setbox1=\hbox{$\scriptstyle #1$}
\setbox2=\vbox{\hrule width 0.8\wd1}
\raise0.5\ht1\hbox{$\scriptstyle{\lower\dp1\box2}\atop\box1$}}
\def\specialbar#1{\setbox1=\hbox{$\scriptstyle #1$}
\setbox2=\vbox{\hrule width 0.8\wd1}
\vbox{\hbox{$\,$\lower\ht1\box2}\hbox{$\,$\raise\ht1\box1}}}

\article
%%%%%%%%%%%%%%%%%%%%%%%%%%%%%%%%%%%%%%%%%%%%%%%%%%%%%%%%%%%%%%%%%%
%defsnew
\def\i{\infty}  %overwrites plain tex defn
\def\L{\Lambda} %overwrites plain tex defn
\def\l{\lambda} %overwrites plain tex defn
   %overwrites plain tex defn
\def\t{\tau}    %overwrites plain tex defn
\def\o{\over}   %overwrites plain tex defn
\def\b{\beta}   %overwrites plain tex defn

\def\f{\phi}         \def\p{\partial}
\def\K{\chi^{-1}}    \def\G{\Gamma}   \def\k{\kappa}
\def\ra{\rightarrow}  \def\kl{\kappa L}       \def\lb{\bar\l}  \def\ft{\phi^2}
          \def\fb{\bar\f_B}
\def\e{\varepsilon}  
\def\gf{\gamma_{\f}}  \def\tb{\bar t}      \def\fbd{\fb'}
\def\ff{\phi^4}
\def\vf{\varphi}
\def\r{\rho}
\def\gft{\gamma_{\f^2}}  

\def\frac#1#2{{{#1}\over{#2}}}
\def\s{\scriptstyle}     \def\ss{\scriptscriptstyle}

\def\kc{\k_c}
         \def\rc{\r_c}
          
\def\T{\cal T}

\def\gl{\gamma_{\l}}

\def\nf{{(5N+22)\o9}}
\def\ns{{(N^2+6N+20)\o36}}
\def\nt#1{{(N+2)\over{#1}}}
\def\ne#1{{(N+8)\over{#1}}}
\def\io{\hbox{$\bigcirc\kern-2.3pt{\bf\cdot}$\kern 2.3pt}}
%tadpole
\def\bub{\hbox{${\bf\cdot}\kern-5pt\bigcirc\kern-5pt{\bf\cdot}$}}
%bubble
\def\itr{\hbox{${\bf\cdot}\kern-6pt\in\kern-2.5pt\ni\kern-6pt{\bf\cdot}\kern
2.5pt$}}
%setting sun
\def\itrp{\hbox{${\bf\cdot}\kern-6pt\in\kern-2.5pt\ni\kern-6pt{\bf\cdot}
\kern-10.5pt/$\kern 5pt}}
%wavefn ren
\def\if{\hbox{${\bf\cdot}\kern-3pt\langle\kern-2.4pt|\kern-2.2pt)$}}
%cone
\def\fe{\hbox{$\bigcirc$}}   %1loop free energy
%mass insert
\def\itro{\hbox{${\bf\cdot}\kern-5pt\bigcirc\kern-5pt{\bf :}$}}

\def\fb{\bar\phi}

\def\Zf{Z_{\phi}}
\def\Zft{Z_{\phi^2}}

\def\Zl{Z_{\l}}

\def\cgt{\xi_{\ss gt}}
\def\cgf{\xi_{\ss g\fb}}

\def\npw{{p^2\o\k^2}=\a_{1},\t=\a_{2}}
\def\npl{SP=\a_{1},\t=\a_{2}}

\def\Ait{A_{4-d}\bub}
\def\Aif{A_{2(4-d)}\left(\if-{1\o2}\bub^2\right)}
\def\Aitrp{A_{2(4-d)}\itrp}
\def\ao{-\ne{6}A_{4-d}\bub}
\def\at{\nf A_{2(4-d)}\left(\if-{1\o2}\bub^2\right)+\nt{9}A_{2(4-d)}\itrp}

\def\el{\e(\bub)}
\def\ftwo{{\Aitrp\o{(\Ait)}^2}}
\def\fone{{\Aif\o{(\Ait)}^2}}
\def\nkol{{\omega(n)\o\k^2L^2}}

\def\omokl#1{{\omega(n_{#1})\o\k^2L^2}}
\def\dq{{d^{d-1}q\o(2\pi)^{d-1}}}
\def\tpn#1{\t+{\omokl{#1}}}
\def\tpnx{\t+{{\omega(n_1,n_2,x)\o\k^2 L^2}}}
\def\oxy{\omega(n_1,n_2,x,y)}

\def\tauxy#1#2{\tau r(x,y)+{{\omega(n_{1},n_{2},x,y)\o\k^2L^2}}}
\def\a{\alpha}

\def\cF{{\cal F}}

\def\de{d_{\ss eff}}
\def\dea{\de^{\ast}}
\def\ef{_{\ss eff}}
%%%%%%%%%%%%%%%%%%%%%%%%%%%%%%%%%%%%%%%%%%%%%%%%%%%%%%%%%%%%%%%%
%The paper proper
\singlespace
\font\bigg=cmbx10 at 17.3 truept
\preprintno{THU-93/14, DIAS-STP-93-19;\quad August '93}
\vskip 0.5truein
\centerline{\bigg ``ENVIRONMENTALLY FRIENDLY''}
\vskip\baselineskip
\centerline{\bigg RENORMALIZATION }
\vskip\baselineskip
\centerline{\bf Denjoe O' Connor}
\centerline{School of Theoretical Physics,}
\centerline{Dublin Institute for Advanced Studies,}
\centerline{10 Burlington Road,}
\centerline{Dublin 4, Ireland.}
\vskip\baselineskip
\centerline{\bf C.R. Stephens }
\centerline{Institute for Theoretical Physics,}
\centerline{Rijksuniversiteit Utrecht, }
\centerline{Princetonplein 5, }
\centerline{3508TA Utrecht, Netherlands.}
\vskip 1truein
{{\bf Abstract:}
We analyze the renormalization of systems whose effective degrees of
freedom are described in terms of fluctuations which are
``environment'' dependent. Relevant environmental parameters considered
are: temperature, system size, boundary conditions, and external
fields. The points in the space of \lq\lq coupling constants'' at which
such systems exhibit scale invariance coincide only with the fixed
points of a global renormalization group which is necessarily
environment dependent. Using such a renormalization group we give
formal expressions to two loops for effective critical exponents for a
generic crossover induced by a relevant mass scale $g$.  These
effective exponents are seen to obey scaling laws across the entire
crossover, including hyperscaling, but in terms of an effective
dimensionality, $d\ef=4-\gl$, which represents the effects of the
leading irrelevant operator.  We analyze the crossover of an $O(N)$
model on a $d$ dimensional layered geometry with periodic, antiperiodic
and Dirichlet boundary conditions. Explicit results to two loops for
effective exponents are obtained using a [2,1] Pad\'e resummed
coupling, for: the ``Gaussian model'' ($N=-2$), spherical model
($N=\infty$), Ising Model ($N=1$), polymers ($N=0$), XY-model ($N=2$)
and Heisenberg ($N=3$) models in four dimensions. We also give two loop
Pad\'e resummed results for a three dimensional Ising ferromagnet in a
transverse magnetic field and corresponding one loop results for the
two dimensional model. One loop results are also presented for a three
dimensional layered Ising model with Dirichlet and antiperiodic
boundary conditions.  Asymptotically the effective exponents are in
excellent agreement with known results. }
\vfill\eject

\beginsection{\bf INTRODUCTION}
Field theory is our most powerful tool
for describing systems with a large number of degrees of freedom, and
in particular in situations where the latter can act collectively. In
one guise or another such systems comprise a large proportion of
current physics. Even though one starts at a certain energy scale with
a ``fundamental'' set of degrees of freedom, which may be very
elementary looking, e.g. QED, one knows that as one progresses to
different (usually lower) energy scales, the original elementary degrees
of freedom, though in principle offering a complete description of the
low energy physics, become, in practice, unusable.

For example, one would be rather perverse to try and describe a phase
transition in a ferromagnet in terms of QED. As is well known the
transition is much better described in terms of a
Landau-Ginzburg-Wilson Hamiltonian (LGW Hamiltonian) based on effective
degrees of freedom associated with a magnetization density $\f(x)$.
This is an implementation of the philosophy of effective field theory,
whereby the theory provides a good description only at energy scales
where the effective degrees of freedom chosen offer a reasonably
faithful representation of the physics.  For instance, if we heated up
the ferromagnet until it became a plasma of nuclei and electrons, then
a description in terms of a LGW effective Hamiltonian would be rather
inappropriate. The difficulty, of course, is that the effective degrees
of freedom of the system are scale dependent, i.e. there is a
``crossover'' in them, and consequently in the physical behaviour of
the system. Thus, if one requires a complete description of the physics
at all scales one must be able to account for this fact. Our aim in
this paper will be to develop further a formalism for describing such
systems using field theoretic renormalization groups (RG). We will for
the most part keep the formalism as general as possible, specializing
in the latter half of the paper to specific crossovers of interest. Our
reason for doing this is that the formalism is applicable to a very
wide class of crossovers. By presenting the general results the reader
is then at liberty to treat a crossover of particular interest to
him/her merely by inserting into the appropriate parts of the formalism
the particular details specific to that crossover. We will
use a language that is hopefully accessible to people with some field
theoretic background in both particle physics and statistical physics
as the applications of the formalism are equally applicable to both
areas.

To illustrate the ubiquity of the concept of crossover behaviour we will
briefly mention (without detailed explanation) some physically
pertinent examples. In QCD, at high energies, a description in terms of
quark-gluon degrees of freedom is appropriate, whilst at low energies
baryon-meson degrees of freedom are more suitable. In four
dimensional finite temperature field theory, for $T\gg m(T)$, where
$m(T)$ is a typical finite temperature mass scale, three dimensional
degrees of freedom are appropriate, whereas for $T\ll m(T)$ they are four
dimensional. In the early universe the effective degrees of freedom are
time dependent due to the cosmological expansion. In fact, for quantum
field theory in curved spacetime where one is concerned with ``in'' and
``out'' vacua, the difference between these vacua, which results in
particle production, can be thought of as being representative of the
difference between ``in'' and ``out'' effective degrees of freedom. In
critical phenomena crossover can be induced by many different effects.
One exciting much interest currently is the effect of randomness, e.g.
a random magnetic field, random impurities etc. In such cases the
effective degrees of freedom depend on  the concentration of the
impurities or the strength of the magnetic field.  Another eliciting
great interest concerns the effects of finite size. This is of
relevance in lattice simulations and also in real experimental
systems.

The physics of crossovers then, concerns systems which exhibit
qualitatively different degrees of freedom at different scales. Our
goal is to qualitatively and quantitatively describe such systems. For an
exact model of course there is no problem, however, even though one has
an exact answer that does not necessarily imply that one has a good
intuitive understanding of the physics. Often with an exact model one
can not see the wood from the trees. Generically one must resort to
approximation techniques. The dominant one, of course, is perturbation
theory, where one must answer the crucial question: what parameter
should one perturb in? Often the difference between getting a
nonsensical versus a reliable answer comes down to being able to pick
and work with a good expansion parameter.

In crossover problems one starts with some interacting Hamiltonian
which will come with a coupling constant, $\l$, which one would be
tempted to use as an expansion parameter when computing the correlation
functions associated with this Hamiltonian. However, the starting
``bare'' (microscopic) Hamiltonian will only offer an accurate
perturbative description when one looks at scales $\k\sim\L$, $\L$
being the UV cutoff (inverse lattice spacing) for the theory.  At
scales $\k\ll\L$, as is well known, the bare parameters offer a
perturbatively useless description. One must therefore perform a
renormalization to a new set of parameters. This is done by defining
correlation functions at a fiducial scale $\k'$, and using them as the
parameters with respect to which one describes physics at a scale
$\k\neq\k'$. As long as the two scales $\k$ and $\k'$ are not too
different, the description of physics at the scale $\k$ in terms of
parameters defined at the scale $\k'$ will be perturbatively reliable.
The crossover from bare degrees of freedom to renormalized ones,
associated with taking the continuum limit or the removal of a
regularizing UV cutoff, is perhaps the most elementary example of a
crossover. In contrast to a scale invariant system where the bare
degrees of freedom and the renormalized ones are relatively simply
related they are quite different when the system is not scale
invariant.  In the presence of a cutoff the bare and renormalized
degrees of freedom are always different as the system cannot be scale
invariant.

So why are the bare and renormalized effective degrees of freedom so
different? Due to the effects of fluctuations; the fluctuations
``dress'' the bare parameters. This dressing in principle can become
infinite if one takes the $\L\ra\i$ limit. For us the crucial
difference is between a large dressing and a small dressing not between
an infinite one and a finite one. Moreover, as far as the RG is
concerned, we feel that there is no particular virtue in emphasizing UV
dressings as opposed to IR ones. In particle physics it is the
former which have received most attention, however, the machinery of
the RG must be invoked any time one has large dressings of one's
parameters, irrespective of from which particular end of the spectrum
the dominant fluctuations arise. Large dressings imply that one is not
tracking the effective degrees of freedom in the problem very well.
Renormalization is a methodology by which one can use the freedom to
reparametrize one's theory to find parameters which track the
effective degrees of freedom more accurately. In the case of bare versus
renormalized parameters, without interactions the bare parameters
cannot get dressed and remain the same. Indeed, it is the presence of
interactions that induces the crossover between them.  The
renormalization carried out was dependent on the interaction strength
$\l$. That is why the renormalization worked in the first place,
because it depended on the parameter that was inducing the crossover.
We could have been perverse and tried to renormalize the sytem in a
$\l$ independent way. The Gaussian theory would have a large dressing
associated with the specific heat, or vacuum energy, for instance. This
would have to be renormalized. After the renormalizations associated
with the Gaussian theory were performed one would still find that
connected correlation functions had large dressings. This is a
reflection of the fact that the renormalization process, using only
counterterms appropriate for the Gaussian theory, was not sufficient to
make sense of the interacting theory.

Generically, in the renormalization process one takes fluctuations
between scales $\L$ and $\k'$, contributing to a given correlation
function, and absorbs them into a redefinition of a parameter, which is
usually a vertex function at the scale $\k'$. To investigate physics at
a scale $\k$, one considers the fluctuations between $\k$ and $\k'$,
explicitly in perturbation theory, in terms of the coupling $\l$
defined at the scale $\k'$ instead of the bare coupling. The important
question of course is what fluctuations are being absorbed into
redefinitions of the parameters?  If we renormalized using only
counterterms associated with the Gaussian theory then only
non-interacting fluctuations are being absorbed into redefinitions. For
the connected correlation functions this is equivalent to not having
absorbed in anything. Obviously this won't help much. In the standard
$\l$ dependent case all the fluctuations associated with interacting
effective degrees of freedom between the scales $\L$ and $\k'$ have
been absorbed. What we do know for a fact is that the effective degrees
of freedom in the system are interacting, therefore one would expect to
get more sense out of a renormalization procedure that explicitly takes
account of this fact. The reader might wonder why we have gone to such
lengths analyzing such a standard procedure, and why on earth one would
ever try to renormalize an interacting theory using renormalization
that was explicitly $\l$ independent. We hope by the end of the paper
the reason will be clear. For the moment we will simply say that a
successful renormalization is one that can take proper account of what
the true effective degrees of freedom of a system are.  In the above
they are $\l$ dependent, therefore a good renormalization should be
$\l$ dependent.

So, an important aspect of renormalization is that it is a methodology
whereby through a suitable redefinition of parameters one can make
perturbative sense out of something that originally was perturbative
nonsense. How sensible the expansion actually becomes, however, is
crucially dependent on what physics the renormalization process can
capture. In the above the renormalization had to capture the fact that
the effective degrees of freedom were interacting. If one considered an
interacting field theory in a three dimensional box of size $L$, one
could renormalize the theory in an $L$ independent fashion. When one
considered physics on scales $\k\sim L^{-1}$ one would find that the
theory was perturbatively ill defined, whereas an appropriate $L$
dependent renormalization made perturbative sense. The reason for this,
of course, is that the effective degrees of freedom in the system are
explicitly $L$ dependent. An $L$ independent renormalization ignores
this important physical fact. The only fluctuations being absorbed into
the renormalized parameters in this case are $L$ independent, no matter
what renormalization scale one chooses. $L$ here is the parameter which
induces the crossover and therefore a good renormalization scheme
should be $L$ dependent.

More generally we will speak of a crossover being induced by a
particular ``environmental'' variable. The idea to understand here is
that the fluctuations in a system ``feel'' out the environment they are
in. For instance in the above case of physics in a box, the size of the
box $L$ is a relevant environmental variable, as clearly effective
degrees of freedom associated with length scales $\ll L$ and $\sim L$
are completely different. The effective degrees of freedom are thus
sensitive to the environment. There are many, many different
environmental parameters describing all sorts of crossover systems.  In
this paper we will consider the effects of a generic environmental
variable $g$, which could represent variously: finite size,
temperature, spin anisotropy, long range interactions, cosmological
constant, Hubble parameter, magnetization, electric/magnetic field and
so on. What we will be showing is how to describe qualitatively and
quantitatively the response of a system to changes in the environment.
We will show that in order to obtain a perturbative description of such
systems one must implement an RG which is explicitly dependent on the
environment. We will call such RGs ``environmentally friendly''. If an
environment ``unfriendly'', that is to say, environment independent, RG
is implemented, we will see that this generically leads to breakdown of
perturbation theory, wherein the theory becomes strongly coupled in
terms of the effective degrees of freedom characteristic of the
environmentally unfriendly RG, and that large perturbative dressings
appear which can actually become divergent.

Crossover behaviour, as mentioned, is quite ubiquitous. It is not our
intention here to give a review of what has been done, consequently we
will be rather selective in our comments, and, more apologetically, in
our references. In critical phenomena it has recieved much more
attention than in particle physics (for an early review see
\ref{1}), though many problems in particle physics are
essentially crossover problems. In principle, basically any laboratory
system will exhibit crossover behaviour in some regime. Some of the
more experimentally accessible ones are: uniaxial dipolar ferromagnets
\ref{2}, systems exhibiting a bicritical point \ref{3},
dimensional crossover in liquid ${\rm He}^4$ \ref{4}, quantum
ferromagnets (\ref{5} and references therein), bulk/surface
crossovers \ref{6}. Dimensional crossover, one of the chief concerns
of this paper, has been studied mainly in the context of finite size
scaling \ref{7}. Much work has been done on the latter, and finite
size effects in general, in the context of lattice simulations; quantum
Monte Carlo methods \ref{8} are also closely related.

{}From a more theoretical standpoint, and more particularly from the
point of view of RG theory, much work has also been done. For a typical
crossover induced by a generic anisotropy parameter $g$, scaling
formulations have been almost invariably based on RG's that are $g$
independent \ref{9}. This automatically leads to critical
and crossover exponents that are defined with respect to the isotropic
$g=0$ fixed point. Such RG's would be incapable, in and of themselves,
of bridging the crossover to the anisotropic $g=\i$ fixed point. What
we mean by this is that the fixed points of such RG's will not yield
all the possible points of scale invariance of the system, in
particular the anisotropic scale invariant system will be
inaccessible.  The fundamental length scale utilized is $\xi_0$, the
correlation length of the isotropic system. It is pertinent to note
that this is not the physical correlation length in the crossover
system.  The desire to make accessible another fixed point besides the
isotropic one has often entailed the matching of asymptotic expansions
around the anisotropic and isotropic fixed points, or the use of high
temperature expansions in conjunction with an ansatz for the scaling
function \ref{10} \ref{11}. Exact models have also played
a role, for instance, the two dimensional Ising model \ref{12}
and the spherical model \ref{13} in the context of finite size
scaling.

An RG approach, due to Riedel and Wegner \ref{14}, and used by
others, utilizes ``model recursion relations''. Here it is assumed
known what the two fixed points of the crossover are, subsequently RG
equations for the scaling fields are postulated that interpolate
between the two fixed points. As there is no reference to the
underlying microscopic Hamiltonian such a ploy can only offer
qualitative information. In addition there are many different crossover
systems that exhibit crossover between exactly the same two fixed
points. Universality implies that two systems that lie in the same
universality class can via a suitable rescaling of variables be shown
to have exactly the same IR properties, e.g. a simple fluid and an
Ising ferromagnet.  For crossovers the asymptotic fixed points do not
determine ``crossover universality classes''. The scaling functions
associated with different crossovers between the same two fixed points
cannot generally be transformed into one another by a change of scaling
variable. However, there is a concept of two systems being in the same
crossover universality class --- a $(d-1)$-dimensional Ising model in a
transverse magnetic field and a $d$-dimensional layered Ising model
with periodic boundary conditions being a case in point. Wilsonian RGs
of the ``momentum shell integration'' type have been frequently
employed with varying success. These methods generally have two main
defects: ease of extension beyond first order, and using too rough an
approximation in the shell integration. Given that many physically
different systems can crossover between the same asymptotic fixed
points it is important to be able to pick out the details of the
crossover curves in order to distinguish between them. By using an
approximation on the momentum shell integration it is easy to blur
such distinctions.

{}From a field theory point of view much less work has been done. This is
mainly because of the preoccupation with renormalization in the context
of UV divergences, though there are one or two notable exceptions. Amit
and Goldschmidt \ref{15} introduced the concept of generalized
minimal subtraction (GMS) in the context of crossover at a bicritical
point. Their results for $\gamma\ef$, however, differ significantly
from those found by Nelson and Domany \ref{16} and independently by
Seglar and Fisher \ref{17} using momentum shell integration, in
particular in that the latter find a characteristic ``dip'' in the
curves. In this case the results of our methodology applied to a
bicritical crossover \ref{18} agree with the latter. We believe
this to be due to a deficiency of GMS which fails to capture the true
behaviour of the leading irrelevant operator which in our crossover
formalism plays a crucial role. GMS was also applied to uniaxial
dipolar ferromagnets in \ref{19}, once again the results of our
analysis \ref{20}\ref{18} are somewhat different.
Schmeltzer \ref{21} calculated $\gamma\ef$  to one loop
for three dimensional quantum ferroelectrics.
Lawrie \ref{22} considered dimensional crossover for
$d$-dimensional quantal and $d+1$ dimensional finite-sized Ising models
for $3<d<4$. Unlike our methods the $\varepsilon$ expansion methodology
he used could not capture the crossover between two non-trivial fixed
points as such an expansion is around the upper critical dimension
which changes across the crossover. Nemirovsky and Freed (\ref{23}
and references therein) used minimal subtraction techniques but failed
to be able to access the full crossover. Field theoretic results for
dimensional crossover in a fully finite geometry or a cylinder have
been obtained \ref{24} but the techniques used have not been
extended to the case of a system with more than one fixed point.  In a
particle physics context some interesting work has been done in the
context of the finite temperature RG \ref{25} (see \ref{26} for
some recent applications of our work to finite temperature field
theory).

In this paper we will present a field theoretic formalism which is
applicable to a very wide class of crossovers associated with a field
theory in a particular environment. We have previously presented some
one loop results in the case of dimensional crossover, where the
environmental variable was $L$, the finite size of the system, both in
the case of above \ref{27} \ref{28} and below \ref{29} the critical
point. In the latter the background magnetization was also a relevant
environmental variable inducing a crossover. We have also considered,
at one loop, uniaxial dipolar ferromagnets and bicritical systems in
\ref{20} and \ref{18}, where the environmental variables were $\a_0$
and $m$ respectively. Here $\a_0$ is the dipole-dipole coupling
strength and $m$ is the mass of the $O(N-M)$ components for the case
$O(N)\ra O(M)$. The main thrust of the present paper is to generalize
the techniques used to a generic environment induced crossover, and in
particular, to extend the methods beyond one loop as there are many
subtleties involved in doing so, not least of which is the application
of resummation techniques. We present two loop expressions for
effective exponents and some scaling functions, which are set up in
just such a fashion that the reader interested in a specific crossover,
may take them, insert the relevant environment dependence in the
Feynman diagrams, crank the handle (analytically if possible,
numerically if not), and spit out the answers. To explicitly
demonstrate the efficacy of the methodology we consider dimensional
crossover for a $d$ dimensional $O(N)$ model on a layered geometry with
periodic, Dirichlet or antiperiodic boundary conditions. We give
explicit, surprisingly simple, expressions for two loop effective
exponents for a four dimensional layered system with periodic boundary
conditions, and corresponding (not so simple) expressions for a three
dimensional layered system. We also give results for Dirichlet and
periodic boundary conditions. Two loop results for the effective
exponents of a three dimensional Ising model in a transverse magnetic
field are presented, as are one loop results for the corresponding two
dimensional model. Graphical displays of our results can be found
throughout the paper.

The format of the paper will be as follows: in section 2 we consider
some formal aspects of renormalization, mainly to set notation. In
section 3 we consider the relationship between the RG and the
environment. Specifically, we compare and contrast the field theoretic
and ``Wilsonian'' RGs asking and answering the question of what does
one require of a ``good'' RG, and in particular addressing which RGs
will access all the points of scale invariance of a physical system. We
illustrate these ideas by comparing massless versus massive
renormalization showing that the former yields an RG which has fewer
fixed points than the latter. We discuss precisely why a good RG should
be environmentally friendly.  In sections 4  we derive formal two loop
expressions for the Wilson functions of our generic $g$ dependent
crossover. In section 5 we discuss further the idea of an
environmentally friendly versus environmentally unfriendly RG, showing
that the former offers a more global description of the crossover
sytem, the latter being incapable of accessing some of the points of
scale invariance of the system. We also discuss an RG where the
environment itself ``runs'',  which is useful for addressing  questions
such as: what is the shift in critical temperature due to changes in
the environment?  In section 6 we consider the consequences of an
environmentally friendly RG, defining effective critical exponents and
showing that they obey scaling laws, including effective hyperscaling
with respect to an effective dimensionality which is a measure of the
importance of the leading irrelevant operator across the crossover. We
exhibit formal scaling forms for the crossover equation of state and
crossover coexistence curve in terms of effective exponents. We exhibit
non-linear scaling fields which interpolate across the crossover
becoming at the endpoints of the crossover the associated linear
scaling fields. We discuss the concepts of crossover universality and
crossover universality class. In section 7 we address the important
question of ``what should we perturb in?'' We argue that an appropriate
perturbation parameter is the resummed solution of the $\beta$ function
of the environmentally friendly RG. We emphasize that in the context of
the RG, perturbation theory is carried out at the level of the Wilson
functions, all of which as perturbative expansions should be resummed.
We resum these series using a [2,1] Pad\'e approximant which is
appropriate at the two loop level.  When the Wilson functions are
integrated (exponentiated) the resulting functions should not then be
themselves expanded. We show that an environmentally friendly RG is
most potent when {\bf all} environment dependent fluctuations are
absorbed into reparametrizations of the parameters.  In section 8 we
discuss the floating fixed point which yields a good approximation to
crossovers via the solution of the crossover $\beta$ function as an
algebraic equation instead of as a differential equation. In section 9
we present two loop diagramatic representations for effective exponents
and scaling fields. These can be treated as ``black box'' type
expressions where one inserts the particular environmental dependence
in the Feynman diagrams and reads off the corresponding exponent. In
section 10 we consider one and two loop results for an $O(N)$ model on
a layered geometry with periodic boundary conditions paying special
attention to some models of interest. We also discuss the effects of
boundary conditions on the crossover. In section 11 we consider the
quantum/classical crossover in a $d$ dimensional Ising model in a
transverse magnetic field. We present both one and two loop results. In
section 12 we draw some conclusions and make some speculations. There
are two appendices containing various expressions for diagrams needed
in the calculations.

\beginsection{\bf FORMAL RENORMALIZATION}
In this section we present a
formal approach to the problem of renormalization.  We begin with the
action (LGW Hamiltonian) \eqlabel{\action} $$S[\vf_{B}]=\int
d^{d}x\left[{1\o2}(\p\vf_{B})^2+{1\o2}m^2_{B}\vf_{B}^2+
{1\o2}t_{B}(x)\vf_{B}^2
+{\l_{B}\o4!}\vf_{B}^{4}-H_B(x)\vf_B+g_BO_B\right]\no$$ which
represents the ``microscopic'' theory.  The term $g_{B}O_{B}$ is used
to abstractly refer to an ``anisotropy'' in the system, $O_{B}$ being
an operator conjugate to the coupling $g_{B}$. We will think of
$g_{B}=0$ as representing an isotropic system. Physically this
anisotropy could be of diverse origin.  The general methodology
espoused in this paper is not confined to sytems described by the
Hamiltonian (\docref{action}), though the latter is general enough for
most of our considerations. From time to time, however, we will extend
our attention to a general Hamiltonian who's parameters live on an
abstract ``space of coupling constants'' $\cal M$. Each of the
parameters in (\docref{action}) is associated with a coordinate
direction in this space.
We will for convenience tend to use the terminology
associated with viewing (\docref{action}) as the Hamiltonian of a
magnetic system.

The partition function $Z$ is defined by the functional integral
$$Z=e^{-W[J_{B}+H_B,t_{B}+m_B^2]}=\int[d\vf]e^{-S[\vf_{B}]+\int
d^{d}xJ_{B}(x)\vf_{B}(x)}\no$$ where $W$ is the generating functional
of connected Greens functions (free energy for prescribed sources).
$J_B$ is to be understood here as some formal source, differentiation
with respect to which generates the connected Greens functions
(correlation functions), and which will after generation of the
correlation functions be put to zero. $H_B$ will represent a real
magnetic field in the critical phenomena context. The effective action
(free energy for prescribed magnetization), $\G[\phi_B,t_B+m_B^2]$, is
the generating functional of one particle irreducible $N$-point vertex
functions and is given by the Legendre transform \eqlabel{\effaction}
$$\G[\f_{B},t_{B}+m^2_{B}]=W[J_{B}+H_B,t_{B}]+\int d^{d}x
(J_B(x)+H_B(x))\f_{B}(x)\no$$ where $\f_B=<\vf_B>_{\ss
J,H}=\fb_B+<\vf_B>_{\ss J}$, $\fb_B$ being the magnetization when
$J=0$. The vertex functions are then defined in terms of functional
derivatives of $\G[\f_{B},t_{B}+m_B^2]$ with respect to $\f_{B}$ and
$t_{B}$.  Specifically $$\G[\f_{B},t_{B}+m_B^2]=
\sum_{N=0}^{\infty}{1\o
N!}\G_{B}^{(N)}(\fb_B,t_B+m_B^2)(\f_B-\fb_{B})^{N}$$
$$\G_B^{(N)}(\fb_B,t_B+m_B^2)= \sum_{M=0}^{\i}{1\o
M!}\G_{B}^{(N,M)}(\fb_B,m_B^2)t_B^M$$ where implicit multiple
integrations over position or momenta are understood.

Note that $\fb_{B}=0$ above the critical temperature when $H_{B}=0$,
and hence the quantities $\G^{(N)}$ are independent of $\fb_{B}$ in
this domain. For $H_B\neq0$, or below the critical temperature, there
is a non-zero spontaneous magnetization $\fb_B$. Deviations from this
magnetization are responses to the externally applied source $J_{B}$,
they again will be set to zero once the vertex functions are obtained.
The equation of state is given by \eqlabel{\gob}
$$\G_{B}^{(1)}=H_B\no$$ when $J_B=0$.  This is the basic equation of
the effective action and serves to determine the background
magnetization $\fb_{B}$.

The calculation of correlation functions more often than not requires
the introduction of an ultraviolet (UV) cutoff (physical or
otherwise).  An infrared (IR) cutoff is frequently also necessary.  We
renormalize by imposing normalization conditions, though for the moment
we will be somewhat nebulous about the precise normalization point.
Two quantities required in this procedure are the wave function
renormalization constant $Z_{\f}$, where $\f=Z_{\f}^{-{1\o2}}\f_{B}$,
and the composite operator renormalization constant $Z_{\ft}$.  The
relation between the bare and renormalized vertex functions is
\eqlabel{\grdef} $$\G_{B}^{(N,M)}=\Zf^{-{N\o2}}\Zft^{-M}\G^{(N,M)}\no$$
our convention being that the unsubscripted quantity is the
renormalized one.  The parameters of the theory $t_{B}$ and $\l_{B}$
are multiplicatively renormalized, the relation between the bare and
renormalized quantities being $$t=\Zft^{-1}t_{B}\qquad
\l=\Zl\l_{B}\no$$ $m_B^2$ must be additively renormalized,
$m^2=m_B^2+\delta m_B^2$.

The renormalization factors are determined by the following generic
normalization conditions \eqlabel{\normt}
$$\left.\G^{(2)}\right|_{\npw} =(\a_{1}+\a_{2})\k^2\no$$
\eqlabel{\normzf} $$\left.{\p\G^{(2)}\o\p p^2}\right|_{\npw}=1\no$$
\eqlabel{\normzft} $$\left.\G^{(2,1)}\right.|_{\npl}=1\no$$
\eqlabel{\normzl} $$\left.\G^{(4)}\right|_{\npl}=\l\no$$ where
$\tau={t\over\k^2}$,  and $SP$ denotes $p_i\cdot
p_j={\k^2\over4}(4\delta_{ij}-1)$, $p_{i}$ being the momentum entering
through the $j$th external leg, and $\k$ is an arbitrary
renormalization scale.  For the moment we will not specify the values
of the various parameters involved in the normalization conditions.
The alert reader might wonder why there is no condition corresponding
to a renormalization of $g_B$.  The answer is that in this paper we
take it to be a RG invariant (i.e. a non-linear scaling field in the
sense of Wegner \ref{30}), though this in no way restricts the
generality of our approach. We define the correlation length via the
relation \eqlabel{\corlngt}
$$\xi^2={\int d^{d}x{ x}^2 G^{(2)}(x,0)
\over
 2d \int d^{d}x G^{(2)}(x,0) }\no$$ where
$G^{(2)}(x,0)=\left.{{\delta^2W\o\delta J(x)\delta
J(0)}}\right|_{J=0}$.
The conditions (\docref{normt}) and
(\docref{normzf}), together with $\a_{1}=0$, imply that $\xi$ defined
by (\docref{corlngt}) is $t^{-\frac12}$ at the normalization point.

The wavefunction renormalization constant $\Zf$ is obtained from
(\docref{normzf}) by observing that since $\Zf$ is independent of
$p$,
and $\G_{B}^{(2)}=\Zf^{-1}\G^{(2)}$, then
$$\left.\p_{p^2}\G^{(2)}_{B}\right|_{\npw}=\Zf^{-1}\no$$ Similarly
(\docref{normzft}) and (\docref{grdef}) imply that
$$\left.\Zft^{-1}=\Zf\G^{(2,1)}_{B}\right|_{\npl}\no$$ whilst
(\docref{normzl}) and (\docref{grdef}) imply that
$$\left.\Zl=\Zf^2{\G^{(4)}_{B}\o\l_{B}}\right|_{\npl}\no$$

The RG
equation  can be viewed as a simple consequence of the fact that the
bare theory is independent of the arbitrary renormalization scale $\k$
at which we choose to define our parameters.
Thus \eqlabel{\gronb}
$$\k{d\o d\k}\G^{(N)}_{B}=0\no$$
Using the relation between the bare
and renormalized vertex functions (\docref{gronb}) becomes
\eqlabel{\rgrn}
$$\k{d\G^{(N)}\o d\k}-{N\o2}\gf\G^{(N)}=0\no$$ where
\eqlabel{\gamf}
$$\gf={1\o\Zf}\k{d\Zf\o d\k}\no$$
Equation
(\docref{rgrn}) has the formal solution
$$\G^{(N)}(\k)
=e^{-{N\o2}\int_{\k_{0}}^{\k}\gf(x){dx\o x}}\G^{(N)}(\k_0)\no$$
We can
further expand (\docref{rgrn}) by noting that there is implicit
dependence on $\k$ through the renormalized couplings, and possible
explicit dependence also. The differential equation then becomes
\eqlabel{\rge}
$$\left(\k{\p\o{\p\k}}+ \beta{\partial\over{\partial
\l}}+ \gamma_{\f^2}t{\partial\over {\partial t}} -{1\over2}{\gamma^{
}_{\f}}{\Bigl[} N+\fb{\partial\over{\partial \fb}}{\Bigl]}\right)
\Gamma^{\scriptscriptstyle{(N)}}=0\no$$
with \eqlabel{\gamft}
$$\gft={1\o\Zft^{-1}}\k {d\Zft^{-1}\o d\k}\no$$ \eqlabel{\gaml}
and
$$\b(\l)=\k{d\l\o d\k}\no$$
where $${\b(\l)\o\l}=\gl={1\o\Zl}\k{d\Zl\o
d\k}\no$$
$\l$ being the dimensionful coupling constant.  The functions
$\gf$, $\gft$ and $\gl$ are the Wilson functions.   Note that a term
$\p\o\p g$ cannot appear due to the assumed non-renormalization of $g$.
If we had chosen to parametrize the anisotropy by a coupling other than
a non-linear scaling field,  however, (\docref{rge}) would then be
augmented by a term $\gamma_{g}g{\p\o\p g}$.  Equation (\docref{rge})
in conjunction with the dimensional analysis equation \eqlabel{\diman}
$$\rho{d\G^{(N)}\o d\rho}={N({d\o2}-1)}\G^{(N)}\no$$ can be used to
relate the correlation functions at two different scales.

\beginsection{\bf RENORMALIZATION GROUP AND THE ``ENVIRONMENT''}
\subsec{\bf Comparison of Field Theoretic and ``Wilsonian'' RG's}
Before proceeding further let us consider the meaning of the RG
equation (\docref{rge}). The RG,  in the field theoretic context
\ref{31}, expresses the invariance of physical quantities under a
change of parameter. More generally it  can be viewed as a
reparametrization invariance of a differential equation and its
boundary conditions \ref{32}.  It is an exact symmetry  and
therefore is a
true invariance group of field theory.  It is not unlike other
reparametrization invariances. Viewed passively it expresses a
triviality. Its power is that it can be used in conjunction with
(\docref{diman}) to express a relationship between a parametrized
system at two different scales. In contrast the RG transformation
developed by Wilson and Kadanoff \ref{33} is a mapping between
probability distributions \ref{34}, realized as a mapping between
renormalized Hamiltonians, and implemented by a ``coarse graining''
procedure such as momentum shell integration \ref{35}. This
Wilsonian RG is in fact a semi-group and only asymptotically represents
an invariance of the system.  That the two different approaches have
led to essentially the same conclusions is a highly non-trivial
result.  The advantage of the Wilsonian methodology is that it is
conceptually very clear, whilst the advantage of the field theoretic
approach is that it is much easier to implement systematically, and
perturbatively. The general motivation for using an RG approach is as a
tool in understanding how systems with many degrees of freedom behave
differently at different scales, one of the most fundamental facts
about our universe being that it is full of ``scales''.

In the Wilson approach it is well known that there are many different
types of RG, e.g. momentum shell integration, block spinning, majority
rule etc.  (see \ref{36} for a recent review of the Wilsonian
approach.) They are all realizable on some sufficiently large space of
probability distributions, and correspond to maps from measures to
measures.  Usually, however, the Wilsonian RG is taken to be a map from
Hamiltonians to Hamiltonians, and realizable as a flow on a
sufficiently large parameter space ${\cal M}$.  The general intuitive
idea behind these transformations is that they are mappings (almost
invariably approximate) between different effective degrees of freedom,
represented by an effective Hamiltonian associated with different
scales.  If, for a particular system, each different scheme could be
implemented, then universal quantities ought to be independent of the
particular scheme used. An example of such a quantity would be a
critical exponent, which is associated physically with a system that is
almost scale invariant and is determined by a point in ${\cal M}$ which
is associated with the scale invariant system (usually scale invariance
also implies conformal invariance \ref{37}). If the action of the RG
operator is to be a ``good'' representation of the action of the
dilatation operator then the fixed points of the RG transformation
should be coincident in $\cal M$ with the points of scale invariance.
That is not to say that they must have the same coordinates, but that
the intrinsic geometry of the RG flows and the dilatation flows should
be the same. For instance, if there were five points of scale
invariance in $\cal M$,  whereas the RG used only had three fixed
points, then one could not think of the latter as being a very
successful representation of scale changes. Additionally, the
properties of the fixed points should be independent of the ``coarse
graining'' procedure one uses, i.e. of the particular choice of RG.

Many systems, as a function of the parameters of the Hamiltonian
describing the system, can exhibit scale invariance at more than one
point of $\cal M$.  Generally one then expects to see a crossover
between different asymptotic regimes of the theory as governed by the
various fixed points, hence, one is generically looking at the
interpolation between two or more scale invariant field theories. If
one considered a ``coarse graining'' procedure, such as momentum shell
integration, for a system which possesses another length scale
$g^{-1}$, other than the correlation length $\xi$, it is clear that the
qualitative nature of the iterations of the RG should change as one
considers momentum shells with $k\ll g$ and $k\gg g$. If one starts
iterating at a scale $\Lambda\gg g$, then as the iteration proceeds
into the IR, the RG flow will pass close to the $g=0$ fixed point (it
will only actually hit it for $g=0$) before proceeding on to the $g=\i$
fixed point. What is manifest is that a ``physically sensible coarse
graining'' procedure will show up the qualitative change in the
effective degrees of freedom as one considers different regimes of the
iteration. The action of the ``physically sensible coarse graining''
follows in a reasonably faithful fashion the action of scale changes,
hence one more generally may reasonably expect that there exists an  RG
that can successfully describe the  change in behaviour of systems as
one changes scale, mass scale, temperature scale, momentum scale etc.
That the Wilson RG has been immensely successful is beyond question.
Our purpose in the following is to construct such a field theoretic
RG.

\subsec{\bf Massive Versus Massless Renormalization}
Let us now turn
back to the question of the field theoretic RG. If one considered a
massive $\l\ff$ theory in three dimensions, one sets up the field
theoretic RG formally in the same manner as shown in section 2.  If one
wishes to investigate the theory under changes in scale, and capture
the points of scale invariance in ${\cal M}$, then one should look for
fixed points of a suitable RG. There immediately arises the important
question of what set of parameters will one's RG transformation depend
on. This was not such an issue in the Wilsonian approach.  There, in
principle the RG
transformation naturally depends on all the parameters of the
effective Hamiltonian, in principle an infinite number, though usually
it is sufficient to only consider it as a function of the relevant and
marginal couplings, and these are usually finite in number.

In perturbative renormalization in field theory the story is somewhat
different. In this setting there is a division of the role played by
the Wilsonian RG transformation into two parts, represented by
equations (\docref{rge}) and (\docref{diman}). If one accepts the
philosophy that renormalization is solely a way of accounting for UV
divergences, then one has a great deal of freedom as to what parameters
one's counterterms should depend on, as the counterterms are
effectively released from any dependence on IR scales. The idea then is
to find the simplest form of (\docref{rge}) by making use of this
freedom. The ultimate version of this type of approach  is minimal
subtraction, where one chooses as counterterms only those parts of
diagrams that survive in the extreme UV limit.  For the Hamiltonian
(\docref{action}), for instance, one knows that in the extreme UV
counterterms that are mass independent are as good as mass dependent
counterterms,  if one's purpose is solely to perturbatively take the UV
cutoff $\Lambda$ to infinity. This is because mass is an irrelevant
coupling in the UV.

In our three dimensional example, only two fixed point are
perturbatively accessible using mass independent schemes, the Gaussian
and Wilson-Fisher fixed points with zero mass.  If one uses mass
dependent schemes, another fixed point becomes accessible --- the
infinite mass (temperature) Gaussian fixed point \ref{38}. In
fact these statements go beyond perturbation theory --- if mass
independent renormalization is used then the infinite mass Gaussian
fixed point cannot be seen from the RG flow of the Wilson functions, if
mass dependent renormalization is used then it can.  In other words, in
the mass independent renormalization the infinite mass Gaussian fixed
point is not captured by the differential generator of (\docref{rge}),
but relegated to a perturbative analysis of the vertex functions. Of
course  one can do such straight perturbation theory near the infinite
mass limit, and in that sense there was no necessity to implement the
RG. The reason one can get away with it, in the case of a mass
operator, is that the critical exponents associated with the infinite
mass Gaussian fixed point are mean field exponents, and therefore there
was nothing extra to exponentiate, however, one is not always so
lucky.  Typically, if one  tries to track the theory back into the IR
having used a minimal subtraction scheme, in the presence of an
additional mass scale, $g$, perturbation theory will break down, and
some other procedure will be necessary. It is only the $g$ dependent RG
which is capable of giving globally valid perturbative information.

More generally, if one has a field theory parametrized by a set of
parameters, $P\equiv\{g^i\}$, corresponding to a point in ${\cal M}$,
it might occur that different subsets of the parameters, relevant for
describing the theory at  different scales, are taken into one another
by the RG flow on ${\cal M}$. If one's renormalization depends only on
a subset of the parameters, one is restricting one's flow to take place
only in a subspace ${\T}$ of ${\cal M}$.  The resultant RG, $RG_{\T}$,
depends only on a subset $K$ of the parameters, and the RG flows take
place only on ${\cal T}$. If any of the $P-K$ parameters are relevant
in the RG sense, then the true RG flows of the theory, $RG_{\cal M}$,
thought of as true scale changes, will wish to flow off ${\T}$ into
${\cal M}$.  However, the use of $RG_{\T}$ does not allow for such
flows.  Such a state of affairs would be shown up by the perturbative
unreliability of the results based on $RG_{\T}$. If none of the
parameters $K$ are relevant then there should be no problem.  However,
one can only say what parameters are relevant when one knows the full
fixed point structure of the theory! In principle it is obviously
better to work with $RG_{\cal M}$. If a certain parameter was important
then one has made sure that its effects are treated properly, and if it
wasn't then that will come out of the analysis. There can be no danger,
except for extra work, from keeping a parameter in, but there can be
severe problems if it is left out.

The message then is that in the field theoretic context one's choice of
renormalization can be quite crucial, some points of scale invariance
in ${\cal M}$ being inaccessible with respect to certain RG's.  One
might be rather worried by this, given that physics should be
renormalization scheme independent. One must be careful to make a
distinction between the points in ${\cal M}$, where the system under
consideration is scale invariant, and those points which are fixed
points of a particular RG.  They are not necessarily the same. If one
has chosen an RG which is a good representation of scale changes then
they will be.  We are not saying that there are conformally invariant
systems that can only be accessed utilising certain RG's. What we are
saying is that there exist RG's which are sufficient {\bf alone} to
describe the system.  On the other hand, there are others that must be
supplemented by extra non-perturbative information. For our massive
field theory, the Wilson functions derived using mass dependent
renormalization are sufficient on their own to access the mean field
fixed point, whereas for mass independent schemes one has to supplement
the Wilson functions with extra information from some other source,
e.g. summing up possibly infinite sets of Feynman diagrams, in order to
access the fixed point.

\subsec{\bf The RG Should be Environment Dependent}
We can now perhaps
take a somewhat broader view. In the above we discussed the fact that
there were inequivalent field theoretic RG's, in the sense that the
sets of fixed points of these RG's were not the same.  Intuitively why
is this, and why, in particular, does it happen in the field theoretic
RG? Consider a physical system as we observe it at different scales.
If we start at a very small scale $\Lambda^{-1}$, characterizing the
system by a set of degrees of freedom (bare parameters), then try to
describe the effective physics at some much larger scale $\k^{-1}$, it
is almost invariably true, in a system with many degrees of freedom,
that the physics at the scale $\k$ is very complicated in terms of the
physics at the scale $\L$. More often than not a better prescription is
in terms of effective degrees of freedom, more appropriate for scales
$\sim\k$. This is the whole raison d'etre behind effective field
theory.  Of course, experimentally one will find that the effective
degrees of freedom are very different at different scales. In the
above, the effective degrees of freedom at scales $\k\gg m$ are
essentially massless with power law correlations, whereas for $\k\ll m$
they are essentially decoupled.  This is a statement about what one
would see experimentally, different sorts of correlations in different
asymptotic regimes. Clearly one would like one's calculational schemes
to capture this experimental fact --- different effective degrees of
freedom at different scales. Though the Wilson RG was originally
developed to tackle systems with just the opposite problem---effective
degrees of freedom that are the same at all scales (!)---a ``good
coarse graining'' procedure will pick up the change in effective
degrees of freedom. Similarly, we will show that a ``good'' field
theoretic RG can follow the changing effective degrees of freedom.

We mentioned above that the field theoretic formalism is
calculationally much simpler, but have gone into some detail about
potential pitfalls in using it. In the rest of the paper we will see
these pitfalls discussed in a much less heuristic fashion than here and
also see how they can be avoided.  The goal then is clear: to develop
field theoretic RGs who's fixed points encompass all the points of
scale invariance for a particular system.  The above discussion of
massless versus massive field theoretic RGs is just an illustrative
example, our main concern in this paper will be dimensional crossover,
however, similar considerations hold much more generally as we will now
discuss.

What one considers to be fluctuations in a system are not unique.
Fluctuations are usually defined with respect to some ``background'',
or in the terminology we will adopt here, some ``environment''. The
environment can have many different properties. One of the first things
one might ask is what space (or spacetime) does the environment live
in. Here we want to be more general and consider the space itself as
part of the environment. Thus if one works in infinite flat space in
three dimensions, $R^3$, or in a ``box'' of size $L$ ($L^3$), then the
properties of the space should be counted as part of the environment.
The reason is obvious. Fluctuations of a field are qualitatively
different in $R^3$ and $L^3$. If one's only concern is UV divergences
then certainly fluctuations with momenta $k\gg L^{-1}$ are
qualitatively the same on either $R^3$ or $L^3$, however, for momenta
$k\sim L^{-1}$ the qualitative difference is great, and for $k\ll
L^{-1}$ it is absolutely profound as such modes don't even exist on
$L^3$!  Clearly then the effects of the environment are very scale
dependent. If one renormalizes one's field theory on $L^3$ using only
minimal subtraction say, then, the information one gets out from the
consequent RG will be information appropriate to the field theory on
$R^3$, not $L^3$. Hence we will say that minimal subtraction is not
environmentally friendly on $L^3$, the environmental variable here
being $L$.  One could have, of course, an anisotropic box with
different size sides, then one would have three environmental
variables. A proper description of the fluctuations from a RG
standpoint would demand that the RG depended on these three variables.
We could consider a space with curvature (constant or not), this would
add additional environment dependence. This would be of some relevance
in the early universe, where the cosmological constant can also be
regarded as an environmental variable.  The IR behaviour of quantum
fields in inflation would only be accessible using RG methods using an
RG that depended on the cosmological constant.

So, one of the most important types of environmental variable comes
from properties of the space one works in. A second type comes from
what sort of background fields are present. A gravitational background
field is obviously related to the above question of what space one is
working in.  One could also consider the fluctuations of a charged
field say, coupled to a background electric or magnetic field. The
spectrum of fluctuations in a uniform magnetic field $B$, for instance,
is completely different to that for $B=0$. The $B$ field sets a length
scale, and fluctuations with wavelengths very large or small relative
to it are very different.  In fact one can think of the $B$ field as
making the fluctuations live in a ``box'' \ref{39}. The
particular background could of course be much more complicated than a
simple constant field. One example would be the Abrikosov flux lattice,
a background of monopoles another, the latter possibly being of great
relevance for QCD. In other words when one is looking at QCD, then the
way quarks and gluons behave in a background of monopoles will be very
different to their behaviour in no background. The standard
renormalization of QCD, using minimal subtraction, is really only
appropriate for the situation where there is no background, or one is
looking at length scales much shorter than the characteristic scale of
the background, i.e. at high energy --- asymptotic freedom --- where
quarks and gluons are a good representation of the effective degrees of
freedom.

In the above we have given some examples of environment and maintained
that the environment on certain length scales will greatly influence
the nature of the effective degrees of freedom which are a description
of the system. Quite often the environmental variables can be thought
of as setting a fixed length scale, or set of scales, which lead to
fluctuations in the presence of this environment that are qualitatively
different at different scales. In this paper we will be exhibiting a RG
methodology which can describe such changes. However, the above example
of QCD illustrates another important point, and that is that in many
cases in practise, one might not be able to make a very clean
distinction between fluctuations and  environment as they interact
dynamically. Not only does the environment influence the fluctuations
but also the fluctuations feed back into the environment. Quite often
this feedback is negligible, however, when it is not then there needs
to be a feedback mechanism between the RG as a function of the
environment and the environment as a function of the fluctuations. A
simple  example of such a state of affairs is  $\l\phi^{4}$  below the
critical point. In such a situation the fluctuations influence the
background field and one can investigate this explicitly via the
equation of state. It is the equation of state that tells one how the
fluctuations are reacting back on the environment, the environment in
this case being the background magnetization. For more complicated
systems there will be more ``equations of state'', these must be solved
in conjunction with the environment dependent RG equations to get a
closed system.  Obviously more often than not this will not be easy.

We hope the idea of environmentally friendly renormalization is
understandable and hope the reader by the end of the paper will see the
necessity of it.

\beginsection{\bf FORMAL PERTURBATION THEORY}
\subsec{\bf Determination of Renormalization Constants}
In this section we will present formal
diagramatic series for the correlation functions and the
renormalization constants. We will assume the generic normalization
conditions of section 2, but for now will not specify the precise
parameter/environment dependence of them. The following considerations
will be largely independent of such things and therefore applicable to
a large class of crossover problems. The expressions apply to
crossovers where the number of symmetry components of the order
parameter doesn't change, however,  the changes needed to encapsulate
this larger class of crossovers are small \ref{18}.

The effective action (free energy) (\docref{effaction}) is given by
$$\G=S+\Sigma$$ where $S$ is the classical action (mean field
Hamiltonian) and $\Sigma$ represents the contribution due to
fluctuations. The diagramatic series for the effective action can be
constructed in a loop expansion which gives a series ordered by the
number of loops. For the $O(N)$ model under consideration the
quadratic part of the Hamiltonian is defined by an operator
$$A^{ab}=A_1P_{1}^{ab}+A_2P_{2}^{ab}\no$$ where the two projection
operators $P_1$ and $P_2$ are
$$P_{1}^{ab}={\f_{B}^{a}\f_{B}^{b}\o\f_{B}^{2}}\quad{\rm and}\quad
P_{2}^{ab}=\delta^{ab}-{\f_{B}^{a}\f_{B}^{b}\o\f_{B}^{2}}\no$$ We
denote the inverse of $A^{ab}$
$$G^{ab}=G_{1}P_{1}^{ab}+G_{2}P_{2}^{ab}$$ For the case of dimensional
crossover which we  will consider in section 11
$$A^{ab}=(\square+m_{B}^2+t_{B}+{\l_{B}\o2}\f_{B}^{2})P_{1}^{ab}
+(\square+m_{B}^2+t_{B}+{\l_{B}\o6}\f_{B}^{2})P_{2}^{ab}\no$$ which
gives $$G_{1}={1\o\square+m^2_{B}+t_{B}+{\l_{B}\o2}\f_{B}^2}\quad{\rm
and} \quad G_{2}={1\o\square+m^2_{B}+t_{B}+{\l_{B}\o6}\f_{B}^{2}}\no$$
where $\square$ is the Laplace operator in the geometry of interest.
The effective action to two loops is given by $$\eqalign{\G =&\int
dx{1\o2}[\f_{B}(\square+m^2_{B}+t_{B})\f_{B}(x)+{\l_{B}\o4!}\f^{4}_{B}(x)]
+{1\o2}Tr\ln[A_{1}]+{1\o2}(N-1)Tr\ln[A_{2}]\cr &+{\l_{B}\o4!}\int dx
[3G_{1}^{2}(x,x)+2(N-1)G_{1}(x,x)G_{2}(x,x)+(N^2-1)G^2_{2}(x,x)]\cr
&\qquad-{\l_{B}^2\o36}\int dx\int dy
\f_{B}(x)[3G_{1}^{3}(x,y)+(N-1)G_{1}(x,y)G_{2}^{2}(x,y)]\f_{B}(y)\cr}\no$$

The contributions to the respective $\G^{(N)}$ can be obtained by
differentiating with respect to $\f_{B}(x)$ then setting
$\f_{B}=\fb_{B}$, where  $\fb_{B}$ is the solution of
$\G^{(1)}_{B}(\fb_{B})=H_B$.  The $\G^{(N,M)}$ can further be generated by
differentiation with respect to $t_{B}(x)$.  Specializing now to
constant magnetic field $H_{B}$; the magnetization $\fb_{B}$ will also
be constant for a homogenous and isotropic  environment, which we will
now assume for simplicity.  The effective potential (free energy
density) to two loops is then given by
$$\eqalign{{\cal F}
&={1\o2}(m^2_{B}+t_{B})\fb_{B}^{2}+{\l_{B}\o4!}\fb^{4}_{B}
+{1\o2}tr\ln[A_{1}]+{1\o2}(N-1)tr\ln[A_{2}]\cr &+{\l_{B}\o4!}{1\o
V}\int dx
[3G_{1}^{2}(x,x)+2(N-1)G_{1}(x,x)G_{2}(x,x)+(N^2-1)G^2_{2}(x,x)]\cr
&\qquad-{\l_{B}^2\o36}{1\o V}\int dx\int dy
\fb_{B}[3G_{1}^{3}(x,y)+(N-1)G_{1}(x,y)G_{2}^{2}(x,y)]\fb_{B}\cr}\no$$
where $V$ is the volume of the system and $tr={Tr\o V}$.

We will now  restrict our considerations to situations where
$\fb_{B}=0$,
which corresponds to being above the critical temperature with $H_B=0$,
once again this in no way restricts the generality of our methodology
\ref{29}. In this case we also have $G_{1}=G_{2}=G$, and the vertex
functions are proportional to $O(N)$ symmetric tensors. We will
concentrate on the coefficient of the relevant tensor eliminating the
tensorial structure altogether for simplicity.  Thus adopting a
suggestive diagramatic notation (see appendix A for an explanation and
for the actual expressions in the dimensional crossover problem) and
working with dimensionless diagrams by pulling out powers of an
arbitrary scale $\k$, one finds the one loop contributions are given by
$$\cF_{1}={N\o2}\k^d\fe\no$$
$$\G_{1}^{(2)}={(N+2)\o6}\l_{B}\k^{d-2}\io\no$$ $$\G_{1}^{(4)}
=-{3\o2}({N+8\o9})\l_{B}^{2}\k^{4-d}\bub\no$$
The two loop contribution is
$$\cF_{2}={\l_{B}\o4!}N(N+2)\k^{2d-4}\io^2$$ $$\G_{2}^{(2)}
=-{\l_{B}^{2}\o36}\k^{2d-6}(N+2)\left({(N+2)}\bub\io+2\itr\right)\no$$
and
$$\G_{2}^{(4)}={\l_{B}^{3}\o36}\k^{2(d-4)}\left((N^2+6N+20){\bub}^{2}
+2(N+2)(N+8)\itro\io+4(5N+22)\if\right)\no$$

Putting the zero, one and two loop contributions together, we are in a
position to construct the transformation to renormalized parameters,
however, our main concern here is to generate the renormalization
constants so that the Wilson functions may be calculated.  An important
observation is that for an amputated diagram it is only the number of
insertions that is relevant not the operator inserted. We have not
explicitly exhibited the different momentum combinations that
contribute to $\G^{(4)}$ as separate diagrams, the sum over the
different combinations of momenta being implicit. The diagrams as
written are therefore to be understood as averaged over the Mandelstam
variables, which at the symmetric point all become equal. Thus the
diagramatic series formally take on their symmetric point form. The
reader should however be careful to remember this point when
implementing expressions for the formal series at a non-symmetric point.

Adding the different loop contributions together gives
$$\cF={N\o2}\k^d\left(\fe+{\l_{B}\o12}\k^{d-4}{(N+2)}\io^2\right)\no$$
$$\G_{B}^{(2)}=p^2+m_{B}^{2}+t_{B}+{\l_{B}\o6}(N+2)\k^{d-2}\left(\io
-{\l_{B}\k^{d-4}\o6}\left((N+2)\bub\io+2\itr\right)\right)\no$$
\eqlabel{\gtotl} $$\eqalign{\G_{B}^{(2,1)}&=1-{\l_{B}\k^{d-4}\o6}(N+2)
\left(\bub-{\l_{B}\o3}\k^{d-4}(N+2)\itro\io\right)\cr
&\qquad+{\l_{B}^{2}\o36}(N+2)\k^{2(d-4)}
\left({(N+2)}\bub^{2}+6\if\right)\cr}\no$$
$${\p\G_B^{(2)}\o\p
p^2}=1-{\l_{B}^{2}\o18} \k^{2(d-4)}(N+2)\itrp\no$$ and \eqlabel{\gftl}
$$\eqalign{\G_{B}^{(4)}&=\l_{B}-{\l_{B}^2\o6}(N+8)\k^{d-4}
\left(\bub-{\l_{B}\o3}\k^{d-4}(N+2)\itro\io\right)\cr
&\qquad+{\l_{B}^{3}\o36}\k^{2(d-4)}\left((N^2+6N+20){\bub}^{2}
+4(5N+22)\if\right)\cr}\no$$ where we have regrouped the two loop
tadpole terms with the corresponding one loop terms.

The Wilson functions are derived from the renormalization constants
$\Zf$, $\Zft$ and $\Zl$, however, as mentioned in section 2 an additive
mass renormalization must also be performed, which corresponds in
critical phenomena language to a shift in ``reference'' temperature
from the mean field critical temperature to some other, usually the
``true'' critical temperature, if such exists.  We begin with
$$\G^{(2)}_B(p,m_B^2+t_B)=p^2+m^2_{B}+t_{B}+\Sigma^{(2)}_{B}(p,m_B^2+t_B)\no$$
where $\Sigma^{(2)}_{B}$ is the contribution from fluctuations.  There
are two undetermined parameters in $\G^{(2)}$, $m_{B}^2$ and the scale
of $t_{B}$. We find these by imposing two normalization conditions. The
first to be determined is the additive renormalization given by
$m^2=m_B^2+\delta m_B^2$, where $m^2$ is the value of $\G^{(2)}$ at
$t=0$ and $p=0$.  We will in fact take $m^2=0$. Therefore we determine
$\delta m_B^2$  by requiring that \eqlabel{\delmb} $$\delta
m_B^2=\Sigma^{(2)}_{B}(0,\delta m_B^2)\no$$
An important point to make here is that both $m_B^2$ and $\delta m_B^2$ are
$p$ and $t$ independent, a fact that we will use to our advantage later.
Condition \docref{delmb}, perturbatively for $d<4$, is very difficult to
treat to the presence of IR divergences. The requirement that
$\G^{(2)}$ be zero when $t=0$ and $p=0$ can only be used, of course, if
the system exhibits critical behaviour in the specific environmental
setting under study.  We then have, as usual, that $t_{B}$ is a linear
measure of the deviation from the critical temperature.

Recall that \eqlabel{\taylor}
$$\G_{B}^{(2)}(p,m_{B}^2+t_{B})=\sum_{M}{1\o
M!}\G^{(2,M)}(p,m_{B}^2)t_{B}^{M}\no$$ This implies that
$$0=\G_{B}^{(2,0)}(0,m_{B}^2)$$ which is our specification of $\delta
m_{B}^2$, however, we can equally well write (\docref{taylor}) in the
form $$\G_{B}^{(2)}(p,m_{B}^2+t_{B})=\sum_{M}{1\o
M!}\G_{B}^{(2,M)}(p,m_{B}^2+\L^2){(t_{B}-\L^2)}^{M}$$ where we have
just chosen to Taylor expand around a different point, $\L$ here being
some fixed scale.  This latter equation implies that at $t_{B}=\L^2$
$$\G^{(2)}_{B}(0,m_{B}^{2}+\L^2)=\G^{(2,0)}_{B}(0,m_{B}^2+\L^2)$$ All
this says is that we can choose any fiducial (RG invariant) point to
measure deviations in temperature/mass with respect to. It just so
happens that if there exists a critical point the critical temperature
is in some sense a preferred point around which to measure deviations.

Another condition which could be used to determine $\delta m_B^2$ is
\eqlabel{\condition}
$$\G^{(2)}(0,t(\kc)=\kc^2)=\kc^2\no$$
which corresponds to $\a_1=0$, $\a_2={\kc^2\o\k^2}$.
Here the scale $\kc$ is a fixed scale being the value of the renormalization
scale which satisfies the equation $t(\kc)=\kc^2$, $t(\kc)$ being the running
temperature evaluated at the scale $\kc$.  Equation (\docref{condition}) gives
$\delta m_B^2$ to be a
solution of \eqlabel{\deltamB} $$\Zf^{-1}\kc^2=m^2+\delta
m_B^2+\Zft^{-1}\kc^2+\Sigma(0,m^2+\delta m_B^2+\Zft^{-1}\kc^2)\no$$ If
there exists a true critical point, then as the critical temperature is
approached, one finds that, as $\kc\ra0$,  the above condition
becomes equivalent to (\docref{delmb}).

The condition (\docref{delmb}) is to be  solved perturbatively to yield
the ``shift'' from the mean field critical temperature to another
critical temperature. The particular temperature one shifts to depends
on the values of the environmental parameters, such as $g$, entering
the normalization condition. This is natural because $T_c$ generically
depends on environmental variables, therefore as the environment
changes, the relevant $T_{c}$ to which it is appropriate to shift to
changes accordingly. The shift, although intersting, is not something we
will consider in this paper.
The upshot of all the above is that $\itro\io$ in (\docref{gtotl}) and
(\docref{gftl}) cancel with the mass counterterm associated with the
propagators in the one loop diagrams $\io$ and $\bub$.  In fact a
similar  cancellation will occur order by order in perturbation
theory.

Using the condition $\G^{(2)}(p=0,t=0)=0$ gives
\eqlabel{\rengt}
$$\eqalign{&\ \ \ \ \G_{B}^{(2)} =p^2+ t_{B}\cr
&\qquad-{\l_{B}^{2}\o36}(N+2)\k^{2d-6}
\left\{{(N+2)}(\bub\io-\bub|_{\ss 0}\io|_{\ss 0})
+2(\itr-\itr|_{\ss 0})\right\}\cr}\no$$
where the subscript $0$ indicates $p=t_{B}=0$.
However, as mentioned previously this has potential problems with IR
divergences. An analogous expression which is free of this problem
follows from (\docref{deltamB}).

$\G^{(4)}_B$ and
$\G_{B}^{(2,1)}$ at the generic normalization points of section 2 are
\eqlabel{\gfb}
$$\eqalign{\G^{(4)}_B=&\l_{B}-\ne{6}\l_{B}^{2}\k^{d-4}\bub\cr
&+\nf\l_{B}^3\k^{2(d-4)}\if+\ns\l_{B}^3\k^{2(d-4)}\bub^2\cr}\no$$ and
\eqlabel{\gbtotl}
$$\G_{B}^{(2,1)}=1-\nt{6}\l_{B}\k^{d-4}\bub+\nt{6}\l_{B}^2\k^{2(d-4)}
\left(\if+\nt{6}\bub^2\right)\no$$
Remember that the diagrams in these expressions are now to be evaluated
at the appropriate normalization point, explicit expressions for the case
of a layered geometry can be found in the appendices.

Differentiating (\docref{rengt}) with respect to $p^2$, then setting
the condition (\docref{normzf}) yields in the two loop approximation
\eqlabel{\zf} $$\Zf=1+\nt{18}\l_{B}^{2}\k^{2(d-4)}\itrp\ +\dots\no$$
{}From (\docref{gbtotl}) one finds
$$\eqalign{\Zf^{-1}\Zft^{-1}=&1-\nt{6}\l_{B}\k^{d-4}\bub\cr
&+{{(N+2)}^2\o36}\l_{B}^2\k^{2(d-4)}\bub^2+\nt{6}\l_{B}^2\k^{2(d-4)}\if}$$
and hence \eqlabel{\zft} $$\eqalign{
\Zft^{-1}=&1-\nt{6}\l_{B}\k^{d-4}\bub+{{(N+2)}^2\o36}\l_{B}^2\k^{2(d-4)}
\bub^2\cr
&\qquad+\nt{6}\l_{B}^2\k^{2(d-4)}\left(\if+{1\o3}\itrp\right)\cr }\no$$

The renormalized four point function, including wavefunction
renormalization yields
$$\eqalign{\G^{(4)}_B&=\l_{B}-\ne{6}\l_{B}^2\k^{d-4}\bub\cr
&+{\l_{B}^3\o9}\k^{2(d-4)}\left((5N+22)\if+{(N^2+6N+20)\o4}\bub^2+(N+2)\itrp
\right)\cr} \no$$ therefore \eqlabel{\zl} $$\eqalign{
\Zl=&1-\ne{6}\l_{B}\k^{d-4}\bub\cr
&+\l_{B}^2\k^{2(d-4)}\left(\nf\if+\ns\bub^2+\nt{9}\itrp\ \right)\cr}\no$$
It is obviously trivial to obtain the renormalization constants in
terms of the  renormalized coupling. We leave them in terms of the bare
coupling as this coupling that must be kept constant when evaluating
the Wilson functions.

\subsec{Diagramatic Expansion of the Wilson Functions}
Having derived the renormalization constants we are now in a position to
calculate the Wilson functions. From the definition of $\gft$
(\docref{gamft}), using (\docref{zft}) one finds
$$\eqalign{\gft=&-\nt{6}(A_{(4-d)}\bub)\k^{d-4}\l\cr+ &\nt{6}\left(
A_{2(4-d)}\left(\if-{1\o2}\bub^2\right)
+{1\o3}A_{2(4-d)}\itrp\right)\k^{2(d-4)}\l^2\cr}\no$$ where we have
defined $$A_{\nu}=\k^{\nu}\k{d\o d\k}\k^{-\nu}=\k{d\o d\k}-\nu$$ and
we now work in terms of the renormalized coupling $\l$. Similarly, from
the definition of $\gf$ (\docref{gamf}), and using (\docref{zf}) one
gets  $$\gf=\nt{18}(A_{2(4-d)}\itrp\ )\k^{2(d-4)}\l^2\no$$ and
finally from (\docref{gaml}), using (\docref{zl}) one obtains for $\gl$
$$\eqalign{\gl=&-\ne{6}(A_{(4-d)}\bub)\k^{(d-4)}\l\cr &+\left(\nf
A_{2(4-d)}(\if-{1\o2}\bub^2)+\nt{9}A_{2(4-d)}\itrp
\right)\k^{2(d-4)}\l^2\cr}\no$$

{}From the definition of $\gl$ and its relation to the $\b$ function
one obtains the $\b$ function of the dimensionful coupling
$$\eqalign{\k{d\l\o d\k}=&-\ne{6}(A_{(4-d)}\bub)\k^{(d-4)} \l^2\cr
&+\left(\nf A_{2(4-d)}(\if-{1\o2}\bub^2)+\nt{9}A_{2(4-d)}\itrp
\right)\k^{2(d-4)}\l^3\cr}\no$$
We now introduce the dimensionless
coupling constant $\lb=\l\k^{d-4}$, the $\b$ function of which is
\eqlabel{\dimlesslam}
$$\eqalign{\k{d\lb\o d\k}=&-(4-d)\lb-\ne{6}(A_{(4-d)}\bub) \lb^2\cr
&+\left(\nf
A_{2(4-d)}(\if-{1\o2}\bub^2)+\nt{9}A_{2(4-d)}\itrp\ \right)\lb^3\cr}\no$$
We stress that these expressions for the Wilson functions are
totally finite for $d\leq4$.

We now introduce another coupling \ref{27} by performing a coordinate
transformation on $\cal M$. The new coupling $h$ , which we dub the
floating coupling, is defined so as to make the coefficient of the
quadratic term in $\b(h)$ unity. From (\docref{dimlesslam}) this is achieved
via the redefinition $h=\ao \lb$.  Thus the $\beta$-function for this
coupling is
$$\beta(h)=-\el
h+h^2+{4\left((5N+22)\Aif+(N+2)\Aitrp\right)
\o{\left((N+8)\Ait\right)}^2}h^3\no $$
where
$$\el=4-d-\k{d\ln A_{4-d}\bub\o d\k}\no$$
Note that in the limit $N\ra\infty$ the $\beta$-function for this coupling
becomes exact at the quadratic order.  The
condition $\b(h)=0$ defines what we call the floating fixed point, the
significance of which will be seen shortly.  Note that in fixed
dimension flat space for massless propagators, or where the mass is
taken to be the scale of our normalization point, $\Ait$ and $\Aif$ are
just numbers. Finally we introduce for later convenience two
combinations of diagrams, $f_1$ and $f_2$ defined by
$$f_{1}=-\fone\no$$ $$f_{2}=\ftwo\no$$

\beginsection{\bf ENVIRONMENT DEPENDENT RENORMALIZATION}
\subsec{\bf ``Running'' Mass/Temperature or Momentum}
In sections 2
through 4 we have set up the machinery for investigating crossovers in
a formal fashion and without specializing to a particular one. At this
point we want to put some flesh on the bare bones of the formalism.
Remember that our ultimate goal is to be able to describe systems that
exhibit qualitatively different effective degrees of freedom at
different scales.  This necessarily implies that the effective degrees
of freedom are sensitive to the environment.  In section 3 we had a
preliminary discussion of this in the context of comparing and
contrasting the field theoretic RG with the Wilson RG. The previous
sections were the embodiment of the  former implemented perturbatively.
So how does one see the concept of different RG's and enviroment
dependence coming out here? Obviously it is the fluctuations of the
system that are sensitive to the environment and it is these very
fluctuations that we are cavalierly calling effective degrees of
freedom. It should be clear then that it is the actual functional forms
of the Feynman diagrams themselves that exhibit this dependence.

So how do different RG's appear?  This depends crucially on one's
normalization conditions. If, as discussed in section 2, there exists
an important mass scale $g$, then physics at scales $\k\ll g$ and
$\k\gg g$ will be very different. We said earlier that if one's goal is
to describe systems at different scales then the RG is in principle an
ideal tool for the investigation. A good RG will be one who's fixed
points represent all the points of scale invariance of the physical
system in $\cal M$. If we chose renormalization conditions which were
explicitly $g$ independent, i.e. we used the generic normalization
conditions of section 2 with $g=0$, for example \eqlabel{\gts}
$$\left.\G^{(2)}(p,t,\l,g=0,\k)\right|_{\npw}
=(\a_{1}+\a_{2})\k^2\no$$
and analogously for the other normalization
conditions, then the Wilson functions would be found to be independent
of $g$. This we emphasize is not just a perturbative failing. If the
renormalization schemes used are $g$ independent then so are the
resulting Wilson functions and that is that. If we use $g$ dependent
renormalization conditions, however, then the resulting Wilson
functions will be explicitly dependent on ${g\o\k}$.

To be more concrete: consider $g$ to be a fixed ``physical'' scale
(i.e. it doesn't get renormalized as mentioned in section 1). We will
choose the normalization conditions \eqlabel{\gta}
$$\Gamma^{(2)}(p=0,t=\k^2,\lb,g,\k)=\k^2\no$$ \eqlabel{\gwf}
$${\p\Gamma^{(2)}\over\p p^2}(p,t=\k^2,\lb,g,\k)|_{\ss p=0}=1\no$$
\eqlabel{\gfg}
$$\Gamma^{(4)}(p=0,t=\k^2,\lb,g,\k)=\lb\k^{4-d}\no$$
\eqlabel{\gco}
$$\Gamma^{(2,1)}(p=0,t=\k^2,\lb,g,\k)=1\no$$
Condition (\docref{gta}), together with the multiplicative renormalization
of $t$, implies that $t$ is proportional to $T-T_c(g)$, i.e.
that one is measuring
temperature relative to the $g$ dependent critical point (we are
assuming that the system can exhibit critical behaviour for any value
of $g$).  If an analogous condition with $g=0$ had been used,
temperature would be consequently measured relative to $T_c(0)$, the critical
temperature of the isotropic system. As these conditions are explicitly
$g$ dependent, the RG equation (\docref{rge}) for an $N$-point vertex
function takes the form \eqlabel{\grge} $$\left(\k{\p\o\p\k}
+\beta(\lb,{g\o\k}){\p\o\p \lb} +\gamma_{\phi^{2}}(\lb,{g\o\k
})t{\p\o\p t} -{1\over2}\gamma_\phi(\lb,{g\o\k
})\left(N+\fb{\p\o\p\fb}\right)\right) \Gamma^{(N)}=0\no$$ where now
the Wilson functions $\beta$, $\gf$ and $\gft$ not only depend on
$\lb$, the dimensionless renormalized $\ff$ coupling, in the standard
fashion, but also on ${g\over\k}$. Notice that \docref{gta}-\docref{gco}
imply that the parameter which is generating the flow here is the renormalized
mass/temperature parameter. We could equally well of course have used
external momentum to generate the flow.

The RG equation can be solved in the
standard manner using the method of characteristics to yield
\eqlabel{\rgsol} $$\eqalign{\Gamma^{(N)}(p_i,t,\fb,\lb,g,\kappa)
&=(\kappa\rho)^{d-N{(d-2)\over 2}}{\rm
exp}\left(-{N\over2}\int_{1}^{\rho}\gamma_{\phi}(\lb(x),{g\over \kappa
x}) {dx\over x}\right)\cr &\Gamma^{(N)}({p_i\over\rho\kappa},
{t(\rho)\over\rho^2\kappa^2},{\lb(\rho)\fb^2(\rho)\o2{(\rho\k)}^{d-2}},
\lb(\rho),{g\over\rho\k},1)\cr}\no$$ where $t(\rho)$, $\lb(\rho)$, and
$\fb(\rho)$; the running temperature, coupling and magnetization
respectively, satisfy \eqlabel{\lbg} $$\rho{d\lb(\rho)\over
d\rho}=\beta(\lb(\rho),{g\over\k\rho})\no$$ \eqlabel{\lbt}
$$\rho{dt(\rho)\over d\rho}=\gft(\lb(\rho),{g\over\k\rho})t(\rho)\no$$
\eqlabel{\lbf} $$\rho{d\fb(\rho)\o
d\rho}=-{1\o2}\gf(\lb(\rho),{g\over\k\rho})\fb(\rho)\no$$ With a $g$
dependent renormalization condition such as (\docref{gfg}) the $\beta$
function for the dimensionless coupling will generically have the form
\eqlabel{\hgrun} $$\rho{d\lb(\rho)\over
d\rho}=-(4-d)\lb(\rho)+a_2({g\over\k\rho})\lb^2(\rho)+
\sum_{n=3}^{\i}a_n({g\over\k\rho})\lb^n({\rho}) \no$$ Thus we have
immediately $g$ independent and $g$ dependent RGs.

One can now ask how are they different?  What is the criterion? Well,
let's think of the fixed points of the groups.  How does one find them
in the field theoretic formalism? One looks for zeros of the $\b$
functions. The fixed points of (\docref{lbg}-\docref{lbf}) are fixed
points of the $g$ dependent RG.  Fixed points of scale transformations
are found by looking for fixed points of the dimensionless couplings
$\lb$, $\tb={t(\rho)\o\rho^2\k^2}$ and the dimensionless field
$\fbd={\fb(\rho)\o{(\rho\k)}^{{d\o2}-1}}$.  For the $g$ independent RG
the equation (\docref{hgrun}) is $$\rho{d\lb(\rho)\over
d\rho}=-(4-d)\lb(\rho)+a_2(0)\lb^2(\rho)+
\sum_{n=3}^{\i}a_n(0)\lb^n({\rho}) \no$$ The $a_n$ are all just numbers
so one gets out a fixed point to lowest order $\lb^{\ast}={(4-d)\o
a_2(0)}$ (the higher orders can be found in the standard fashion
\ref{40}\ref{41}). Remember that a fixed point of the RG does not just
consist of a zero of one of the $\b$ functions, but actually of all of
them. In the present case this means one must also consider the
equation for the running temperature/mass and the running
magnetization. Using the renormalization condition (\docref{gts}) when,
$g=0$, implies that $t(\rho)=0$ is a fixed point. One can formally
solve (\docref{lbt}) to get $t(\rho)=t{\rm exp}\int\gft$. Now,
$t(\rho)=0$ implies $t=0$ due to the multiplicative renormalizability
of $t$. The $g$ independence of the normalization condition
(\docref{gta})
implies that $t=T-T_c(0)$. Thus, when $\fb=0$, the fixed point being
found here is the non-trivial fixed point of the isotropic system. An
unstable Gaussian fixed point at $T=T_c(0)$, and a stable Gaussian fixed
point at $T=\i$, are also accessible to this $g$ independent RG. Clearly
no anisoropic fixed point is accessible.

Turning now to the $g$ dependent RG one might think naively that to
lowest order $t(\rho)=0$ (i.e. $T=T_c(g)$) and $\lb(\rho)={(4-d)\o
a_2({g\o\k\rho})}$ is a fixed point.  However, it is hardly a fixed
point as the last condition defines a function of $\rho$ and
therefore cannot be preserved under the RG flow.  Let us change
variables to the floating coupling $h=a_2({g\over\k\rho})\lb(\rho)$,
the $\b$ function of which is
$$\beta(h)=-\e({g\over\k\rho})h(\rho)+h^2(\rho)+
\sum_{n=3}^{\i}{a_n({g\over\k\rho})\over
(a_2({g\over\k\rho}))^2}h^n({\rho})\no$$
where
$\e({g\over\k\rho})=4-d-{d{\rm ln}a_2\over d{\rm ln}\rho}$.  To find
the fixed points of the system one must remember that there is really
one more $\b$ function equation. The dimensionless anisotropy
$g\o\k\rho$ actually plays the role of a new coupling in the problem,
$\bar g$. The flow equation for it is $$\rho{d\bar g\o d\rho}=-\bar g$$
The fixed points of all three equations are to lowest order (we omit
consideration of the Gaussian fixed points) \eqlabel{\fpo} $$\bar
g=0,\ \  h=\e(0), \ \ t(\bar g=0)=0\no$$ and \eqlabel{\fpt} $$\bar
g=\i,\ \  h=\e(\i), \ \ t(\bar g=\i)=0\no$$ The first one of course is
the same fixed point accessible to the $g$ independent RG. The second
one, however, is totally inaccessible to this RG.  If under favourable
circumstances one has a good idea of what the effective degrees of
freedom associated with the $g=\i$ fixed point are, then one might
impose normalization conditions at $g=\i$. Once again the consequent RG
would be $g$ independent, only this time the fixed point accessed would
be (\docref{fpt}) not (\docref{fpo}) as in the case of the $g=0$  RG.

The question naturally arises as to the role of the minimally
subtracted RG, which is loosely related  to the field theoretic $\e$
expansion. It is used quite ubiquitously in particle physics. The
answer is that it is the RG associated with the $g=0$, $t=0$ fixed
point. It may be used to assist in defining the microscopic theory, in
a fashion that is independent of an UV cutoff. However, in reality this
fixed point  plays no more preferred role than any of the others.  Our
formulation makes this manifest.

The RG that is capable of encapsulating both fixed points is the $g$
dependent RG. It is of course by no means unique. Just as in the
non-crossover case where different renormalization schemes can exhibit
the same physics, e.g.  minimal subtraction and massless normalization
conditions, so here there will be different schemes. The key is that
they will differ only by relatively unimportant reparametrizations of
the crossover variables.  These differences are the field theoretic
analog of the differences between block spinning, decimation,
blockspinning on different lattices etc. in the Wilson type RG.
In the space of coupling constants, $t$, $h$ and $g$, the physical system
exhibits many points of scale invariance, two of which are
(\docref{fpo}) and (\docref{fpt}). One of these is accessible to a
$g=0$ RG, one to a $g=\i$ RG, and only for a $g$ dependent RG are both
accessible. In this sense the $g$ dependent RG is global in the space
of couplings whereas the others aren't.  The reader should understand
that we are not saying that using a $g$ independent RG precludes
accessing the other fixed point, it is rather that the $g$ independent
RG {\bf on its own} cannot access it, i.e. the points of scale
invariance and the fixed points of the $g$ independent RG are not the
same. If one is able to supplement the $g$ independent RG with extra
information then perhaps the other fixed point could become
accessible.  Almost certainly this will require the extra information
to be non-perturbative with respect to the $g=0$ fixed point.

\subsec{\bf ``Running'' the Environment}

In the last section we considered RG flow generated by
mass/temperature or momentum. Here we will consider briefly the
concept of RG flows  generated by the environment itself.
Previously the environment was held fixed as far
as the RG flow was concerned. One could examine what happened when
one changed the environment by examining RG flows with respect to
this new changed environment, one could not easily however go directly from
one environment to another by the action of the flow itself. For the
generic anisotropy $g$ this means that a particular value of $g$ fixes a
curve, one then examines RG flow along this curve. Another value of $g$
yields another, distinct curve along which one can also examine RG flows.
It is possible however to also generate an RG flow in the ``direction''
of the environment itself. This can generically be done by implementing
normalization conditions with respect to some ``fiducial'' environment.
We will assume that the parameter describing the environment does not itself
renormalize and therefore there is no distinction between bare and renormalized
$g$.
For example consider the normalization conditions
\eqlabel{\egta}
$$\Gamma^{(2)}(p=0,t,\l,g=g_{\ss 0})=t\no$$ \eqlabel{\egwf}
$${\p\Gamma^{(2)}\over\p p^2}(p,t,\l,g=g_{\ss 0})|_{\ss p=0}=1\no$$
\eqlabel{\egfg}
$$\Gamma^{(4)}(p=0,t,\l,g=g_{\ss 0})=\l\no$$
\eqlabel{\egco}
$$\Gamma^{(2,1)}(p=0,t,\l,g=g_{\ss 0},\k)=1\no$$
The fiducial environment, $g_{\ss 0}$, is abitrary therefore
the bare vertex functions must be independent of $g_{\ss 0}$. Thus
$$\left(g_{\ss 0}{\p\o\p g_{\ss 0}}
+\beta(\l,{g_{\ss 0}\o {|t|}^{1\o2}}){\p\o\p \l} +\gamma_{\phi^{2}}(\l,{g_{\ss
0}\o {|t|}^{1\o2}
})t{\p\o\p t} -{1\over2}\gamma_\phi(\l,{g_{\ss 0}\o {|t|}^{1\o2}
})\left(N+\fb{\p\o\p\fb}\right)\right) \Gamma^{(N)}=0\no$$ where
the Wilson functions $\beta$, $\gf$ and $\gft$ depend on
$\l$ and ${g_{\ss 0}/ {|t|}^{1\o2}}$. The absence of a term $\gamma_{g}g{\p\o\p
g}$ is accounted for by the non-renormalization of $g$. More generally such a
term would be necessary.

The RG equation can be solved in the
standard manner using the method of characteristics to yield
\eqlabel{\ergsol} $$\eqalign{\Gamma^{(N)}(p_i,t,\fb,\l,g,g_{\ss 0})
&=(g_{\ss 0}\rho)^{d-N{(d-2)\over 2}}{\rm
exp}\left(-{N\over2}\int_{1}^{\rho}\gamma_{\phi}(\l(x),{g_{\ss 0}x\over
{|t|}^{1\o2}})
 {dx\over x}\right)\cr &\Gamma^{(N)}({p_i\over\rho g_{\ss 0}},
{t(\rho)\over{(\rho g_{\ss 0})}^2},{\l(\rho)\fb^2(\rho)\o2{(\rho g_{\ss
0})}^{2}},
\l(\rho),{g\o g_{\ss 0}\rho},1)\cr}\no$$ where $t(\rho)$, $\l(\rho)$, and
$\fb(\rho)$; the running temperature, coupling and magnetization
respectively, satisfy \eqlabel{\elbg} $$\rho{d\lb(\rho)\over
d\rho}=\beta(\l(\rho),{g_{\ss 0}\r\over {|t|}^{1\o2}})\no$$ \eqlabel{\elbt}
$$\rho{dt(\rho)\over d\rho}=\gft(\l(\rho),{g_{\ss 0}\r\over
{|t|}^{1\o2}})t(\rho)\no$$
\eqlabel{\elbf} $$\rho{d\fb(\rho)\o
d\rho}=-{1\o2}\gf(\l(\rho),{g_{\ss 0}\r\over {|t|}^{1\o2}})\fb(\rho)\no$$
It should be clear here that in this RG, as distinct from that of the last
section, it is the environment itself that is running. One would implement
the same RG strategy however, i.e. use the arbitrariness in the scale $\r$
to choose a value that allows for a perturbative evaluation of the right
hand side of (\docref{ergsol}). The appropriate condition to set is $g=g_{\ss
0}\r$.
The beauty of running the environment is that, by an appropriate choice of
initial condition for the characteristic equations, one can relate parameters
associated with the environment $g$ to those associated with the environment
$g=0$. Thus one can answer questions about the ``shift'' in a perturbatively
controllable manner using RG techniques. It should also be clear that we could
combine the two RG's and run the environment and mass/temperature together. We
will return to these important
matters in a future publication.

\beginsection{\bf FORMAL SCALING FORMS}
In this section, working still
with an abstract anisotropy mass scale $g$, we will investigate the
formal consequences of using a $g$ dependent RG.  Consider first the
solution (\docref{rgsol}) of the RG equation (\docref{grge}), where we
work now with the floating coupling directly \eqlabel{\gnrho}
$$\eqalign{ \Gamma^{\scriptscriptstyle{(N)}}(t,\fb,h,g,\k)&
=(\k\rho)^{{N+d-{1\over2}Nd}}\cr &\exp\left({-{N\over2}\int_{1}^{\rho}
\gamma_{\f}(h(x, {g\o x}){dx\over x}}\right)
\Gamma^{\scriptscriptstyle{(N)}}{{\left({{t(\rho)}\over (\k\rho)^2}, {
{\l\fb^2(\rho)} \over 2{(\k\rho)^{2} }},
{h(\rho)},{g\o\k\rho},1\right)}}\cr}\no$$ RG invariance implies that
the right hand side of (\docref{gnrho}) is independent of $\rho$,
therefore we are at liberty to choose a value of it that suits us. We
will fix it using the condition \eqlabel{\trho} $$t(\rc)=\rc^2\k^2\no$$
$\rc$ being the particular value of the RG running scale which
satisfies this equation. With the generic conditions (\docref{normt})
and (\docref{normzf}), when $\fb=0$, one finds $\rho^2\k^2=\xi^{-2}$,
$\xi$ being the correlation length.  We must then ask which correlation
length, as this is sensitive once again to what normalization
conditions are used. $g$ independent conditions would lead to the
isotropic correlation length. Here we will assume that $g$ dependent
renormalization is used and therefore the appropriate correlation
length is $\xi_{gt}$, where $t=T-T_c(g)$. We append a $t$ to it also
because of the assumption that $\fb=0$ in the normalization conditions,
i.e we are using the correlation length in the symmetric (disordered)
phase.

We could have used other conditions to determine $\rc$ such as
\eqlabel{\btccor} $${\lambda(\rc)\over2}\fb^2(\rc)=\rc^2\k^2\no$$
\eqlabel{\phys} $$t(\rc)+{\lambda(\rc)\over2}\fb^2(\rc)=\rc^2\k^2\no$$
The former with appropriate normalization conditions for $\G^{(2)}$ and
${\p\G^{(2)}\o\p p^2}$ corresponds to a parametrization in terms of
$\xi_{\ss g\fb}$, the correlation length in the ordered phase when
$T=T_c(g)$. If $g$ independent renormalization was used then the
corresponding correlation length would be that in the ordered phase of
the isotropic system at $T=T_c(0)$. The condition (\docref{phys}) would
yield a parametrization in terms of the ``true'' correlation length of
the system, either above or below $T_c(g)$. From a purely formal
standpoint any of the correlation lengths will do, they all correspond
to non-linear scaling fields, however, as we shall see, a particular
correlation length is naturally associated with a particular field
theoretic RG, and within that RG certain fixed points of the system
might be inaccessible, hence the utility of that correlation length
from an RG standpoint is diminished. The optimum correlation length to
use should be the true physical correlation length in the system which
is associated with the RG that depends on all environmental
parameters.  Naturally, working with this correlation length might
computationally be quite difficult. One is always at liberty to omit a
parameter from the renormalization to try to simplify matters, however,
by so doing one risks making a particular crossover in $\cal M$
inaccessible. Generically a breakdown in perturbation theory will then
take place, indicating that one is trying to access a region of $\cal
M$ which is inaccessible to the particular perturbation theory one is
implementing.

An extremely useful way to parametrize a crossver is via the
introduction of effective critical exponents \ref{14}, which are
natural generalizations of the standard critical exponents associated
with one particular fixed point.  We define the effective critical
exponent for the correlation length \eqlabel{\nuef}
$$\nu_{eff}=-\left.{d{\rm ln}\cgt\o d{\rm ln}|t|}\right\vert_{g}\no$$
With the condition (\docref{trho}) one sees that \eqlabel{\nuefa}
$${\nu\ef}={1\o2-\gft}$$ One can also define an effective exponent for
the susceptibility \eqlabel{\eqn5} $$\gamma_{eff}=\left.{d{\rm ln}\K\o
d{\rm ln}|t|}\right\vert_{g}\no$$ where $\K=\G^{(2)}(t,\fb,h,g,\k)$ is
the inverse susceptibility.  An effective exponent $\eta\ef$ is defined
naturally via \eqlabel{\eqn6} $$\eta_{\ss eff}=2-\left.2{d{\rm
ln}\Gamma^{(2)} \over d{\rm ln}p^2}\right\vert_{t=0,g}\no$$ which is
equivalent via the condition (\docref{normzf}) to $\eta_{\ss
eff}=\gf(g/p)$. This $\gf$, however, is exactly the same one that
appears in (\docref{lbf}), only its argument is different.  For the
exponents $\delta$ and $\b$ associated with the ordered phase one can
define effective exponents \ref{29}
\eqlabel{\eqnd}
$${\delta\ef}=\left.{{d{\rm ln}H}\over{d{\rm
ln}\fb}}\right\vert_{t=0,g} \no$$ and \eqlabel{\eqnb}
$${\beta\ef}=\left.{d{\rm ln}\fb\over d{\rm ln}|t|}\right\vert_{g}\no$$
the latter being defined on the crossover coexistence curve. Finally,
one can introduce an effective exponent for the specific heat
\eqlabel{\eqn9} $${\alpha\ef}=\left.{d{\rm ln}\G^{(0,2)}\o d{\rm
ln}|t|}\right\vert_{g}\no$$

There is one last concept we need before proceeding further.  Consider
in fixed dimension $d$, without $g$ dependence, how $\G^{(4)}$ varies
with temperature near a fixed point \eqlabel{\omef} $${d{\rm
ln}\G^{(4)}\over d{\rm ln}|t|}=(4-d-2\eta)\nu\no $$ We use here the
scaling behaviour of the four point coupling to get information about
the dimensionality $d$ of the system. One can introduce the concept of
an effective dimensionality, $\de$, through the natural analog of
(\docref{omef}) \eqlabel{\omeff} $$\left.{d{\rm ln}\G^{(4)}\over d{\rm
ln}|t|}\right\vert_{g}=(4-\de-2{\eta\ef}){\nu\ef}\no$$ $\de$ is in fact
simply related to $\gl$, $\de=4-\gl$, and can therefore be thought of
as a measure of how important the leading irrelevant operator $\ff$
is.  In the vicinity of a particular fixed point $\gl$ plays very
little role, merely governing the corrections to scaling about that
fixed point.  In a small neighbourhood of this fixed point these are
negligible. In the case of a crossover this will not be true as it is
the corrections to scaling that actually interpolate from one fixed
point to another.  This is why $\gl$ takes on an important role.  In
fact, for crossovers where the different asymptotic regimes correspond
to systems with different upper critical dimensions, $\de$, in the
absence of transients,  can correctly give the change in upper critical
dimension as one interpolates between the two fixed points.

{}From the $g$ dependent RG equations for the various vertex
functions one can derive relations between these effective exponents:
\eqlabel{\sexplaw}
$${\gamma\ef}={\nu\ef}(2-{\eta\ef})\no$$
\eqlabel{\dexplaw}
$${\delta\ef}=\left({\de+2-\eta\ef\o\de-2+\eta\ef}\right)\no$$
\eqlabel{\bexplaw}
$${\beta\ef}={{\nu\ef}\o2}(\de-2+{\eta\ef})\no$$
\eqlabel{\shexplaw}
$${\alpha\ef}=2-{\nu\ef}{\de}\no$$ and combinations thereof. (RG
derivations of these relations can be found in
\ref{28}, \ref{18} and \ref{29}. As long as $g\ll\k$ and $\xi\gg\k^{-1}$,
where $\k$ is a typical microscopic scale, these exponent relations
will be universal, below the upper critical dimension.  Just as in the
standard case where the scaling relations imply that there are really
only two independent critical exponents, so here there are really only
two independent effective exponents.  The difference with the
crossovers considered here is that it is necessary to know one more
function, $\gl=4-\de$, which represents the effects of the leading irrelevant
operator.  In the case of a crossover for which the dimensionality does
not change, e.g. crossover in a bicritical system, the effects of the
leading irrelevant operator can be subsumed into an $N_{\ss eff}$, which
is a measure of the effective number of components of the order
parameter \ref{18}.

Utilizing (\docref{trho}) in (\docref{gnrho}) yields
\eqlabel{\prescalingform}
$$
\eqalign{\G^{(N)}(t,\fb,h,g,\k)=&\xi_{gt}^{\ss{{Nd\over2}-N-d}}
e^{-{N\o2} \int_1^{\ss{1\o\xi_{gt}\k}} \gf{dx\over
x}}\G^{(N)}\left(1,{\l(g\xi_{gt})\fb^2(g\xi_{gt})\xi_{gt}^2},
h({g\xi_{gt}}),{g\xi_{gt}},1 \right)\cr} \no$$
where using the running
equations (\docref{lbg}) and (\docref{lbf})
$${\l(g\xi_{gt})\fb^2(g\xi_{gt})}={\l\fb^2}{\rm
exp}\int_1^{\ss{1\o\k\xi_{gt}}} (\gl-\gf){dx\o x} \no$$
We can rewrite (\docref{prescalingform}) in terms of the effective
exponents
\eqlabel{\gnaa}
$$\Gamma^{(N)}=\cgt^{\ss{{Nd\over2}-N-d}}e^{-{\frac
N2}\int_\k^{\ss{\cgt^{-1}}} {\eta\ef} {dx\over x}}{\cal
F}_t^{(N)}\left(\fb^2e^{-{\int_\k^{\ss{\cgt^{-1}}}}
(\de-2+\eta\ef){dx\over x}},{g\cgt}\right)\no$$ where ${\cal
F}_t^{(N)}$ is a universal function. Similarly, in terms of $\cgf$ one
finds \eqlabel{\gnab} $$\G^{(N)}=\cgf^{\ss{{Nd\over2}-N-d}}e^{-{\frac
N2}\int_\k^{\ss{\cgf^{-1}}} \eta\ef{dx\over x}}{\cal
F}_{\fb}^{(N)}\left(te^{-{\int_\k^{\ss{\cgf^{-1}}}
{1\over\nu\ef}{dx\over x}}},{g\cgf}\right)\no$$ In the various
integrals in (\docref{gnaa}) and (\docref{gnab}) one can change
variables via the exponent definitions, exponent laws, and the
conditions fixing $\rho$.  For example using (\docref{nuef}) and
(\docref{nuefa}) one finds
$$\int_\k^{{\ss{\cgt^{-1}}}}({d\ef}-2+{\eta\ef}){dx\over x}
=\int_\k^t({d\ef}-2+{\eta\ef}){\nu\ef}{dt'\o t'}\no$$ Thus
$$\G^{(N)}=e^{\int_1^t(N+d-{\frac N2}(d+{\eta\ef})){\nu\ef}{dt'\over
t'}} {\cal F}^{(N)}_t\left(\fb e^{-{\int_1^t\beta\ef{dt'\over
t'}}},ge^{-\int_1^t \nu\ef{dt'\over t'}}\right)\no$$ Similarly, one can
also find that $$\G^{(N)}=e^{\int_1^{\fb}(N+d-{\frac
N2}(d+{\eta\ef})){\nu\ef\over\beta\ef} {d{\fb}'\over {\fb}'}}{\cal
F}_{\fb}^{(N)} (te^{-{\int_1^{\fb}{1\over\beta\ef}{d{\fb}'\over
{\fb}'}}}, ge^{-\int_1^{\fb}{\nu\ef\over\beta\ef}{d{\fb}'\over
{\fb}'}})\no$$ A particularly interesting case is the equation of state
which can be written as $$H={\rm
exp}\left({\int_1^{\fb}\delta\ef{d{\fb}'\over {\fb}'}}\right){\cal G}
(te^{-{\int_1^{\fb}{1\over\beta\ef}{d{\fb}'\over {\fb}'}}},
ge^{-\int_1^{\fb}{\nu\ef\over\beta\ef}{d{\fb}'\over {\fb}'}})\no$$
where $\cal G$ is a universal function. The crossover coexistence curve
is $$t={\rm exp} \left({\int_1^{\fb}{1\over\beta\ef}{d{\fb}'\over
{\fb}'}}\right)\no$$

Thus using the effective exponents one can derive natural non-linear
scaling fields (we use the term non-linear here to refer to the fact
that they are non-linearly related to the linear scaling fields
associated with the individual fixed points not as in the sense of
Wegner that they are eigenfunctions of the dilatation operator).  The
natural non-linear scaling fields that appear above are the various
correlation lengths, $\xi_{gt}={\rm exp}\int^t_1\nu\ef(x){dx\o x}$,
$\cgf$, etc. and $g$ (these are also non-linear scaling fields in the
sense of Wegner).  Note that the non-linear scaling field $\cgt$,
interpolates between the linear scaling fields $t^{-{\nu(0)}}$ and
$t^{-{\nu(\i)}}$ associated with the $g=0$ and $g=\i$ fixed points in
the limits $g\ra0$, $\xi_{gt}\ra\i$ ($g\xi_{gt}\ra0$) and
$g\xi_{gt}\ra\i$ respectively.  In the equation of state there is
another non-linear scaling field
$te^{-{\int_1^{\fb}{1\over\beta\ss\ef}{d{\fb}'\over {\fb}'}}}$ which
interpolates between the two linear scaling fields
$t\o{\fb}^{\ss 1/\b(0)}$ and $t\o{\fb}^{\ss 1/\b(\infty)}$ in the limits
$g\ra0$, $\xi_{\fb}\ra\i$ ($g\xi_{\fb}\ra0$) and $g\xi_{\fb}\ra\i$
respectively.

We hope the general pattern is clear. The natural non-linear scaling
fields for the crossover can be found by taking the natural linear
scaling fields for one of the fixed points, writing it in exponential
form, e.g. $t^{\nu}=e^{\int_1^t\nu{dx\o x}}$, then replacing the
exponent with the effective one, e.g. $t^{\nu}\ra
e^{\ss{\int^t_1{\nu\ef}}}$. Note the effective exponent {\bf cannot} be
taken outside the integral to recover a form $t^{\nu\ef}$. In this way
one also arrives at the concept of effective crossover exponents.
Standardly a crossover exponent involves the ratio of two eigenvalues
of the RG operator linearized around the isotropic fixed point. The
generalization of this involves exponentiating the integral of the
difference of two effective exponents. Naturally, one must derive what
these effective exponents are. This will be the task of most of the
rest of the paper. The key is that they should be evaluated using an
appropriate RG, i.e. one that is appropriate to the crossover in
question --- one that is environmentally friendly!

Finally, in this section we would like to discuss universality in the
context of crossover behaviour. We have stated that in the regime where
all length scales are much bigger than the lattice spacing $a$, that
the scaling functions ${\cal F}^{(N)}$, ${\cal G}$,  etc. and the
effective exponents will all be universal functions. The meaning of
universality here is just the standard one. For example, two layered
Ising models with different lattice structures will exhibit precisely
the same crossover curves (up to a trivial constant rescaling) as long
as the finite size of the system $L\gg a$. However, the crossover
curves for a layered Ising model of size $L$ but with different
boundary conditions for instance will not be the same. Neither can they
be made the same by any rescaling.

The environment affects the infrared behaviour of the theory and
therefore the universality class. Universality in the non-crossover
case is generically governed only by the dimensionality of the system
and the symmetry of the order parameter. The environment gives us extra
``labels'' for delineating different crossover universality classes.
The label ``boundary condition'' for instance allows us to classify,
say, three dimensional Ising models with one finite dimension into
different classes. The effective exponents are different for different
crossover universality classes. However, the effective exponent laws
have a much wider universality, being valid irrespective of crossover
universality class. For example, the effective exponents obey scaling
laws even when one of the fixed points is associated with the upper
critical dimension. Logarithmic corrections to scaling are naturally
incorporated in the effective exponents, as indeed are power law
corrections coming from the leading irrelevant operator. Furthermore,
in the simplified context of crossover from the Wilson-Fisher to
Gaussian fixed points it was found \ref{42} that effective
exponents obeyed ``all the thermodynamic scaling relations (e.g.,
$\alpha{\ef}+2\beta{\ef}+\gamma{\ef}=2$) but not those of
hyperscaling''. In fact if one generalizes $d$ in this crossover to
$4-\gl$, we find that all scaling laws including hyperscaling are
obeyed by the effective exponents. Clearly the notion of $\de$ is
inappropriate here, however, the effects of the leading irrelevant
operator must still be accounted for. This is achieved by the
incorporation of $\gl$ into the scaling laws as above. In the context
of the general type of crossover discussed in this paper we have seen
that the scaling laws are always obeyed.

We are making an issue of these considerations here for the following
reason: often it is the case that the asymptotic scaling behaviour at
both ends of a crossover is known, i.e. the fixed points and critical
exponents at either end are known. Given this knowledge Riedel and
Wegner \ref{14} wrote down Wilson functions based purely on the
requirement that the function should allow for an interpolation between
the two known fixed points. In light of our discussion of universality
we regard this as a potentially dangerous procedure.  To understand
exactly why consider the Gaussian fixed point and the Wilson-Fisher
fixed point for a three dimensional Ising model. These fixed points are
the extremal fixed points for several crossover systems:  three
dimensional uniaxial dipolar ferromagnets, three dimensional Ising
model in a transverse magnetic field, four dimensional layered Ising
model (with possible different boundary conditions) as well as the
standard Gaussian $\ra$ Wilson-Fisher crossover of a three dimensional
Ising model. If one adopts the approach of constructing a crossover
curve by simply demanding that the two fixed points are captured in the
RG flow then there is no way of distinguishing all these physically
different crossovers.  They are induced by different
environmental variables which enter the theory in different ways.  Only
via an environmentally friendly RG can one distinguish the different
behaviours of these systems and see whether they fall into different
crossover universality classes.

In the context of a Wilsonian RG, such as momentum shell integration it
is quite possible to obliterate real distinctions between crossover
curves by implementing the momentum shell integrations using too rough
an approximation, or by over emphasizing the high momenutum shells.
For instance, Hertz \ref{43} when considering the
crossover between quantum and classical critical behaviour, derived
approximate RG equations which were identical in form to those for the
Gaussian Wilson-Fisher crossover, by ignoring the relevance of the gap
in Matsubara frequencies relative to the momentum shell being
integrated over, which difference is significant at the classical end
of the crossover. As mentioned previously these two crossovers lie in
two crossover universality classes.

The reader might indeed wonder if the concept of environment is so
stringent as to make every crossover fall into its own universality
class (apart from the standard trivial changes in lattice structure
etc. one can engender). This is not so. For instance, in the above
cited examples, it is known \ref{44}\ref{43}\ref{8} that a four
dimensional layered Ising model with periodic boundary conditions, and a
three dimensional Ising model in transverse magnetic field, are
equivalent. In the language used here they are in the same crossover
universality class. We will show in section 11 that they ehhibit the
same effective exponents.

\beginsection{WHAT SHOULD WE PERTURB IN?}
In the last section we once
again worked at a formal level, here we would like to discuss the
explicit implementation of perturbation theory. We have explained that
the main idea in crossovers is trying to quantitatively describe
systems that exhibit different degrees of freedom at different scales.
If one cannot solve a model exactly one must resort to an approximation
procedure. Perturbation theory is a ubiquitous one. However, there is
always the perennial question --- perturbation theory in what? In
applications of Wilsonian RGs this is often a very difficult question.
In the field theory approach it looks much simpler, however, our
discussion of the different field theoretic RGs should give one pause
for thought. If one uses $g$ independent renormalization to eliminate
UV problems then one discovers two things: firstly, that direct
perturbation theory fails badly in the regime $g\xi\gg 1$. New
``divergences'' appear, which at a given order in perturbation theory
are generically of the form ${(g\xi)}^n$ as $g\xi\ra\i$. Secondly, if
one uses a $g$ independent RG one finds a running coupling which, in
the same regime, becomes very strong. For example, for $\l\ff$ theory
on $S^1\times R^3$ one finds $$\l(L\k)={\l\o1+{3\l\o16\pi^2}{\rm
ln}{L\k}}$$ where $\k$ is the momentum/mass scale of interest. Clearly
in the IR regime, $L\k\ra0$, one is entering a strongly coupled regime
reminiscent of what occurs in QCD. The words ``divergences'' and
``strong coupling'' lead invariably to the invocation of
``non-perturbative'' techniques. It sometimes seems that what is meant
by the latter is something which cannot be formulated perturbatively in
terms of any coupling, such as doing a lattice simulation. We believe
that quite often the case is overstated. To us, at least in many cases,
``divergences'' and ``strong coupling'' are symptomatic of the fact
that one is implementing a perturbation theory which is not capturing
the correct qualitative nature of the effective degrees of freedom.
One of the main purposes of this paper is to show that more often than
not these deficiences can be overcome perturbatively by finding an
appropriate environment dependent expansion parameter.

In the case of a theory with an $O(N)$ symmetry, if one uses a $1\o N$
expansion, one finds that if a $g$ independent renormalization is used,
the $1\o N$ expansion breaks down in the $g\xi\ra\i$ limit, except when
$N\ra\i$ before $g\xi$ does.  This is the spherical model limit, which
is exactly solvable. One could also consider summing up sets of Feynman
diagrams such as is done in summing up the daisy diagrams (Hartree/Fock
approximation) in finite temperature field theory \ref{45}.
Naturally, this is always a tricky proposition as certain diagrams that
are dominant in one regime might not be so dominant in another.  One
must still also address the question of how to renormalize.
Additionally, one faces the problem that any crossover accessed by the
resummation might not be the one of interest. This is in fact what
happens in finite temperature field theory where the resummation of
daisy diagrams in an attempt to get an improved description of a phase
transition in a relativistic field theory merely accesses a mean field
fixed point instead of the fixed point associated with the transition.

The key to understanding whether a particular perturbation parameter is
suitable or not is to see how it gets dressed by fluctuations. If an
effective coupling is small in one regime there is no guarantee that it
will be small in some other.  That is why one implements
renormalization in the first place. One could consider a bare coupling
to be small consistently if one was interested in physics at scales
$\sim\L$. However, when one is interested in scales $p\ll\L$ one finds
that the coupling gets large perturbative dressings, so one
renormalizes to a point $\k\sim p$. In this case one has taken all the
fluctuations between $\k$ and $\L$, absorbed them into $\l_B$, and
called it $\l_{\k}$, which is presumed to be a perfectly reasonable,
finite, ``observable'' coupling. The best thing to do of course is
consider an infinitessimal dressing --- this yields the $\b$ function,
the fixed points of which correspond to those points where further
dressing does not change the coupling. One of the main keys then to
getting a reasonable perturbative parameter is having a good handle on
the dressing of the coupling.

Let us now ask by what fluctuations the coupling is getting dressed?
The fluctuations that represent the effective degrees of freedom at the
scale of interest of course. Usually we do not have an exact
representation of these fluctuations so we represent them
perturbatively. Now, one might think it quite perverse to dress a
renormalized parameter with fluctuations that were not a good
representation of the effective degrees of freedom of the system,
however, this is exactly what is commonly done, as if for example one
uses a $g$ independent RG. Such a procedure is not terrible at all
scales of course. When $\k\gg g$ the effective degrees of freedom are
effectively $g$ independent and therefore for considering physics at
these scales it would not be a bad approximation.  The fluctuations are
environment dependent and therefore the dressing of the coupling should
be environment dependent. If one implements an environment dependent
renormalization then one's effective expansion parameter is the running
coupling which, for example, is a solution of (\docref{lbg}) in the $g$
dependent case. The fixed points of (\docref{fpo}) are (\docref{fpt}),
neither of these is a good expansion parameter throughout the
crossover.

One is not guaranteed of course that the running coupling will be small
throughout the crossover. This will in fact be the case in this paper
where the coupling $\sim 1$ in certain asymptotic regimes.  One could
take the philosophy that only the lowest order terms should be believed
in giving the solution, then the higher order terms are used merely to
generate an iterated form of the lowest order solution. The preferable
thing though is to use a resummation procedure. For our two loop
results we use a [2,1] Pad\'e approximant \ref{40} to find
\eqlabel{\pade} $$\beta(h)=-\el h+{h^2\o 1+Fh}\no$$ where
$$F=-{4\left((5N+22)\Aif+(N+2)\Aitrp\right)
\o{\left((N+8)\Ait\right)}^2}$$
The explicit coupling obviously depends
on the crossover in question.  Naturally one would prefer to work to
more than two loops, there is nothing in principle in our formalism
besides tedious computation to prevent this from being done.  So the
idea then of implementing perturbation theory is the following:  one
first of all implements it in terms of the running coupling constant
$h$ generated from a $g$ dependent RG. The actual $h$ used is then the
resummed solution of the $\b$ function equation. It is important to
realize that in the case of the crossover this will be a function, not
a pure number, though for a given $\rho$ and $g$, which can be
arbitrary, it is just a particular number.

So, what other subtleties does the RG engender? Consider the solution
of the RG equation (\docref{rgrn}) for $\G^{(2)}$ in the disordered
phase $$\G^{(2)}(t,h,g,\k)=(\rho\k)^2e^{-\int_1^{\rho}\gf(x){dx\o x}}
\G^{(2)}({t(\rho)\o \rho^2\k^2},h(\rho),{g\o\rho\k},1)\no$$ Just
because one can derive a coupling constant $h(\rho)$ that is not too
badly behaved through the crossover does not imply that a perturbative
expansion of $\G^{(2)}$ will be sensible, the reason being that
$h(\rho)$ is only one component in the expansion. $h(\rho)$ multiplies
functions of $g\o\rho\k$ and $t(\rho)\over\rho^2\k^2$. Even if $h$ is
not too large these functions may become large in certain physical
regimes, e.g.  ${t\o g^2}\ra0$. If this is the case (and this will be
the case in crossovers) then one must invoke the full power of the RG
by using the arbitrariness in the renormalization scale $\rho$ to find
a point $\rho_c$ where the functions are not large. If at this point
one  also finds that $h(\rho_c)$ is not too large then one can make
real progress. Obviously $\rho_c$ will be explicitly $g$ dependent. The
value of $\rho_c$ chosen will usually be determined by the fact that
one is interested in a regime where the correlation length is large
relative to $g^{-1}$ say. In such a regime perturbation theory breaks
down in a fairly generic fashion.  The thing to do, as is well known,
is to map, using the freedom of choice in the renormalization scale, to
a correlation length that is well out of the critical regime. What
correlation length must one map?  We discussed this matter somewhat in
the last section. If one uses a $g$ independent RG then the natural
correlation length is $\xi_0$, i.e. the correlation length in the
isotropic system. Using such an RG, however,  will make certain regions
of $\cal M$ inaccessible. The better correlation length to use is the
physical one, naturally associated with the $g$ dependent RG, which is
global and can access all the fixed points of $\cal M$. A good mapping
point, for instance, would be determined by
$t(\rho_c)=\rho_c^2\k^2$,
whereupon $\rho_c$ satisfies \eqlabel{\cond}
$$\rho_c^2={t\o\k^2}e^{\int_1^{\rho_c}\gft(x){dx\o x}}\no$$ where
$t=(T-T_c(g))$, i.e a $g$ dependent renormalization has been used.  It
is this key equation which must be solved perturbatively.

What do we mean by solving it perturbatively? Ultimately what one is
expanding perturbatively are the characteristic functions of the RG, in
particular the quantities $\gft$, $\gf$ and $\gl$. One determines them
as series in the solution of the resummed $\b$ function differential
equation. One should in fact, based on a democratic treatment of all
the characteristic functions, Pad\'e resum the series for $\gft$ and
$\gf$ as well. To two loops, obviously $\gf$ cannot be resummed. The [2,1]
Pad\'e resummed expression for $\gft$ is
$$\gft=\left({N+2\o N+8}\right){h\o{1-{6\o(N+8)}(\fone+{1\o3}\ftwo)h}}\no$$
For small values of $N$ one finds that the
effective exponents derived from [2,1] Pad\'e resummed  Wilson functions
differ little from those derived from a using resummed $\beta$-function for the
floating coupling with the perturbative series for the effective exponents
in this coupling.  See Fig.10 for an illustration of this in the context of
dimensional crossover. The former prescription caprures the exactly solvable
$n=\i$ limit and therefore seems  prefered.
Correlation functions are related to exponentials
of the Wilson functions, it would be quite against the spirit of
the RG to start
perturbatively expanding the exponentials, otherwise there would be
little point in using the RG in the first place. This is one of the
problems with trying to use $\varepsilon$ expansion methods to
calculate scaling functions. The question is always: if something is to
a power of $\varepsilon$ should one $\varepsilon$ expand it? In
(\docref{cond}) one should expand $\gft$ in the integrand to a given
order $h^n$, $h$ being the resummed solution of the $\b$ function
differential equation to order $h^{n+1}$. {\bf One should not expand
the exponential}. To lowest order for instance, one would solve
$$\rho_c^2={t\o\k^2}{\rm exp}\left[ \left({N+2\o
N+8}\right){\int_1^{\rho_c}h(x,{g\o x}){dx\o x}}\right]\no$$ where $h$
is a solution of $$\rho{dh\o d\rho}=-\e(\bub)h+h^2\no$$ More often than
not one will have to resort to numerical techniques to solve the
equation that fixes $\rho$, however, we will see shortly that as long
as one captures the essential physics of the crossover with an
appropriate RG then one's final answers are rather robust under changes
in solving the condition.

Besides the ambiguity in using the $\varepsilon$ expansion to determine
scaling functions, i.e. should or shouldn't one expand something which
is to a power of $\varepsilon$, there are two other defects. First of
all, if one makes a definite choice in deciding to expand everything
in $\varepsilon$, as is done in finding the equation of state
\ref{46}, one will find that the resulting
expression is only thermodynamically valid in a restricted region of
the thermodynamic state space. Secondly, if one is considering a
crossover wherein the upper critical dimension changes from one
asymptotic region to another, then an $\varepsilon$ expansion will be
inadequate, being able to capture the singularity at one end but
totally failing to capture the singularity at the other. In
principle,
in such circumstances a $1\o N$ expansion could work, if used in
conjunction with an appropriate environmentally friendly
renormalization. However, it is well known that the $1\o N$ expansion
gives poor results for small values of $N$, is difficult to implement
at higher orders, and additionally such an expansion would be useless if
there was a change in the number of symmetry components of the order
parameter during the crossover, e.g. for the bicritical crossover. We
believe our perturbation technique to be the only one on offer
applicable to crossovers where both the dimensionality and/or the
number of components of the order parameter can change.

In the above we are exploiting the arbitrariness in the RG scale to
allow us to map to a point where perturbation theory can be trusted. We
now discuss how one can exploit the various renormalization conditions
in order to facilitate calculations. We have already seen that the type
of condition used, in the sense of dependence or independence of the
environment, is quite crucial to achieving a description of crossover
systems.  There are, as pointed out, different, inequivalent
representations of the RG.  Having determined the RG that is most
suitable, and having determined the RG scale that allows perturbation
theory to be implemented across the crossover, one can make further
simplifications by trying to put as much of the physics as possible
into the characteristic functions alone.

What this means is that for an object like the inverse susceptibility,
choosing a renormalization condition like (\docref{gta}), implies
\eqlabel{\sus} $$\G^{(2)}(1,h(\rho_c),{g\o\rho_c\k},1)=1\no$$
which gives
\eqlabel{\susc} $$\G^{(2)}(t,h,g,\k)={\rm
exp}\int_1^t\left({2-\gf\o2-\gft}\right){dx\o x}\no$$
Clearly through
the choice of normalization condition all the physics has been put into
the two functions $\gf$ and $\gft$; i.e.
$\G^{(2)}(1,h(\rho_c),{g\o\rho_c\k},1)$ did not need to be
perturbatively expanded. According to the framework outlined above, a
perturbative treatment of (\docref{susc}) entails expanding
$\beta$, $\gf$ and $\gft$ as a power series in $h$ to a certain order, Pad\'e
resumming the resulting series, and
then putting in $h$ as the resummed solution of the $\b$ function. We
could have used a condition other than (\docref{sus}), such as
\eqlabel{\mless}
$$\Gamma^{(2)}(0,h(\r),{g\o\r\k},1)=0\no$$. In such
circumstances
$\G^{(2)}(1,h(\rho_c),{g\o\rho_c\k},1)$ would also be perturbatively
expanded in $h$, hence not all the physics would be in the
characteristic functions.  One would obtain an expression of the form
\eqlabel{\diff}
$$\G^{(2)}(t,h,g,\k) ={\rm
exp}\left({\s{\int_1^t\left({2-\gf\o2-\gft}\right){dx\o x}}}\right)
\left(1+h(\rho_c)B_1({g\o\k\rho_c}) +h^2(\rho_c)B_2({g\o\k\rho_c})+
\cdots\right)\no$$ where $B_1$ and $B_2$ are dependent on the precise
normalization condition one uses. For the massless normalization condition
(\docref{mless}) $B_1$ contains a contribution from the difference between
the massive and massless tadpole. At the one loop level we could have
exploited the arbitrariness in the renormalization scale $\r$ to set
$\G^{(2)}(1,h(\rho_c),{g\o\rho_c\k},1)=1$. In this case $\gft$ will
not be the same as found using the condition (\docref{sus}). A quantity
such as an effective exponent, however, will be invariant
under such a change in normalization point. Unfortunately this would not work
beyond the one loop level, for $d<4$, due to the presence of IR divergences
in $B_2$, as was pointed out in section 4.
If one wanted to
calculate a scaling function, such as the susceptibility above, the
functions $B_1$ and $B_2$ would, for instance, carry information about the
difference
between the GMS method used in \ref{15} and \ref{19}, and using a normalization
condition, as we have used in our
method. Although they might not be IR divergent, as emphasized above, this
difference is not easily exponentiated, so
that in the calculation of a scaling function GMS will yield less
information than using normalization conditions. This is one disadvantage
of GMS versus our method.

It is useful to give an analog expression diagramatically to lowest
order in the coupling $\lb$ $$\G^{(2)}(t,h,g,\k) ={\rm
exp}\left({\ss{\int_1^t\left(1-\nt{6}(A_{(4-d)}\bub)\lb\right){dx\o
x}}} \right)\left(1+{\lb\o6}(N+2)({\io}'-{\io})+ \cdots\right)\no$$ The
quantity $({\io}'-{\io})$ is a measure of what has been ``left out'' of
the RG and would be the analog of ${\rm ln}({t\o\k^2})$ in four
dimensions, which is the difference to lowest order between the
normalization condition $\G^{(2)}(t)=t$ and minimal subtraction. The RG
equation tells us how to perform an infinitessimal dressing of a
quantity.  By different choices of counterterm one can dress the
quantity with different fluctuations, $g$ dependent, $g$ independent
etc. Among the $g$ dependent schemes themselves though, as we are
seeing here, one can dress things differently. Should one dress things
using the whole of a Feynman diagram, or what we think to be the most
important part of it, e.g. the mass independent part, or....? The RG
only tells us about dressings, but because we can integrate an
infinitessimal dressing, one can, in principle, derive a huge amount of
information from it, as this integrated dressing gets exponentiated in
the process of solving the RG equation. What is not included as
dressing cannot be exponentiated, therefore one potentially loses a
large amount of predictive power. Manifestly the best thing do do then
is dress up as much as possible, and particularly in an environmentally
friendly manner.

\beginsection{\bf THE FLOATING FIXED POINT}
In this section we will
introduce the concept of the floating fixed point.  We said in section
6 that to find the fixed points of the RG one had to find the fixed
points of all the $\b$ functions, including $\b(\bar g)$.  We exhibited
the two relevant fixed points. If we had just examined $\b(\lb)$ in
isolation, and set $\b(\lb)=0$, to lowest order one finds
$\lb={d-4\o\ne{6}(A_{(4-d)}\bub)}$. In the limit ${g\o\rho\k}\ra\i$ one
finds that $\lb\ra0$. This implies that it wasn't very useful to think
of the algebraic zero of the $\b$ function as a fixed point as it did
not cross over between the right values (a more detailed discussion of
this can be found in \ref{27}).  If we use the floating coupling,
however, and consider $\b(h)=0$, one finds that the solution of this
equation interpolates between the correct fixed points of the
crossover.  The condition $\b(h)=0$ defines the floating fixed point
$h^{\ast}$.  To lowest order $h^{\ast}=\el$. To higher order one could
try to iterate a solution, i.e take this solution, put it into the
$h^3$ term in $\b(h)$, and resolve the algebraic equation $\b(h)=0$. To
the next order the iterative solution is
$$h^{*}(z)=\e(z)+{4\o{(N+8)}^{2}}\left((5N+22)f_{1}(z)
-(N+2)f_2\right)\e(z)^2+O(\e(z)^3)$$ Finally one can Pad\'e resum the
floating fixed point to obtain
$$h^{*}(z)={\e(z)\over
1-{4\o{(N+8)}^{2}}\left((5N+22)f_{1}(z) -(N+2)f_2\right)\e(z)}$$

The utility of using the floating fixed point is that to find it one
need only solve an algebraic equation instead of a differential
equation.  The floating fixed point is guaranteed to give the correct
asymptotic answers, i.e. the floating fixed point coincides at $\bar
g=0$ and $\bar g=\i$ with the isotropic and anisotropic fixed points
which are the fixed points of the system of differential $\b$ function
equations. Naturally the solution of the differential equation and the
algebraic equation will differ through the crossover. This difference
is found (in all cases examined so far) to be relatively small, so the
floating fixed point can be thought as giving a good approximation to
the true crossover $h$.

One can go through the entire analysis of sections 6 and 7 expanding
everything around the floating fixed point in analogy to expansion
around a normal fixed point.  For example,
the non-linear scaling field $\cgt$
expanded around the floating fixed point becomes
$$\cgt=A(t,g){\rm
exp}\left(-\int_1^t{\nu\ef}^{\ast}{dt'\o t'}\right)\no$$ where
$A(t,g)={\rm exp}\int_1^t({\nu\ef}^{\ast}-{\nu\ef})$ is a correction to
scaling factor around the floating fixed point and $\nu^{\ast}$ is the
effective correlation length exponent evaluated at the floating fixed
point.  Remember that ${\nu\ef}^{\ast}$ is a function of $t$ and $g$,
therefore one cannot pull it out of the integral as if it were a
constant. This can only be done asymptotically where $\nu^{\ast}$
becomes the exponent at the true fixed points. If we had expanded
around the isotropic fixed point we would find $$\cgt=B(t,g){\rm
exp}\left(-\int_1^t\nu(0){dt'\o t'}\right)\no$$ where $B(t,g)={\rm
exp}\int_1^t(\nu(0)-{\nu\ef})$ is a correction to scaling factor
around the isotropic fixed point. The key difference is that $A$ is
small throughout the crossover and approaches zero in the asymptotic
regimes, whereas $B$ goes to zero near the isotropic fixed point but
becomes singular near the anisotropic fixed point due to the fact that
it must contain a singularity of the form $t^{-(\nu(\infty)-\nu(0))}$
in order that the isotropic singularity be replaced by the
anisotropic one.
Thus using the floating fixed point corrections to scaling
can be made small throughout the crossover. Importantly, in our
formalism the corrections to scaling around the floating fixed point
can also be calculated explicitly.  One can define effective exponents,
${\alpha\ef}^{\ast}$, ${\beta\ef}^{\ast}$, ${\gamma\ef}^{\ast}$,
${\delta\ef}^{\ast}$,  ${\nu\ef}^{\ast}$ and an effective dimension
$\dea$ associated with the floating fixed point and find that they obey
exact analogs of the equations (\docref{sexplaw}-\docref{shexplaw}).
Similarly for all the non-linear scaling fields, there are floating
fixed point analogs which differ from the one's already defined only by
small corrections to scaling throughout the crossover.

\beginsection{\bf DIAGRAMMATIC REPRESENTATION OF EFFECTIVE EXPONENTS
AND SCALING FIELDS\hfil}
In the last sections we tried to convey the utility
of the field theoretic RG if implemented judiciously. In this section
we wish to derive perturbative expressions for the effective exponents
and some of the scaling fields mentioned in section 6. Things will
still be formal to the extent that we do not specialize yet to a
particular crossover.  The reason for doing this is that in our
diagramatic notation it is clear that any crossover can be treated by
putting into the diagrams the effects of the particular environmental
parameters that are inducing that crossover, whatever they may be. We
have already given the expression for the Wilson functions
diagramatically so we just need to construct the effective exponents
from them and $\de$.  Note that here we do not Pad\'e resum $\gft$
and we implement the scaling laws perturbatively in $h$. These
expressions will be useful for comparative purposes
and as we shall see are quite reliable for small values of  $N$. Analogous
expressions can be obtained when $\gft$ is also resummed, in which case
the effective exponents should be constructed from these resummed quantities
without further expansion.

One finds to two loops \eqlabel{\etaefd}
$${\eta\ef}=2{(N+2)\o{(N+8)}^{2}}\ftwo h^2\no$$ \eqlabel{\nuefd}
$$\eqalign{{\nu\ef}={1\o2}+&{(N+2)\o 4(N+8)}h+\cr
&{(N+2)\o8{(N+8)}^2}\left(N+2+12\fone+4\ftwo\right)h^2\cr}\no$$
\eqlabel{\gamefd} $$\eqalign{{\gamma\ef}=1+&{(N+2)\o 2(N+8)}h+
{(N+2)\o4{(N+8)}^2}\left(N+2+12\fone)\right)h^2\cr}\no$$ Using
$\de=4-\el-{\b(h)\o h}$ one finds \eqlabel{\deffd} $${d\ef}=4-h-{4\o
{(N+8)}^{2}}\left((5N+22)\fone+ (N+2)\ftwo\right) h^2\no$$ and hence
\eqlabel{\alphd} $${\alpha\ef}={(4-N)\over 2(N+8)}h +
\left({(4-N)(N+2)+16(N+8)\fone\o 4(N+8)^2}\right)h^2\no$$
\eqlabel{\betad} $${\beta\ef}={1\over2} -{3\over
2(N+8)}h-\left({3(N+2)+2(7N+38)\fone\o 4(N+8)^2}\right)h^2\no$$
\eqlabel{\deltd}
$${\delta\ef}=3+h+{1\o2}\left(1+8{(5N+22)\o{(N+8)}^2}\fone\right)h^2\no$$
If one solves the Pad\'e resummed differential equation (\docref{pade})
and inserts it in these expressions then one has the effective
exponents for a general class of crossovers. For the Pad\'e resummed
floating fixed point $$h^{\ast} =\el{\left({\s
1+{4\o(N+8)^2}}\left({\ss
(5N+22)\fone+(N+2)\ftwo}\right){\s\el}\right)}^{-1}\no$$ Thus one finds
$${\eta\ef}^{\ast} ={\s 2{(N+2)\o(N+8)^2}\ftwo{\el}^2} {\left({\s
1+{4\o(N+8)^2}}\left({\ss (5N+22)\fone
+(N+2)\ftwo}\right){\s\el}\right)}^{-2}\no $$
$$\eqalign{{\nu\ef}^{\ast}=&{1\o2}+{(N+2)\o 4(N+8)} \el{\left({\s
1+{4\o(N+8)^2}}\left({\ss
(5N+22)\fone+(N+2)\ftwo}\right){\s\el}\right)}^{-1}\cr
&+{(N+2)\o8{(N+8)}^2}{\s(N+2+12\fone+4\ftwo){\el}^2}\cr &\times
{\left({\s 1+{4\o(N+8)^2}}\left({\ss
(5N+22)\fone+(N+2)\ftwo}\right){\s\el}\right)}^{-2}\cr}  \no$$ With the
effective dimension at the floating fixed point, $\dea=\el$, one can
easily derive all the other effective exponents at the floating fixed
point. We leave this as an exercise for the reader.

Consider now the non-linear scaling fields
$\cgt=e^{-\int^t_1{\nu\ef}{dt'\o t'}}$ and
$x=te^{-{\int_1^{\fb}{1\over\beta\ef}{d{\fb}'\over {\fb}'}}}$.
Diagramatically $$\eqalign{\ & \ \ \cgt=\cr &{\rm
exp}\left(-\int^t_1{\ss \left({1\o2}+{(N+2)\o 4(N+8)}h+
{(N+2)\o8{(N+8)}^2}\left(N+2+12\fone+4\ftwo\right)h^2 \right){dt'\o
t'}}\right)\cr}\no$$ and $$x=t\ {\rm exp}\left(-2\int_1^{\fb} \left({\s
{1+{3h\o(N+8)}+}\left[{\ss 3(N+12)
+2(7N+38)\fone}\right]{\s{h^2\o2{(N+8)}^2}}}\right)\right) \no$$ In
terms of the Pad\'e resummed floating fixed point $$\eqalign{\cgt=\ &{\rm
exp}\left(-\int^t_1\left({\s {1\o2}+{(N+2)\o 4(N+8)}\el}
 {\left({\s 1+{4\o(N+8)^2}}\left({\ss (5N+22)\fone
+(N+2)\ftwo}\right){\s\el}\right)}^{-1}\right.\right.\cr &
+{\s{(N+2)\o8{(N+8)}^2}(N+2+12\fone+4\ftwo){\el}^2}\times\cr
&\left.\left.{\left({\s 1+{4\o(N+8)^2}}\left({\ss (5N+22)\fone
+(N+2)\ftwo}\right){\s\el}\right)}^{-2} \right){\s {dt'\o
t'}}\right)\cr}\no$$ and
$$\eqalign{x=\ &t\ \exp\left(-2\int_1^{\fb}\left({\s 1+{3\o(N+8)}\el}
{\left({\s 1+{4\o(N+8)^2}}\left({\ss (5N+22)\fone
+(N+2)\ftwo}\right){\s\el}\right)}^{-1} \right.\right.\cr &  +{\s
{\el\o 2{(N+8)}^2}(3(N+12)-2(7N+38)\fone)}\times\cr
&\left.\left.{\left( {\s 1+{4\o{(N+8)}^2}}\left({\ss (5N+22)\fone
-(N+2)\ftwo}\right){\s\el}\right)}^{-2}  \right)\right)\cr}\no$$

It is impressive to see from these expressions just how much
diagramatic information has been extracted using the RG.  The reason we
have exhibited these cumbersome looking expressions, is that they can
serve as ``black boxes'' for evaluating crossover functions. For any
crossover falling within the general class treated, in our two loop
perturbative analysis, the reader may take the expressions given,
insert the appropriate propagator into the diagrams, crank the handle,
either numerically or analytically, and obtain the corresponding
crossover scaling information.  Now it is time to turn to a specific
example of a crossover to show all the ideas and formalism we have
presented at work.

\beginsection{\bf EXPLICIT DIMENSIONAL CROSSOVER RESULTS}
\subsec{\bf Periodic Boundary Conditions: One Loop Results}
The specific crossover we will consider here is that of a $d$
dimensional layered system satisfying periodic boundary conditions (or
in particle physics language $\l\ff$ theory on $R^{d-1}\times S^{1}$).
We consider, to begin with, an $O(N)$ vector model.  We will be
concerned here with $T>T_c(L)$, where $T_c(L)$ is the critical
temperature of the finite size system, and $d\leq4$. We will also
assume here that the finite system does exhibit a critical point,
though the formalism is equally applicable to the case where such does
not exist.  The environment here is the thickness of the system, which
we denote by $L$, and the periodic boundary conditions. The results
below are equally applicable to relativistic finite temperature after
the trivial substitution $L=1/T$ and $\xi_{L}=1/m_{T}$.

Let us begin by examining the one loop expressions \eqlabel{\betahol}
$$\beta(h)=-\e(\r\k L)h+h^2+O(h^3)\no$$ \eqlabel{\gftNol}
$$\gamma_{\f^{2}}={(N+2)\o (N+8)}h+O(h^2)\no$$
$\gf=0$ to one loop. The
solution of (\docref{betahol}) is \eqlabel{\betahNol}
$$h(\kl\rho)={e^{-\int_{1}^{\rho}\e(\kl x){dx\o x}}\o{h_0^{-1}
-\int_1^{\rho}e^{-\int^x_1\e(\kl x'){dx'\o x'}}{dx\o x}}}\no$$
Choosing
$\alpha_1=0$, $\alpha_{2}={\rc^2}$, in our normalization schemes we
have $\rc\k=\xi_L^{-1}$. Thus we have $h$ in terms of
$z={L/\xi_{L}}$, and an initial coupling at the scale $\kappa$, or
equivalently $z_{0}=\kl$.  Substituting this into the one loop
contribution for $\gft$ we have
\eqlabel{\gftNol} $$\gft(z)={(N+2)\o
(N+8)}{e^{-\int_{z_{0}}^{z}\e(x){dx\o x}}\o{h(z_{0})^{-1}
-\int_{z_0}^{z}e^{-\int^{x}_{z_{0}}\e(x){dx'\o x'}}{dx\o x}}}\no$$
Now, in $d<4$ there is a non-trivial fixed point of the bulk system,
and hence  if we impose the initial condition on the RG flow at
$z_{0}=\i$, i.e. at zero correlation length, and choose a finite
renormalized coupling $h(\i)$, we will be probing the universal part of
the crossover. This is not a necessity but merely allows us to isolate
this universal part. Often, one is also interested in corrections to
scaling, if $z_0$ is not chosen to be infinite then these are
automatically included in the expressions, and as emphasized in section
7 the effective exponent scaling laws are still valid in this case. One
finds
\eqlabel{\separetrix}
$$h(z)=-{A_{4-d}\bub\o\bub}\no$$
which depends on only the variable $z$, and interpolates between the
bulk and reduced fixed points. This is the equation of the
separatrix between the two fixedpoints in the one loop approximation.
It is not difficult to verify that (\docref{separetrix}) satisfies
the one loop $\beta$ function equation. Note that the component diagrams are
finite for $d<4$ but for $d=4$ $\bub$ diverges, it is therefore impossible to
eliminate the dependence on $z_0$ in this case.

With periodic boundary conditions we have
$$\e(z)=5-d-(7-d){{\displaystyle\sum_{n=-\i}^{\i}{4\pi^2n^2\over z^2}
\left(1+{4\pi^2n^2\over z^2}\right)^{d-9\over2}}
\over{\displaystyle\sum_{n=-\i}^{\i}\left(1+{4\pi^2n^2\over
z^2}\right)^{d-7\over2}}}\no$$
and the universal one loop floating
coupling
$$h(z)=(5-d){{\displaystyle\sum_{n=-\infty}^{\infty}}{(1+{({2\pi n\o
z})}^2)}^{(d-7)\o2}
\o{\displaystyle\sum_{n=-\infty}^{\infty}}{(1+{({2\pi n\o
z})}^2)}^{(d-5)\o2}}\no$$
{}From these expressions the  effective exponents can be evaluated.
Rather than writing the expressions down explicitly, as we will be
presenting shortly the full two loop expressions we will restrict
ourselves here to making some specific comments about particular
dimensions.

Conveniently for $d=3$, $\e(z)$ and $h(z)$ are expressible in terms of
elementary functions. We have explicitly \eqlabel{\epstd}
$$\e(z)=1+{z^2\coth({z\o2})\o\sinh z+z}\no$$
Similarly an explicit form for $h(z)$ is
$$h^{-1}(z)={z\sinh({z\over 2})^2\over\sinh z+z}({1\over h(z_{0})}{\sinh
z_0+z_0\over z_0\sinh({z_{0}\over2})^2} -2{\coth({z_{0}\over2})\over
z_{0}})+{\sinh z\over \sinh z+z}\no$$
We present $h$ plotted against $\ln(1/z)$ in Figure 22, where we choose
$h_{0}=.9$ for $\ln(1/z_{0})=-7.5$. The crossover to meanfield theory is
evident as $z\ra\i$.

Focusing on the universal part
of the crossover by choosing $z_{0}\ra\infty$, with $h(z_{0})$
finite we find \eqlabel{\htd} $$h(z)=1+{z\o\sinh z}\no$$ Since in
the case of the two dimensional end only the $N=1$ model exhibits a
standard second order phase transition we restrict our
considerations to this model. (The techniques used here though can
also be used to treat other values of $N$, in particular dimensional
crossover in a non-linear $\sigma$ model \ref{47}). We thus find
\eqlabel{\gftthreed} $$\gft(z)={1\o3}(1+{z\o {\rm sinh}z})\no$$ As
$\gf=0$ to one loop so is $\eta\ef$. Similarly, substituting for $h$
in (\docref{nuefd}) for $\nu\ef$ to one loop gives
\eqlabel{\nefthreed} $${\nu\ef}={7\o12}+{z\o12 \sinh z}\no$$ As
$z\ra\i$, ${\nu\ef}\ra 0.58$ and as $z\ra0$, ${\nu\ef}\ra 0.67$.
{}From (\docref{gamefd}) our expression for $\gamma\ef$ is
\eqlabel{\gefthreed} $${\gamma\ef}={7\o6}+{z\o6\sinh z}\no$$ which
varies between $1.17$ and $1.33$. To one loop ${d\ef}=3-z/\sinh z$
and varies between $3$ and $2$ as $z$ varies between $\i$ and $0$.
Using the scaling law (\docref{dexplaw}) one gets
${\delta\ef}=4+{z\o \sinh z}$ which varies between $4$ and $5$; and
from (\docref{bexplaw}) one finds ${\b\ef}={1\o3}-{z\o 6\sinh z}$
which varies between $0.33$ and $0.17$; and finally
${\alpha\ef}={1\o6}(1+{z\over\sinh z})$. One can also determine
approximations to the effective exponents by calculating them with
respect to the floating fixed point.  We emphasize once again that
one can independently verify the scaling laws by studying the system
when $T<T_{c}(L)$ (see \ref{29}) and calculating the effective
exponents directly.

In the case of $d=4$ we are not as fortunate in finding expressions in
terms of elementary functions for our one loop results, however, the
expressions involving the sums themselves are not too bad and are
rapidly convergent.  Explicitly \eqlabel{\epsfd}
$$\e(z)=1-3{{\displaystyle\sum_{n=-\i}^{\i}{4\pi^2n^2\over z^2}
\left(1+{4\pi^2n^2\over z^2}\right)^{-5/2}}
\over{\displaystyle\sum_{n=-\i}^{\i} \left(1+{4\pi^2n^2\over
z^2}\right)^{-3/2}}}\no$$ The one loop solution to the $\beta$ function
is given by (\docref{betahNol}).  It is now not possible to set the
initial value of the coupling to a finite value and take the limit of
$z_{0}\ra\infty$. The logarithmic corrections  to scaling in $d=4$
mean that the running coupling will retain a dependence on the the
initial condition, i.e. the scale $\kappa$ will remain even when $L$
drops out. Since the one loop results have been presented elsewhere
\ref{27} we will not repeat them here.

\subsec{\bf Periodic Boundary Conditions: Two Loop Results}
{}From the considerations of the previous sections we can
conclude that the Wilson functions are given by \eqlabel{\betahN}
$$\beta(h,z)=-\e(z)h+h^2-{4\o {(N+8)}^{2}}((5N+22)f_{1}(z)-
(N+2)f_{2}(z)) h^3+O(h^4)\no$$ \eqlabel{\gftN}
$$\gamma_{\f^{2}}={(N+2)\o (N+8)}h-6{(N+2)\o{(N+8)}^{2}}\bigl(f_{1}(z)
-{1\o 3}f_{2}(z)\bigr) h^2+O(h^3)\no$$ \eqlabel{\gfN}
$$\gamma_{\f}=2{(N+2)\o{(N+8)}^{2}}f_{2}(z) h^2+O(h^3)\no$$ where the
functions $f_1$ and $f_2$ applicable for general $d$ for this
particular crossover are given in Appendix A. In Appendix B the
explicit, analytic expressions for the case $d=4$, $\a_1=0$, $\a_2=1$
are presented.  Recalling that  ${\nu\ef}={1/(2-\gft)}$ and
${\eta\ef}=\gf$ we have to two loops that \eqlabel{\nuefN}
$${\nu\ef}={1\o2}+{(N+2)\o 4(N+8)}h+{(N+2)(N+2-12 f_{1}(z)
+4f_{2}(z))\o8(N+8)^2}h^2\no$$
Similarly, from the scaling relation for
$\gamma\ef$ (${\gamma\ef}={\nu\ef}(2-{\eta\ef})$), which we have
established in section 7, we have \eqlabel{\gefN}
$${\gamma\ef}=1+{(N+2)\o 2(N+8)}h +{(N+2)(N+2-12f_{1}(z))\o
4(N+8)^2}h^2\no$$
and from the defining relation
${d\ef}=4-\e(z)-{\beta(h)\over h}$ we have \eqlabel{\defN}
$${d\ef}=4-h+{4\o {(N+8)}^{2}}((5N+22)f_{1}(z)- (N+2)f_{2}(z))
h^2\no$$ Using the scaling law for $\alpha\ef$
(${\alpha\ef}=2-{\nu\ef}{d\ef}$) yields
$${\alpha\ef}={(4-N)\over
2(N+8)} h + {((4-N)(N+2)-16(N+8)f_{1}(z))\over 4(N+8)^2}h^2\no$$
The scaling law for $\beta\ef$
(${\beta\ef}={\nu\ef\o2}({d\ef}-2+{\eta\ef})$) gives us
$${\beta\ef}={1\over2}-{3\over 2(N+8)}h
-{(3(N+2)-2(7N+38)f_{1}(z))\over 4(N+8)^2}h^2\no$$ And finally using
the scaling law for $\delta\ef$
(${\delta\ef}=({d\ef}+2-{\eta\ef})/({d\ef}-2+{\eta\ef})$) gives
$${\delta\ef}=3+h+{((N+8)^2-8(5N+22)f_{1}(z))\over 2(N+8)^2}h^2\no$$

Note that the functions $\e(z)$, $f_{1}(z)$ and $f_{2}(z)$ are
independent of $N$. For a three dimensional layered geometry
($R^2\times S^1$), we find $\e(z)$ interpolates between $2$ and $1$,
$f_{1}(z)$ interpolates between $0.28$ and ${1\o3}$ and $f_{2}(z)$
interpolates between $0.23$ and ${4\over 27}$ as $z$ varies from $0$ to
$\infty$. For a four dimensional layered geometry ($R^3\times S^1$),
$\e(z)$ ranges monotonically from $1$, for $z=0$, to $0$ for
$z=\infty$,
and takes the value ${1\o2}$ for $z\approx 3.3$. The function
$f_{1}(z)$ ranges from ${1\o3}$, for $z=0$ to ${1\o2}$ for $z=\infty$;
$f_{2}(z)$ ranges from ${4\o27}$ for $z=0$ to ${1\o4}$  for $z=\infty$.
Thus the equations (\docref{betahN}), (\docref{gftN}), (\docref{gfN})
interpolate between those for $R^{3}$ and $R^{4}$ as $z$ ranges from
$0$ to $\infty$.  We provide plots of $\e(z)$, $f_{1}(z)$ and
$f_{2}(z)$ in Figure 1 for the layered four dimensional geometry
with periodic boundary conditions ($R^3\times S^1$).

To explicitly implement the two loop results we resort to a Pad\'e
resummation technique. As is well known it sometimes occurs that the
zero of the $\beta$-function disappears at two loops only to return at
three loops. This is a property of the perturbative series and is not
specific to our problem. It is equally true in the fixed dimension
non-crossover case treated by Parisi. We therefore adopt Parisi's
approach and use a [2,1] Pad\'e resummation. As this agrees well with
high temperature series and $\varepsilon$ expansion results there is every
reason to have confidence in it as a good method of capturing the true
nature of the resummed expressions. For three dimensions the two loop
Pad\'e resummed results are in excellent agreement with six
loop resummed results \ref{48}, where the Callan-Symanzik equation in
distinction to our homogeneous RG was used, and the best high
temperature series
\ref{41}.  A Borel transformation
is also an option, we have not as yet followed this route. Our series
are still too short to make this worthwhile. Calculations to higher
orders could be implemented, and it should not be too difficult in the
case of a three dimensional layered geometry to adopt the numerical
techniques of \ref{48}, in conjunction with our diagramatic
approach, to getting numerical curves for the expressions to a similar
accuracy.

We have also resummed the series for $\gamma_{\phi^2}$,
in terms of the Pad\'e resummed coupling and used the scaling laws for the
effective exponents.  The differences between these results and  those
obtained when $\gft$ is not resummed, and the scaling laws not used,
are small for small values of $N$. We plot in Figure 10 a comparison of the one
loop, two loop Pad\'e resummed $\gft$ and non-resummed $\gft$ results for
$\nu\ef$ with $N=3 $. By adopting
the philosophy that all the characteristic functions should be
treated on an equal footing, in other words they should all be
Pad\'e resummed one retains the spherical model limit
as an exact limit of the Pad\'e resummed expressions. After Pad\'e
resumming of the characteristic functions the scaling laws should be
used directly without further perturbative expansion. Our
procedure is therefore to solve the Pad\'e resummed
differential equation
$z{{\rm d}h\over {\rm d}z}=\beta(h,z)$
for $h(z)$ and substitute it into the Pad\'e resummed expressions
for $\gft$, $\gf$ and use the scaling laws to obtain the effective exponents.
We were forced to do this numerically as the $\beta$ function
equation does not have
an analytic solution known to us.

An alternative procedure would have
been to solve the differential equation iteratively. The one loop
equation is solved first, and the solution  substituted into the two
loop portion of the $\beta$ function equation. The resulting
differential equation is then solved.  The procedure is iterated as
successive loops are calculated. We have not adopted this approach,
however, it would be interesting to compare the accuracy of the two
procedures, this could only be reliably done at higher order than two
loops.

The condition $\beta(h,z)=0$ defines from section 7 the floating fixed
point, valuable in describing a system where, in terms of the usual
picture of flow lines associated with the $\beta$-function, the flow
lines are explicitly ``time'' dependent.  Again note that in the
$\alpha_{1}=0$, $\alpha_{2}=\rc^2\k^2$ case $z=\rc\kappa L$. The
floating fixed point, as a solution of an algebraic and not a
differential equation, is manifestly independent of $z_0$. In a
comparison of the running and floating effective exponents, as the
values of $z$ may differ by a constant scale difference, one curve may
be shifted along relative to another.  There is also of course some
mismatch between the solution of the differential equation and the
floating fixed point in the crossover region even after this shift is
accounted for. This is because solving the $\beta$ function
differential equation is not the same as solving the associated
algebraic equation. They do, however, track one another closely (when
the solution of the differential equation is dominated by a non-trivial
fixed point) but not exactly. The fact that the floating fixed point is
independent of initial conditions means there is no dependence on
transients, hence logarithmic corrections to scaling are suppressed,
however, these can be recovered by examining corrections to scaling
around the floating fixed point.

The zero of $\beta(h,z)$ from  (\docref{betahN}) is \eqlabel{\hN*}
$$h^{*}(z)=\e(z)+{4\o{(N+8)}^{2}}\left((5N+22)f_{1}(z)
-(N+2)f_2(z)\right)\e(z)^2+O(\e(z))^2$$ Substituting this value of $h$ into
(\docref{gftN}) and (\docref{gfN}) gives \eqlabel{\gftN*}
$$\gamma_{{\f}^{2}}^{*}(z)={(N+2)\o(N+8)}\e(z) +2{(N+2)\o{(N+8)}^{3}}
\left( (7N+20)f_{1}(z)-(N-4)f_{2}(z) \right)\e(z)^2$$ and
\eqlabel{\gfN*}
$$\gamma_{\f}^{*}(z)=2{(N+2)\o{(N+8)}^{2}}f_{2}(z)\e(z)^2$$
These functions interpolate nicely between the two asymptotic fixed
points at $z=0$ and $z=\infty$, and define finite size scaling
functions around which one can develop a correction to scaling
expansion, in $h-h^*$.  The analogous [2,1] Pad\'e resummed expression
for the floating fixed-point is
$$h^{*}(z)={\e(z)\over
1-{4\o{(N+8)}^{2}}\left((5N+22)f_{1}(z) -(N+2)f_2(z)\right)\e(z)}$$
We leave it to the reader to substitute this into the corresponding
expressions for the effective exponents. Figure 11 shows a comparison
of the Pad\'e resummed floating coupling and floating fixed point for
$d=4$ and periodic boundary conditions.

\subsec{\bf Pad\'e Resummed Two Loop Results for Some Specific Models}
We will now write down explicit results for some models of particular
interest.  The two loop results we consider explicitly here are for
general $d$. The explicit expressions for $f_{1}$ and $f_{2}$ can be
found in Appendix A by setting $\a_1=0$ and $\a_2=1$. Specific
expressions for $d=4$ are contained in Appendix B. As the expressions
are rather unwieldy we choose not to exhibit them here.  We should also
point out here that these two loop Pad\'e resummed results hold
identically for the corresponding finite temperature relativisitic
field theories. We present the direct perturbative results for ease of
comparison with other methods.

For the Ising model, $N=1$, we have \eqlabel{\betah1}
$$\beta(h,z)=-\e(z)h+h^2-{4\o27}\left(9
f_{1}(z)-f_{2}(z)\right)h^3+O(h^4)$$
$$\gamma_{\f^{2}}(h,z)={1\o 3}h-{2\o 9}\left(f_{1}(z) -{1\o
3}f_{2}(z)\right)h^2+O(h^3)\no$$
$$\gamma_{\f}(h,z)={2\o27}f_{2}(z) h^2+O(h^3)$$
We find the resulting  effective exponents to be
$${\nu\ef}={1\o2}+{1\o12}h+{(3-12f_{1}(z)+4f_{2}(z))\o216}h^2$$
$${\gamma\ef}=1+{1\o6}h+{(1-4f_{1}(z))\o36}h^2$$
$${d\ef}=4-h+{4(9f_{1}(z)-f_{2}(z))\o27}h^2\no$$
$${\alpha\ef}(h,z)={1\o6}h+{(1-16 f_{1}(z))\o36}h^2$$
$${\beta\ef}={1\o2}-{1\o6}h+{(10f_{1}(z)-1)\o36}h^2$$
$${\delta\ef}=3+h+{(3-8f_{1}(z))\o6}h^2$$

For $N=0$, which is related to the self avoiding random walk and
polymers, we have
$$\beta(h,z)=-\e(z)h+h^2-{1\o8}\left(11f_{1}(z)-f_{2}(z)\right)h^3+O(h^4)$$
$$\gamma_{\f^2}(h,z)
={1\o16}h-{3\o16}\left(f_{1}(z)-{1\o3}f_2(z)\right)h^2+O(h^3)\no$$
$$\gamma_{\f}(h,z)={1\o4}f_{2}(z)h^2+O(h^3)$$
The corresponding effective exponents are then
$${\nu\ef}={1\o2}+{1\o16}h+{(1-6f_{1}(z)+2f_{2}(z))\o128}h^2$$
$${\gamma\ef}=1+{1\o8}h+{(1-6f_{1}(z))\o64}h^2$$
$${d\ef}=4-h+{(11f_{1}(z)-f_{2}(z))\o8}h^2\no$$
$${\alpha\ef}={1\o4}h+{(1-16(f_{1}(z))\o32}h^2$$
$${\beta\ef}={1\o2}-{3\o16}h+{(38 f_{1}(z)-3)\o128}h^2$$
$${\delta\ef}=3+h+{(4-11f_{1}(z)\o8}h^2$$

For $N=-2$ which is believed to have the same two point correlation
function as the Gaussian model we have
$$\beta(h,z)=-\e(z)h+h^2-{4\o3}f_{1}(z)h^3+O(h^4)$$
$$\gamma_{\f^2}(h,z)=0\no$$
$$\gamma_{\f}(h,z)=0$$
As we see there is a
non-trivial crossover of $\gamma_\l$, this induces a crossover in
$\alpha\ef$, $\beta\ef$ and $\delta\ef$. The other exponents retain
their mean field values, ${\nu\ef}={1\o2}$ and ${\gamma\ef}=1$. We find
$${d\ef}=4-h+{4f_{1}(z)\over3}h^2$$
$${\alpha\ef}={h\over2}-{2f_{1}(z)\o3}h^2\no$$
$${\beta\ef}={1\o2}-{1\o4}h+{f_{1}(z)\o3}h^2$$
$${\delta\ef}=3+h+{(3-8f_{1}(z)\o6}h^2$$

The model with $N=\infty$ \ref{49}, which is related to the
spherical model and the Bose gas, yields \eqlabel{\betahInf}
$$\beta(h,z)=-\e(z)h+h^2\no$$
\eqlabel{\gtInf}
$$\gamma_{\f^{2}}(h,z)=h\no$$
\eqlabel{\gfInf}
$$\gamma_{\f}(h,z)=0\no$$
In this case
the one loop expressions are exact and the effective exponents
interpolate between those of the exact model in $d$ and $d-1$. We find
$${d\ef}=4-h\no$$
the other exponents can then be written in terms of
$d\ef$. Specifically
$${\nu\ef}={1\o {d\ef}-2}\no$$
$${\gamma\ef}={2\o {d\ef}-2}\no$$
The exponent $\beta\ef$ does not exhibit a crossover,
retaining the value ${\beta\ef}={1\o2}$ throughout the crossover.
$${\delta\ef}={{d\ef}+2\o {d\ef}-2}\no$$
The specific heat effective
exponent
$${\alpha\ef}={{d\ef}-4\o {d\ef}-2}\no$$
is no longer singular for
${d\ef}<4$ and the other effective exponents do not exhibit a crossover
for $d>4$, as the only fixed point of the model is at $h=0$.

The results for other physically relevant systems such as the XY-model
($N=2$) and Heisenberg model ($N=3$) can be easily determined from the
general results above.  The latter two models are of particular
interest for $d=3$, but exhibit some special features which go beyond
the scope of the present article, such as the Kosterlitz-Thouless
transition in the former and the absence of spontaneous symmetry
breaking in the latter. Our techniques can be simply adapted to the
latter, in the guise of the non-linear $\sigma$ model, leading to a
$\beta$ function that crosses over between the zero temperature fixed
point as $z\ra0$ and the $N=3$ three dimensional critical point as
$z\ra\i$, further details will be presented elsewhere.

We present our two loop results using Pad\'e resummed Wilson functions in
Figures 2 through 9. In all graphs
the horizontal axis is $\ln(\xi_L/L)$, the different curves correspond to $N=0$
(polymers), $N=1$ (Ising model), $N=2$ (XY-model), $N=3$ (Heisenberg model) and
$N=\i$ (spherical model). The logarithmic corrections to scaling at the bulk
end are clearly visible, their magnitude is as expected from four dimensional
calculations. All curves are with the boundary condition $h=1$ at
$\ln(\xi_{L}/L)=-20$, the value of $h$ at the initial scale parameterizes
different possible crossover curves but all curves asymptote to the same form.
In Figure 4 we plot $\eta\ef$, the exponent which governs the fall off in
critical correlations at $T=T_c(L)$. This exponent is not a monotonic function
of $N$ but attains a maximum for some value of between $N=-2$ and $N=\i$ where
it is identically zero. This is the least accurate of our exponents and the
peak appears to be at $N=1$, though more accurate values for this exponent
suggest it occurs at higher values, probably $N=3$.  Figure 7 shows
$\delta\ef$, note that excepting
the cases $N=-2$ and $N=\i$ the curves are very robust to changes of $N$. The
crossover in the cases $N=-2$ and $N=\i$ arises purely due to the change in
$\de$.
Once again the curves are not montonic functions of $N$ there being a minimum
in
the vicinity of $N=3$.
Figure 8 shows a plot of the effective specific heat exponent $\alpha\ef$ which
measures how the  singular part of the free energy changes as $\G$ or $T$
varies. The extra case $N=-2$ is added here, since, in the case of dimensional
crossover it is distinguishable from the Gaussian model due
to the fact that $\gl$ for the latter is zero whereas for the former it
is non-zero being a measure of the changing effect of the leading
irrelevant operator. Not only does one see
the change in sign of the specific heat exponent as a function
of $N$  but one also sees that the effective specific
heat exponent can change sign as a function of ${\xi_{L}/ L}$. This is quite
pronounced for the XY model which starts off positive,
increases then turns negative
at $\xi_{L}\sim 100\, L$.
It would be interesting, based on the Harris criterion
for the relevance or irrelevance of weak disorder, to see whether disorder
could change from being irrelevant to relevant as a function of size.
In Figure 9 we plot $\gl=4-\de$ which also gives information about the
effective dimensionality of the system. Notice that $\gl$ is very robust to
changes in $N$, varying very little across the entire range of
$N$, $[-2,\i]$. The asymptotic values of the effective critical exponents
and associated quantities in the three dimensional regime  are tabulated below.
\medskip
\vfil
\centerline{\vbox{\tabskip=0pt \offinterlineskip
\def\tablerule{\noalign{\hrule}}
\halign to400pt{\strut#&\vrule#\tabskip=1em plus2em&
  #\hfil& \vrule#& #\hfil& \vrule#&
  #\hfil& \vrule#& #\hfil& \vrule#&
  #\hfil& \vrule#& #\hfil& \vrule#&
  #\hfil& \vrule#\tabskip=0pt\cr\tablerule
&&\multispan{13}\hfil Asymptotic Critical
       Exponents\hfil&\cr\tablerule
&& \omit\hidewidth N \hidewidth&&
 \omit\hidewidth $\gf$\hidewidth&&
 \omit\hidewidth $\gft$\hidewidth&&
 \omit\hidewidth $h$\hidewidth&&
 \omit\hidewidth $\gamma\ef$\hidewidth&&
 \omit\hidewidth $\nu\ef$\hidewidth&&
 \omit\hidewidth $\alpha\ef$\hidewidth&\cr\tablerule
&&\llap{-\ }2 && 0\rlap* && 0\rlap* && 1.800 && 1\rlap* && 0.5\rlap* &&
0.5\rlap* &\cr\tablerule
&&\llap{-\ }1&&0.0200&&0.145&&1.820 && 1.088&&  0.550 && 0.351 &\cr\tablerule
&&   0 && 0.0295&& 0.277 && 1.785 && 1.175&&  0.596 && 0.211 &\cr\tablerule
&&   1 && 0.0329&& 0.388 && 1.732 && 1.257&&  0.639 && 0.083 &\cr\tablerule
&&   2 && 0.0332&& 0.479 && 1.675 && 1.330&&  0.676 &&\llap{-\ }0.029
&\cr\tablerule
&&   3 && 0.0322&& 0.552 && 1.621 && 1.395&&  0.709 &&\llap{-\ }0.126
&\cr\tablerule
&&   4 && 0.0305&& 0.611 && 1.573 && 1.451&&  0.737 &&\llap{-\ }0.211
&\cr\tablerule&& $\i$&&0\rlap*&&1\rlap* &&1\rlap* && 2\rlap*&&1\rlap*&&\llap{-\
}1\rlap* &\cr\tablerule \noalign{\smallskip}
&\multispan{15}* These values are exact.\hfil\cr}}      }
All these values are in very good agreement with corresponding high
temperature series and experimental results. We  believe the entire crossover
curves are of similar accuracy.

\subsec{The Effect of Boundary Conditions on the Crossover.}
Let us now consider the case of other boundary conditions, more
detailed results will be presented in a separate paper. Different
conditions constitute an environmental probe which can be used to
differentiate between systems as they fall into different crossover
universality classes. We begin with Dirichlet boundary conditions, but
will only be interested in ``bulk'' not surface physics. We will
confine our considerations to one loop for simplicity. In the three
dimensional layered system we are again in the fortunate position of
being able to present the one loop expressions in terms of elementary
functions.  The floating coupling is
$$h(y)=(1+{\pi\o y^2}){(1+{2y\o \sinh y}-{2\tanh y\o y})\o (1-{\tanh
y\o y})}\no$$
where $y=(z^2-\pi^2)^{1\o2}$.  It is necessary to consider two branches
of the function $y(z)$, however, even though $y(z)$ has a branch point
$h(z)$ is analytic in $z$.  The floating fixed point is simply
$h^{\ast}=\e(y)$ where
\eqlabel{\dir}
$$\e(y)=1+{3\pi^2\o
y^2}+2(1+{\pi^2\o y^2}) {{({y^2\o\sinh^2y}-{\tanh y\o
y})}\o{(1+{2y\o\sinh 2y}-2{\tanh y\o y})}}\no$$
We will leave the reader to derive the corresponding one loop
expressions for the effective exponents in terms of the floating fixed
point. Clearly they will interpolate between exactly the same
asymptotic values as for the floating coupling derived from integrating
the $\b$ function above.

In the case of anti-periodic boundary conditions the extremal values of
these functions are the same, they do however differ in the crossover
region. Explicitly for $d=3$, to one loop, one finds $h(y)=(1 +
{\pi^2\o y^2})(1 - {y\o\sinh y})$. The floating fixed point is once
again $\el$ where this time
\eqlabel{\anti}
$$\e(y)= 1 + {3\pi^2\o y^2 }- {((y^2 +
\pi^2)\tanh(y/2)\o(\sinh y - y)}\no$$
Note that even though $y$ again goes through a branch point, as $z$
ranges between $0$ and $\i$, both (\docref{dir}) and (\docref{anti})
remain single valued. As was the case with Dirichlet boundary
conditions the effective exponents crossover between the same
asymptotic values, however, the effective exponents between the
asymptotic values are different.

At the one loop level the expressions for general $d$, for Dirichlet
boundary conditions, are
$$\e(z)=5-d-(7-d){{\displaystyle\sum_{n=1}^{\i}{\pi^2(n^2-1)\over z^2}
\left(1+{\pi^2(n^2-1)\over z^2}\right)^{d-9\over2}}
\over{\displaystyle\sum_{n=1}^{\i}\left(1+{\pi^2(n^2-1)\over
z^2}\right)^{d-7\over2}}}\no$$
and for the universal one loop floating coupling
$$h(z)=(5-d){{\displaystyle\sum_{n=1}^{\infty}}{(1+{\pi^2(n^2-1)\o
z^2})}^{(d-7)\o2}
\o{\displaystyle\sum_{n=1}^{\infty}}{(1+{\pi^2(n^2-1)\o
z^2})}^{(d-5)\o2}}\no$$
For antiperiodic boundary conditions one finds
$$\e(z)=5-d-(7-d){{\displaystyle\sum_{n=-\i}^{\i}{\pi^2n(n+1)\over
z^2} \left(1+{\pi^2n(n+1)\over z^2}\right)^{d-9\over2}}
\over{\displaystyle\sum_{n=-\i}^{\i}\left(1+{\pi^2n(n+1)\over
z^2}\right)^{d-7\over2}}}\no$$
and finally for the floating coupling
$$h(z)=(5-d){{\displaystyle\sum_{n=-\i}^{\infty}}{(1+{\pi^2n(n+1)\o
z^2})}^{(d-7)\o2}
\o{\displaystyle\sum_{n=-\i}^{\infty}}{(1+{\pi^2n(n+1)\o
z^2})}^{(d-5)\o2}}\no$$

Plots comparing the effects of different boundary conditions for a three
dimensional layered Ising system can be
found in graphs 12-16. The main interesting feature is in the case of
Dirichlet boundary conditions, where there is a ``dip'' in the curves as one
approaches the three dimensional regime. Thus the curves are not monotonic, the
effective exponents approaching their three dimensional asymptotic values from
below. The origin of this effect lies in a similar characteristic dip present
in the behaviour of $\e$, and is a manifestation of the power law decay of the
effects of the boundary in the case of Dirichlet boundary conditions as
compared
to the exponential decay for periodic boundary conditions. In the case of $d=4$
the dip in $\e$ persists and indicates that for ${L\o\xi_L}\gg1$ the effective
dimensionality is actually above the upper critical dimension! This has some
very
interesting ramifications which we will return to at a later date. In the case
of antiperiodic boundary conditions the effect is in the opposite direction to
that of Dirichlet. For $d=3$ the crossover to two dimensions is completed at
larger
values of $L/\xi_L$ when compared to either Dirichlet or periodic boundary
conditions. At the three dimensional end there are persistent tails relative
to periodic boundary conditions. Once again these are due to the power law as
opposed to exponential decay of the boundary effect. For $d=4$, in distinction
to the case of Dirichlet boundary conditions, $\de$ never exceeds the upper
critical dimension.

\subsec{The Uniaxial Dipolar Ferromagnet}
Another model that falls into the same class as the considerations of
this section is that of a uniaxial dipolar ferromagnet
\ref{19}\ref{20}.
In this case the propagator takes the form
$$p^2+\alpha_{0}{p_z^2\o p^2}\no$$
Here the environmental factor that is followed in the crossover is
$\alpha_0$, the coupling strength of the long range dipole-dipole
interaction between the spins.
Once again the formal structure of the
diagramatic series is the same. The diagrams differ in their dependence
on $z^{-1}=\xi_{\ss\alpha_0}\alpha_0^{1\o2}$ now, hence $\e(z)$, $f_{1}(z)$
and $f_{2}(z)$ are different functions, interpolating however, in the
$\alpha_{1}=0$, $\alpha_{2}=\rc^2\k^2$ prescription, between the values
$1$, ${1\o3}$ and ${4\o27}$ respectively for $z=0$, to finite values
close to those of the four-dimensional problem treated above. The numbers
are different however, reflecting the absence of some angular
integration variables. We have not as yet examined the two loop case in
detail and reserve further comment till a future date.
Let us therefore restrict our considerations to one loop.

We find that
$$\e(z)= {z^2\o 1+z^2}\no$$
The solution of the running coupling equation is given by
$$h={1\o\sqrt{1+z^2}
\left(h_{0}^{-1}{1\o\sqrt{1+z_{0}^2}}+\ln({z_{0}(1+\sqrt{1+z^2})\o
z(1+\sqrt{1+z_{0}^2})})\right)}\no$$
There is an interesting distinction between this situation and that of the
dimensional crossover, in that in the case of the uniaxial problem there is
a universal form for the crossover from the three dimensional fixed point to
the pseudo-four dimensional fixed point. The reason for this is that the true
microscopic structure of the model is three dimensional, and therefore the
logarithmic growth
into the UV does not continue ad infinitum. The universal part of the crossover
curve can be obtained as in the case of three dimensional problems by taking
the limit $z_{0}\ra\i$. We find this universal coupling is given by
\eqlabel{\hunivuniax}
$$h(z)={1\o\sqrt{1+z^2}\ln({1+\sqrt{1+z^2}\o z})}  \no$$
We present a comparison of $\nu\ef$ for the four dimensional
dimensional crossover in a layered geometry with periodic boundary
conditions with that of the uniaxial crossover in figure 23. The
initial coupling for the layered geometry is $h_{0}=1$ for
$\ln(1/z_{0})=-20$ and the curve for the uniaxial problem is the
universal crossover exponent derived from (\docref{hunivuniax}).  In
both cases we plot $\nu\ef={1\o 2-\gft}$, where $\gft$ is retained to
order $h$ without further expansion of the denominator. We choose this
form since it is closer to the two loop results as demonstrated in
Figure 10. Figure 21 clearly shows that the uniaxial crossover and that
of the layered geometry are in different crossover universality
classes.

As shown by Kamien and Nelson \ref{50} recently, critical phenomena of,
single component and binary mixture in directed polymers in an external
field, strands of dipoles in both ferro- and electrorhealogical fluids,
and flux lines in high-$T_c$ superconductors, are all in the same
universality class as three dimensional uniaxial dipolar ferromagnets
and ferroelectrics. It would be interesting to compute the effective
critical exponents for the crossover behaviour in these systems based
on the methodology here.

\beginsection{\bf CROSSOVER IN THE QUANTAL ISING MODEL}
In this section
we wish to consider yet another environmental factor --- $\hbar$! The
real world is quantum mechanical, although this is not always a
dominant feature governing  the behaviour of systems, the macroscopic
world typically being governed by classical laws. However, at low
temperatures quantum effects play an important role, and typically
quantum fluctuations need to be taken into account. When one is dealing
with finite temperature systems there are also thermal fluctuations. An
interesting question therefore, especially in the theory of phase
transitions, is how does one crossover between the two types of
fluctuations in a realistic system.  For very low temperatures one
expects quantum fluctuations to dominate, and at higher temperatures
classical thermal fluctuations.

The explicit model we consider is an Ising ferromagnet in a transverse
magnetic field $\G$, described by the lattice Hamiltonian $${\cal
H}=-{1\o 2}\sum_{ij}K_{ij}S_{i}^{z}S_{j}^{z}-\Gamma
\sum_{i}S_{i}^{x}\no$$ where $S_{i}^{a}$  are the components of spin in
the $a$ direction at lattice site $i$. The spins $S_{i}^{z}$ and
$S_{j}^{x}$ do not commute and it is this non-commutation that gives
the additional complications of the model over that of the usual Ising
model. The model is known to have a second order transition, at a
temperature $T_{c}(\G)$ for $\Gamma \le\Gamma_{c}$, while for
$\Gamma>\G_c$  no ordered phase occurs at any temperature.

As we are only interested in the critical properties of this model here
we will use the LGW Hamiltonian derived by Hertz \ref{43} using the
Stratonovich-Hubbard transformation, but now a time ordered version
where time ordering is with respect to imaginary times between $0$ and
$\beta$. The explicit Hamiltonian is $$\eqalign{H=
  & \int d^dx{\displaystyle\sum_n}\left(\nabla{\f}_{\ss
  n}\nabla{\f}_{\ss -n}+ {1\o2}(m_B^2+g_B+4\pi^2n^2T^2){\f}_{\ss
 n}{\f}_{\ss -n}\right) \cr
  & +{\l T\over{4!}} {\displaystyle\sum_{\ss n_1,n_2,n_3,n_4}}
  \delta({\s n_{\ss1}+n_{\ss2}+n_{\ss3}+n_{\ss4}}){\f}_{\ss
  n_1}{\f}_{\ss n_2}{\f}_{\ss n_3}{\f}_{\ss n_4} \cr}\no$$ where we
have dropped operators irrelevant in the critical region.
$m_B^2+g_B=\G-\G_c$, where $\G_c$ is the critical transverse field in
the mean field approximation. The sum is over the Matsubara frequencies
and $T={\sqrt{\alpha}\o\beta}$, where $\alpha={1\o 4\mu a^2
\Gamma\tanh(\beta\Gamma)}$ and $\beta$ is the inverse temperature. We
have also used $K_{q}=K_{0}- \mu a^2 q^2+O(q^4)$ for the Fourier
transform of the Ising spin-spin coupling. Observe that $\alpha$ is
dimensionless, once $\beta$ and $a$ are assigned the units of length,
since from the original Hamiltonian we deduce that $J$, $\mu$ and
$\Gamma$ have the dimensions of inverse length.

In this problem the anisotropy parameter, or environmental variable, is
$T$, which is essentially just the thermodynamic temperature. Note that
it is entering here in a very different fashion to the dimensional
crossover considered in section 10. There the temperature entered in
the standard way in the quadratic term of the LGW Hamiltonian, i.e. as
$T-T_c(L)$. We now wish to renormalize in an environmentally friendly
manner. We use the conditions (\docref{gta}---\docref{gco}). The
condition (\docref{gta}) implies, with the multiplicative renormalization
of $g$, that $g$ is proportional to $\G-\G_c(T)$, the deviation
from the critical line. Note that in the renormalization used here $T$
does not renormalize, i.e. it is a non-linear scaling field. Our task
now is to calculate the solution of the $\beta$ function and to
subsequently compute the effective exponents using the ``black box''
expressions of sections 5 to 9. The only thing we need to remember is
that the environmentally friendly propagator that enters these
expressions is $G({\bf k},\omega)={1\o({\bf k}^2+{\omega}^2+g)}$ where
$\omega={2\pi nT}$.

Consider first a three dimensional model. From (\docref{pade}) with
$N=1$, and using the above propagator in the diagrams, one finds a
floating fixed point that interpolates between the classical and
quantum fixed points. The quantum fixed point in this case is found to
be zero, but there exist important logarithmic corrections to scaling
which are captured by the floating coupling and are visible in Figures 16-20.
An examination of $\de$ reveals that it interpolates between four and
three. This implies that the effects of quantum fluctuations are such
as to increase the effective dimensionality of the system in the low
temperature regime. The classical fixed point is $h(g=0)=1.732$.
The effective
exponents can be found from the floating coupling expressions of
section 9. Using the two loop Pad\'e resummed coupling and Pad\'e
resummed $\gft$ for this model,
the exponents $\nu\ef$, $\eta\ef$, $\gamma\ef$ and $\alpha\ef$ are
plotted in Figures 16-19. Once again they are in excellent agreement at the
three dimensional end with the results of \ref{48}. We find for
instance $\nu(g=0)=0.639$. The other scaling
exponents can be found from the scaling exponent laws. Needless to say
the expressions found agree precisely with those obtained from a direct
calculation.  Using the expressions in the appendices the explicit two
loop Pad\'e resummed expressions for a two dimensional Ising model in a
transverse field can be found. Note that here we are deriving the
effective exponents at fixed temperature, i.e. we are approaching the
critical line from the $g$ direction. We could equally well have looked
at exponents defined with respect to the ``environmental'' $T$ direction as
discussed in section 5. We will return to this in a future paper.

As far as scaling is concerned the relevant correlation length here for
scaling purposes is $\xi_{T}$ the physical correlation length of the
system. If we had used a temperature independent renormalization the
relevant correlation length would have been the zero temperature
quantum correlation length. One can write the scaling functions in
universal scaling form $$\G^{(N)}={\xi_{T}}^{\ss{{Nd\over2}-N-d}}{\cal
F}^{(N)}(T\xi_{T})$$ In particular for the susceptibility
$\chi^{-1}=\xi_{T}^{-2}{\cal F}^{(2)}(T\xi_{T})$. In terms of effective
exponents $$\chi^{-1}={\rm exp}\left(\int_1^g {\nu\ef}(2-{\eta\ef})
{dg'\o g'}\right)\no$$ Strictly speaking the above functions are only
truly universal for $d<3$. For $d=3$ the logarithmic corrections to
scaling evident in the effective exponents preclude a true scaling
form. However, the scaling form in terms of the non-linear scaling
fields of section 6 is ``universal'' as the corrections to scaling are
captured by these fields in their dependence on initial conditions. For
fixed initial conditions all scaling functions are universal even in
the presence of corrections to scaling.

Turning now to the two dimensional model, at one loop we can give some
nice simple analytic forms.  The crossover coupling in terms of
$z={T\xi_{T}}$ is $$h(z)=1+{z\o\sinh(z)}$$ One finds that
$\gft(z)={1\o3}h(z)$, hence the crossover effective exponents are to
one loop given by $${\nu\ef}(z)={7\o12}+{1\o12\sinh(z)}$$
$${\gamma\ef}(z)={7\o6}+{z\o6\sinh(z)}$$ The effective dimensionality
$\de=\el$ is given by $$\e(z)=1+{z^2\coth({z\o2})\o\sinh z+z}\no$$
which once again clearly indicates a change in effective
dimensionality.  In this case from two up to three in the deep quantum
regime.  The susceptibility scaling function at one loop is simply
$\chi=z^2$.  If one wants the results parametrized in terms of
$g=\G-\G_c(T)$ then one must invert the equation
$$g(z)={z^{2}\o{\left(\tanh(z/2)\right)}^{1/3}}\no$$ For the two loop
results we would have to invert the two loop version of this equation.

If one compares any of the two loop or one loop expressions for
effective exponents here with the corresponding expressions for a
layered Ising model with periodic boundary conditions, one will find
that they have exactly the same functional form. In fact, after the
replacements $g\ra t$, $T\ra L^{-1}$ they are identical. Hence we would
conclude that the $d-1$ dimensional Ising model in a transverse field and
the $d$ dimensional layered Ising model with periodic boundary
conditions lie in the same crossover universality class.

Of course it is well known that there is a mapping between these
models. This was shown by Suzuki \ref{8} using a Trotter product
formula, the extra dimension being of finite size $\beta\hbar$ with
periodic boundary conditions, where $\beta$ is the inverse temperature.
In the above we tried not to emphasize this point for two reasons.
First of all, so that we could make the link at the level of the
effective exponents in the context of which we have discussed and
defined crossover universality classes; and secondly, so that we could
bring to light an important point about our formalism. Suppose that at
a theoretical level we knew nothing about the quantum fixed point, but
that experiment had told us that the critical exponents at zero
temperature were very different to those at the classical fixed point.
How would one go about finding a theoretical description of the other
fixed point, and indeed the entire crossover? Within our
environmentally friendly formalism the answer is clear: one starts at
some ``microscopic'' energy scale $E_m$, one writes down a Hamiltonian
for the ``microscopic degrees of freedom'', {\bf including all relevant
environmental parameters}, one cranks the handle of the environmentally
friendly RG and sees what comes out. If there exists another fixed
point which is reached via the effects of a relevant environmental
parameter, then it will be seen in the environmentally friendly RG
flow. If it does not exist then it will not appear. The point is we do
not need to know a priori about its properties, or even of its
existence, these can all be deduced from the RG. What is equally
important is that one can see when one is not implementing an
environmentally friendly renormalization because perturbation theory
within the context of the unfriendly group will break down. This is
one's pointer as to the fact that one had missed some important
environment dependence.

So, we have been able to deduce effective critical exponents to what we
believe to be a very high degree of accuracy for the three dimensional
quantum Ising model using two loop Pad\'e resummed perturbation theory.
We also gave explicit results at one loop level and formal results at
the two loop level for the two dimensional model. The results we gave
above are of course also valid for non-integer dimensions. One can even
extend the results given to a four dimensional quantum Ising model.
Comparing with some of the past literature we find, in contrast to the
claims of Lawrie \ref{22}, that the Matsubara frequencies do in
fact play an important role in determining the scaling functions for
dimensions $3<d<4$, thereby allowing us to distinguish this crossover
from that of a uniaxial dipolar ferromagnet or of the crossover between
the Wilson-Fisher and Gaussian fixed points. Obviously for a four
dimensional model there is no crossover as noted in \ref{51} in
the context of the ``spherical'' quantum Ising model. All effects then
are due to corrections to scaling, though these can be of great
importance for particle physics models in the context of Kaluza Klein
theory \ref{52}. In distinction to \ref{43} aswell we derive
flow equations that are explicitly dependent on the Matsubara
frequencies.

There are other quantum statistical systems whose crossover behaviour
we can compute simply by reading off the environment dependence from
the LGW Hamiltonian and following our black box methodology. Naturally the
results presented hold identically for quantum ferroelectrics as well as
quantum ferromagnets. We could also,
following the work of Hertz \ref{43}, consider (in his
terminology) itinerant ferromagnets, dirty itinerant ferromagnets and
itinerant antiferromagnets. These will fall into different crossover
universality classes since for these models the propagators are
$$G({\bf k},\omega_{n})={\bf k}^2+m_{0}^2+{|\omega_{n}|\o |{\bf k}|}$$
for the itinerant ferromagnet, while for the dirty itinerant
ferromagnet $$G({\bf k},\omega_{n})={\bf k}^2+m_{0}^{2}+{|\omega_{n}|\o
D_{0} {\bf k}^2}$$ and for the itinerant anti-ferromagnet $$G({\bf
k},\omega_{n})={\bf k}^2+m_{0}^{2}+\tau |\omega_{n}|$$ These will all
give rise to different crossover scaling functions. We would therefore
consider the environmental dependence to be relevant and classify the
corresponding crossovers to be in different crossover universality
classes. Clearly there is a rich vein to be explored in applying our
methods to the quantum-classical crossover.

\beginsection{\bf CONCLUSIONS AND SPECULATIONS}
In this paper we have developed in some detail our approach to field
theoretic renormalization of crossovers. As is well known the independence of
the
bare vertex functions of the normalization point gives an equation
which allows reparametrizations of the vertex functions in terms of new
renormalized parameters, via which one tries to implement perturbation
theory. We have emphasized, however, that one should be very wary since
many reparametrizations may not be very useful for a perturbative
analysis of the problem of interest. Such an analysis is best
implemented in terms of parameters which at a particular scale describe
the relevant effective degrees of freedom at that scale, in preference
to degrees of freedom which are associated with some other disparate
one. We therefore emphasized that the optimum choice of parameters is
very sensitive to the environment, the point being that as the
environment effects the fluctuations, the effective degrees of freedom
are consequently environment dependent.

Our goal was to be able to quantitatively describe scale changes in
systems that exhibited crossover behaviour. Scale changes in general
can be fruitfully examined by analyzing the RG equation in conjunction
with the equation of dimensional analysis. Combining the latter with
our notion of environmentally dependent renormalized parameters gave us
a powerful RG equation using which we could follow with our
reparametrizations how the couplings which were natural at a given
scale changed with scale, and hence how the physical correlation
functions themselves changed. One might roughly think of the philosophy
implemented as: 1) check out the environment you are in at a certain
scale; 2) choose the appropriate parameters for that environment at
that scale; 3) change scale, but only by an infinitessimal amount; 4)
reparameterize to variables which are appropriate for the environment
at the new scale. By so doing one obtains an RG the fixed points of
which, in the space of couplings $\cal M$, are diffeomorphic to the
points of scale invariance of the system. If an RG is used which is
independent of a relevant environmental parameter then this will not be
true. Some of the points of scale invariance will be inaccessible to an
environment independent RG used alone, i.e. these points could not be
seen as fixed points generated by the group with environment
independent Wilson functions. In order to access them one would have to
supplement the RG by extra non-perturbative information.

Sometimes environmental factors can be safely neglected, but the art is
to uncover which ones. If one is interested in a system with two
crossovers induced by two parameters $g_1$ and $g_2$, respectively, and
one is only interested in the crossover induced by $g_1$, then one
could renormalize in a $g_1$ dependent and $g_2$ independent way, and
still capture the desired physics. Only the full $(g_1,g_2)$ dependent
RG will capture the full crossover structure however.  When something
has been neglected that ought not to have been, in all cases looked at
so far, the theory has very civilly informed us, via the total
breakdown in perturbation theory, that something was seriously amiss.
We regard this as a very general property of crossover systems.  We
believe that the canonical incantations: ``divergences'', ``strong
coupling'' and ``non-perturbative'' are, more often than not, all
symptomatic of the failure to take into account the changing nature of
the effective degrees of freedom of a crossover system.

Within the context of an environmentally friendly RG, dependent on a
generic crossover parameter $g$, we went on to examine, in a formal way
to begin with, how our program would be implemented. We derived
expressions for the renormalization constants and the Wilson functions
to two loops, which are valid for a rather wide class of crossover
problems. A particular crossover problem could be simply addressed by
writing down the diagramatic components of the expressions with
propagators appropriate to the crossover in question.

Further consideration of the general properties of environmentally
friendly RGs led us to discover that there were generalizations of the
standard scaling laws relating the effective exponents of crossovers.
These effective exponents also appeared naturally in the non-linear
scaling fields which were the arguments of the scaling functions
characteristic of the crossover. We indicated how one could generate
scaling forms in terms of non-linear scaling fields for the entire
crossover by taking a linear scaling field associated with one fixed
point, exponentiating it, and replacing the exponent with the integral
of the corresponding effective exponent. In the scaling limit when all
relevant crossover length scales are much bigger than the lattice
spacing the effective exponents and scaling functions are universal.
For cases such as crossover in a four dimensional layered system,
scaling functions retain a dependence on the initial coupling.  This is
due to the presence of universal logarithmic corrections to scaling.
There is then no universal crossover scaling curve but rather a family
of such curves which can be parameterized by the possible couplings at
a fixed scale.  Despite the fact that thermodynamic functions do not
exhibit ``true'' scaling behaviour, we find that the effective exponent
laws are still valid, including the logarithmic corrections to
scaling.  The effective exponents as functions define different
crossover universality classes. As in the standard non-crossover case
the existence of scaling laws implies that only two of the effective
exponents are independent. However, in contradistinction to the
standard case an essential ingredient in our analysis was the role
played by the crossover of the Wilson function for the $\phi^4$
operator. The fact that the leading irrelevant operator can change its
degree of irrelevance has made it impossible (to date at any rate) to
follow a crossover with changing upper critical dimension using
$\varepsilon$ expansion techniques. In our formalism this presented no
difficulty.

The crossover was naturally captured in terms of our floating $h$
coupling which interpolated in a smooth way between the asymptotic
fixed points. A natural substitute for $\varepsilon$,
$\varepsilon({g\xi_g})$ appeared in $\beta(h,{g\xi_g})$, the zero of which
could
be expressed in terms of a floating fixed point, $h^{\ast}$, which
could be ordered in powers of this $\e({g\xi_g})$. In this case
floating effective exponents appeared which also obeyed the effective
exponent scaling laws. The floating fixed point had the advantage of
being defined with respect to an algebraic equation rather than a
differential equation, and was therefore easier to compute. The
difference between crossover curves defined with respect to the
floating fixed point and the running coupling as a solution of the
$\beta$ function equation is due to a correction to scaling which is
slowly varying and non-singular throughout the entire crossover.  These
correction to scaling factors are calculable in our formalism. One can
think of the floating effective exponents as giving a reasonably good
approximation to the true effective exponents, the former missing some
of the fine detail of the crossover.

We summarized the crossover in the dimension of the $\phi^4$ operator
in terms of $d\ef$, which can be thought of as being a measure of the
``effective dimensionality'' of the system, $\gl({g\xi_g})=4-d\ef$
being then viewed as a measure of the the deviation of the system from
its upper critical dimension, and a natural generalization of the
correction to scaling exponent $\omega$.  In the case of finite size
crossover in dimensions $d<4$ (in the absence of transients) $d\ef$ can
be regarded as a measure of the true change in dimensionality.  This
interpretation of effective dimensionality was also seen to be quite
natural as $d\ef$ appeared in the scaling laws involving the effective
exponents $\beta\ef$, $\delta\ef$ and $\alpha\ef$ in exactly the same
place as $d$ did in the non-crossover case.  $\dea=\el$ to one loop can
also be thought of as a measuring  the change in dimensionality.

We discussed perturbation theory in some detail showing how and why $h$
or $h^{\ast}$ can be used to order perturbation theory. Expansions in
either $\varepsilon$ or $1/ N$ without environmentally friendly
renormalization are both perturbatively useless for crossovers. Even
with environmentally friendly renormalization an $\varepsilon$ is
unsuitable when the upper critical dimension can changes during the
crossover.  Similarly $1/N$ is equally unsuitable in cases where the
number of components of the order parameter can change. The only valid
choices for all the crossovers considered were $h$ and $h^{\ast}$. We
showed that in the crossovers where these couplings could become
$\sim1$ that our methodology lent itself readily to resummation
methods, giving expressions for the [2,1] Pad\'e resummed floating
fixed point and $\beta$ function equation. We emphasized that in our RG
methodology it was the Wilson functions that were perturbatively
computed. Scaling functions were obtained by integration of the
corresponding characteristic equations. With expressions such as
$\exp{\int( h+h^2...)}$ which result from solving the characteristic
equations, it was important not to try and expand the exponentials. The
job of the RG is to replace perturbation theory directly at the level
of the correlation functions with perturbation theory at the level of
the Wilson functions the subsequent integration (exponentiation) of
which ensures that the information in the exponentials is highly
non-perturbative. Just how non-perturbative is seen readily in the
explicit diagrammatic expressions for the non-linear scaling fields in
section 10. Relative to standard perturbation theory there is clearly a
huge amount of information in such expressions. The moral was that one
should include as much information as possible in the Wilson functions
as this information will be ``exponentiated'' by the RG. What is left
out will not be exponentiated and will have to be treated in direct
perturbation theory.

We then shifted attention away from a generic crossover to some
specific examples. We examined in detail the case of dimensional
crossover.  The environmental variable in this case was the finite size
of a layered geometry. We found that our environmentally sensitive RG
Wilson functions depended on the thickness of the layers $L$. The
immediate effect of the inclusion of $L$ was that the differential
generator associated with the RG now interpolated between two extremal
forms, that appropriate to a bulk system in the ${L/\xi_L}\ra\infty$
limit, and that associated with a reduced system in the ${L/\xi_L}\ra
0$ limit. This had the desirable effect of including both extremal
exponents associated with the respective points of scale invariance,
which were now fixed points of this RG, in a natural way into the
Wilson functions. As a consequence of this, it was easy to see that the
effective exponents interpolated between those of the extremal fixed
points and furthermore to verify that the effective exponent laws were
valid throughout the crossover region.

In the context of specific two loop results, we faced the issue that
the non-trivial fixed point of the $\beta$ function disappears at this
order. We circumvented the problem by following Parisi \ref{40}, and
using a $[2,1]$ Pad\'e approximant to re-sum the series. The resulting
$\beta$ function equation was integrated numerically, and the solution
substituted into the resummed perturbative series for the Wilson
functions. In the four to three dimensional crossover the logarithmic
corrections to scaling at the four dimensional end were seen to
naturally crossover to  power law scaling at the three dimensional
end.  Asymptotic effective exponents are in excellent agreement with
the Callan-Symanzik/fixed dimension results of Baker et al. \ref{48}
and also with corresponding high temperature series and experimental
results \ref{41}.

A complimentary expansion which can be included in this approach is the
${1\o N}$ expansion. This is an appropriate one in the case of
dimensional crossover since it does not rely on the uppercritical
dimension remaining fixed. This should prove useful for examining
dimensional crossovers as long as the layer dimension remains larger
than two. We have, therefore, included the $N=\infty$ results, in which
case the results obtained interpolate between the exact spherical model
results for four and three dimensions, again exhibiting the crossover
from logarithmic to power law behaviour, as found by Barber and Fisher
\ref{13}. For polymer type systems ($N=0$) we found similar good
agreement with accepted results. For completeness we presented the
results for the $N=-2$ model the experimental significance of which we
are, however, unaware. It had the curious property of having only a
crossover in the anomalous dimension of the $\phi^4$ operator, due to
the changing upper critical dimension of the model. Perhaps some
experimental model can be found in accord with this. We found
specifically that the exponents $\alpha{\ef}$, $\beta{\ef}$ and
$\delta\ef$ exhibit a crossover even though $\nu\ef$, $\gamma\ef$ and
$\eta\ef$ retain their mean field values.

Boundary conditions are another natural environmental factor which
affect the critical fluctuations of the system. We briefly compared the
differences between periodic, Dirichlet and antiperiodic boundary
conditions on the dimensional crossover, above the transition
temperature, and sufficiently far from any boundaries. In the three
cases the formal structure of the RG is identical. We concluded that
the effect of periodic and antiperiodic was similar, there is an
additional shift, however, of the critical temperature, in the
anti-periodic case and the functions $f_{1}({L\o\xi_L})$ and
$f_{2}({L\o\xi_L})$ are slightly different, but have the same
asymptotic, large argument values. In the case of Dirichlet boundary
conditions, we found, interestingly, that there is a ``dip'' in the
crossover curves, hence they are not monotonic. The origin of this
effect lay in a similar characteristic dip present in the behaviour of
$\e$, and was a manifestation of the power law decay of the effects of
the boundary in the case of Dirichlet boundary conditions as compared
to the exponential decay for periodic boundary conditions. In the case
of $d=4$ the dip in $\e$ persisted and indicated that for
${L\o\xi_L}\gg1$ the effective dimensionality was actually above the
upper critical dimension!

The problem of crossover in uniaxial dipolar ferromagnets was similarly
briefly examined. Here the environmental factor is the dipole dipole
interaction, which modifies the form of the propagator. The diagramatic
series is the same, however the crossover functions have a qualitative
difference which persists to the quasi-four dimensional end. A true
universal crossover curve exists in which all dependence on the initial
coupling to dissapears, in contrast to the dimensional crossover
problem in a four dimensional layered system.

We considered also an Ising model in a transverse magnetic field. For
the three dimensional model we derived two loop Pad\'e resummed
expressions for the effective exponents describing the quantum to
classical crossover.  For the two dimensional model we derived very
simple expressions for the effective exponents. The fact that the
effective exponents for the three dimensional model were identical with
those for a four dimensional layered Ising mdel with periodic boundary
conditions showed that these two models lay in the same crossover
universality class. We compared and contrasted our results with others
that exist in the literature and briefly discussed how our results
could easily be adapted to other quantum/classical crossovers.

Because of the generality of our approach and the ubiquitousness of
crossover behaviour this work has natural extensions to many other
systems.  To list but a few that readily come to mind: one would like
to incorporate the effect of surface Hamiltonians. One would also like
to be able to include the effect of more complicated backgrounds, such
as the effect of the earth's gravitational field in the case of binary
fluids. A background of instantons, or quasi-particles, or vortices for
more general Hamiltonians are similarly natural directions in which to
extend the current analysis. The effect of these we would expect to be
most naturally encapsulated in a position dependent RG (the same would
be true for the application of the methodology to a generic curved
space necessary for doing early universe physics). The inclusion of
dynamical effects would likewise be a natural thing to do. One would
like to see the crossover of dynamical scaling effective exponents as
the neighbourhood of different fixed points is reached, the RG
trajectory could naturally be linked to time evolution, an idea which
could be very useful in black hole formation or inflation where time
gets linked to scale in a very natural way. In general one would expect
to have a direction, position, and time dependent RG which responded to
the anisotropic and time dependent nature of a generic environment. In
the neighborhood of a given fixed point, however, the universality of
that fixed point should become apparent as the associated degrees of
freedom begin to fluctuate critically. There is also a natural rich
mathematical structure apparent in our formulation, associated with the
geometry of the space of coupling constants $\cal M$, which we plan to
illucidate further in a future publication.  Another interesting area
which caught our attention recently is that of microphase separation
\ref{53}, which describes the critical fluctuations in diblock
copolymers \ref{54} and chemically crosslinked two-component networks
\ref{55}. Here the propagator of interest is ${1/(k^2+m^2+{C\o k^2})}$
where $m^2$ is the deviation from the critical value in the Flory
interaction parameter, and $C=R_G^4$, $R_G$ being the radius of
gyration, in the case of diblock copolymers, and  is the elastic
constant in the case of crosslinked two component networks. In our
black box methodology all one has to do to determine the effective
critical exponents is to substitute the above propagator into the
diagramatic expressions we have given.

Our approach has natural applications in a broad range of areas, from
cosmology to turbulence, where RG ideas have proved useful. The generic
features we have outlined will be similar. The diagramatic structure is
determined from the underlying Hamiltonian/action. The appropriate
renomalization should then be addressed, taking due care to
incorporate the environmental factors necessary for the most efficient
treatment of the processes under study. One might be led by effective
Hamiltonians which give good phenomenological descriptions of the
physics in different regimes. If one is sufficently insightful the RG
chosen would be the one which tracked the trajectory between the
microscopic theory and the effective theory. Being so insightful is of
course frequently impossible. The beauty of our approach is that one
does not in principle need to know the other end. The response of the
theory to its environment will tell you if you monitor it sufficently
closely. This in principle could be of help in theories where one does
not know the appropriate degrees of freedom at one end of the crossover
such as quantum gravity.

\noindent {\bf Acknowledgements:}
CRS was supported by a FOM fellowship and is also grateful to D.I.A.S.
for financial support. DOC is grateful to the University of Utrecht for
travel support. We thank Brian Dolan for his careful reading of the
manuscript and helpful conversations. We have benefited from
conversations variously with the following:  Pierre van Baal, Edouard
Br\'ezin, Michael Fisher, Filipe Freire,  Gerard `t Hooft, Bei Lok Hu,
Tom Kibble, Ian Lawrie, John Lewis, Charles Nash, Mark Novotny,
Lochlainn O'Raifeartaigh, Anatoly Patrick, Joe Rudnick, Dimitri Shirkov
and Rafael Sorkin.

\beginappendix{EVALUATING TWO LOOP DIAGRAMS IN A
$d$ DIMENSIONAL LAYERED GEOMETRY}
The task of this section is to evaluate the diagrams entering the RG
equations in the dimensional crossover problem.  We first list
some basic formulae that prove useful.  The introduction of Feynman
parameters in combining two propagators amounts to
$${1\o D_{1}^{a_1}D_{2}^{a_2}}={\G(a_{1}+a_{2})\o\G(a_1)\G(a_2)}
\int_{0}^{1}dx {x^{a_{1}-1}(1- x)^{a_2-1}\o
{\left(xD_{1}+(1-x)D_{2}\right)}^{a_{1}+a_{2}} }
\no$$
A second useful identity is
$$\int{d^{n}q\o (2\pi)^{n}}{1\o{\left[ q^2+\mu^2\right]}^\nu}
={\G(\nu-{n\o2})\o\G(\nu)(4\pi)^{n/2}}\,{1\o\mu^{\nu-{n\o2}}}\no$$

We begin by evaluating the diagram $\bub$.  Our diagramatic notation
here is that the dots on a diagram represent the location of insertions
in that diagram.  The insertions could arise from different sources.
$\bub$ for instance arises at $O(\l)$ in $\G^{(2,1)}$ and at $O(\l^2)$
in $\G^{(4)}$.
$$\bub={1\o\kl}\sum_{n}\int_{q}{1\o[q^2+\t+\nkol][(k-q)^2+\t+\nkol]}\no$$
where
$$\int_{q}f(q)=\int_{R^{d-1}}\dq f(q)\no $$
and $k$ can be composed of several external momenta depending on the
form of the insertion.  For periodic boundary conditions
$\omega(n)={(2\pi n)}^2$, $n=0,\pm 1,\pm 2\dots$. We consider only $n=0$ on
the external lines.  On introducing a
Feynman parameter, performing the integration over $q$, and taking
advantage of the translation invariance of the range of this integral,
$\bub$ becomes
$$\bub={\G({5-d\o2})\o\kl(4\pi)^{(d-1)/2}}\sum_{n}\int_{0}^{1}dx
{1\o{[x(1-x)k^2+\t+\nkol]}^{(5-d)/2}}\no$$
For our generic normalization point with symmetric point momentum
$$p_{i}\cdot p_{j}={1\o4}(4\delta_{ij}-1)\a_{1}\no $$
(the $p_i$ are the external momenta entering the vertices which combine to form
$k$), and mass point
$$\t=\a_{2}\no $$ we have
$$\bub={\G({5-d\o2})\o\kl(4\pi)^{(d-1)/2}}\sum_{n}\int_{0}^{1}dx
{[x(1-x)\a_{1}+\a_{2}+\nkol]}^{(d-5)/2}\no$$
Evaluating the action of $A_{4-d}=\k{d\o d\k}-(4-d)$ we get
$$A_{4-d}\bub(\kl)=-{\G({7-d\o2})\o\kl(4\pi)^{(d-1)/2}}
\sum_{n}\int_{0}^{1}dx
{x(1-x)\a_{1}+\a_{2}\o{[x(1-x)\a_{1}+\a_{2}+\nkol]}^{(7-d)/2}}\no$$

For the symmetric point normalization conditions,
and zero mass,  we have $\a_{1}=1$ and $\a_{2}=0$.
Thus the standard symmetric point prescription yields
$$\bub^{sp}(\kl)={\G({5-d\o2})\o\kl(4\pi)^{(d-1)/2}}\sum_{n}\int_{0}^{1}dx
{[x(1-x)+\nkol]}^{(d-5)/2}\no$$
and
$$A_{(4-d)}\bub^{sp}(\kl)=-2{\G({7-d\o2})\o\kl(4\pi)^{(d-1)/2}}
\sum_{n}\int_{0}^{1}dx
{x(1-x)\o{[x(1-x)+\nkol]}^{(7-d)/2}}\no$$
For a mass normalization condition at zero symmetric point momentum,
we are working at $\a_{1}=0$ and $a_{2}=1$, in which case $\bub$ becomes
$$\bub^{t}(\kl)={\G({5-d\o2})\o\kl(4\pi)^{(d-1)/2}}\sum_{n}
{[1+\nkol]}^{(d-5)/2}\no$$
and
$$A_{(4-d)}\bub^{t}(\kl)=-2{\G({7-d\o2})\o\kl(4\pi)^{(d-1)/2}}\sum_{n}
{[1+\nkol]}^{(d-7)/2}\no$$
The advantage of the mass point prescription is that there is one
less Feynman parameter integral to do.

We see that $A_{(4-d)}\bub$ is in fact finite when $d<5$, and  from
the considerations of section 4 determines the 1-loop RG equations.
A useful observation is that the asymptotic limits of all of these
expressions for large and small $\kl$ can be obtained very simply
from one another. Observe first, that in the limit $\kl\ra0$,
only the $n=0$ term contributes, thus this limit is obtained by
retaining only the $n=0$ term in the above expressions. In contrast
in the limit $\kl\ra\infty$ the sum over $n$ becomes an integral,
which can be thought of as
another momentum integral times a factor of $\kl$.
Thus we can account for this limit by multiplying the expression
in the limit $\kl=0$ (obtained by retaining only the $n=0$ term)
by $\kl$ and replacing $d-1$ by $d$. This is equivalent to viewing
the $\kl\ra0$ limit of the $d$ dimensional layered geometry as the
limit $\kl\ra\infty$ of a $d-1$ dimensional layered geometry. The
limiting expressions don't know the difference. Of course, one can
obtain the asymptotic expressions by  direct methods if so desired,
which allows the computation of leading asymptotic corrections.
For our purposes the expressions themselves and their asymptotic
limits are sufficient. Using the above reasoning we see that
the asymptotic limits are
$$A_{(4-d)}\bub(\kl)=-2{\Gamma({7-d\o2})\o\kl{(4\pi)}^{(d-1)/2}}
\int_0^1dx{[x(1-x)\alpha_{1}+\alpha_{2}]}^{(d-5)/2}$$
for $\kl\ra0$
and
$$A_{(4-d)}\bub(\kl)=-2{\Gamma({6-d\o2})\o{(4\pi)}^{d/2}}
\int_{0}^{1}dx{[x(1-x)\alpha_{1}+\alpha_{2}]}^{(d-4)/2}$$
for $\kl\ra\infty$.

In the RG equations the quantity
$$\e(\kl)=4-d-\k{d\o d\k}\ln[A_{4-d}\bub]\no $$
is of interest. This expression is clearly finite and, by our rules
for relating the asymptotic values, interpolates between $4-d$ and
$3-d$ as $\kl$ ranges from $0$ to $\infty$.

Next we turn to the evaluation of the ``cone'' diagram
$\if$, which appears in the two loop
renormalization of $\G^{(4)}$ and $\G^{(2,1)}$, and is given by
$$\eqalign{\if&={1\o{(\kl)}^2}\sum_{n_1,n_2}\int_{q_{1}}\int_{q_{2}}
{1\o[q_1^2+\tpn{1}][q_2^2+\tpn{2}]}\cr
&\times{1\o [(k_{1}-q_1)^2+\tpn{1}][(k_2+q_1-q_2)^2+\tpn{12}]}\cr}\no$$
where $n_{12}=n_1-n_2$, and
the diagram is evaluated with zero external discrete momenta. Note
that the diagram only depends on the two momentum variables
$k_{1}$ and $k_{2}$ since we must have overall momentum
conservation and there are only three vertices.

We proceed to evaluate this diagram by first combining the second
and last propagators, performing the change of variables
$q_2\ra q_2+x(k_2+q_1)$, to obtain
\eqlabel{\parx}
$$\eqalign{&\int_{q_{2}}{1\o[q^2+\tpn{2}][(k_2+q_1-q_2)^2+\tpn{12}]}\cr
&\qquad={\G({5-d\o2})\o{(4\pi)}^{(d-1)/2}}\int_{0}^{1}dx
{1\o{[x(1-x)(k_2+q_1)^2+\tpnx]}^{(5-d)/2}}\cr}\no$$
where
$\omega(n_{1},n_{2},x)=\omega(n_{12}) x +\omega(n_{1}) (1-x)$.
Next, grouping the other two propagators we obtain
\eqlabel{\parz}
$$\eqalign{&{1\o[q_1^2+\tpn{1}][(k_{1}-q_1)^2+\tpn{1}]}\cr
&\qquad=\int_{0}^{1}dz
{1\o{[(q-z k_{1})^2+z(1-z)k_{1}^2+\tpn{1}]}^2}\cr}\no$$
Combining (\docref{parx}) and (\docref{parz}),
translating the $q_1$ integrand, and
introducing a third Feynman parameter $y$ to combine the two
remaining propagators, yields
$$\eqalign{\if&={\G({9-d\o2})\o{(\kl)}^2{(4\pi)}^{(d-1)/2}}\sum_{n_1,n_2}
\int_{0}^{1}dx\int_{0}^{1}dy\int_{0}^{1}dz
\int_{q_{1}}\cr
&\qquad{{[x(1-x)]}^{(d-5)/2}(1-y)y^{(3-d)/2}
\o{[(1-y)q_1^2
+y{(k_3+zk_{1}+q_1)}^2+z(1-z)k_{1}^2+
\tau r(x,y)+{\omega(n_{1},n_{2},x,y)\o\k^2L^2}]}^{(9-d)/2}}
\cr}\no$$
where $r(x,y)={y\o x(1-x)} +(1-y)$ and
$\omega(n_{1},n_{2},x,y)={\omega(n_{1},n_{2},x)\o
x(1-x)}y+\omega(n_{1})(1-y)$.
Finally, performing the integration over $q_1$ we have
$$\eqalign{\if&={\G(5-d)\o{(\kl)}^2{(4\pi)}^{d-1}}\sum_{n_1,n_2}
\int_{0}^{1}dx\int_{0}^{1}dy
\int_{0}^{1}dz\Psi^{4}_{n_{1},n_{2}}\cr}\no $$
where
$$\eqalign{\Psi^{4}_{n_{1},n_{2}}=
&{{[x(1-x)]}^{(d-5)/2}(1-y)y^{(3-d)/2}
\o{\left[(1-y)\{y{(zk_{1}+k_2)}^2
+z(1-z)k_{1}^2\}+\tau r(x,y)+
{\omega(n_{1},n_{2},x,y)\o\k^2L^2}\right]}^{(5-d)}}
\cr}\no$$
At our generic normalization point,
$k_{1}^2=\a_{1}$, $k_{2}^2={3\o4}\a_{1}$, $k_{1}k_2=-{1\o2}\a_{1}$
and $\tau=\alpha_{2}$; thus we find
$$\eqalign{&\ \ \ \if^{sp}=\cr
&{\G(5-d)\o(\k L)^2(4\pi)^{(d-1)}}\sum_{n_1,n_2}
\int_0^1dx\int_0^1dy\int_0^1dz
{[x(1-x)]^{(d-5)/2}(1-y)y^{(3-d)/2}
\o{\left[s(y,z)\a_1+r(x,y)\a_2
+{\omega(n_1,n_2,x,y)\o\k^2L^2}\right]}^{(5-d)}}
\cr}\no$$
where $s(y,z)=z(1-z)(1-y^2)+{3\o4}y(1-y)$. This gives
$$\eqalign{A_{2(4-d)}\if&=-{\G(6-d)\o{(\kl)}^{2}(4\pi)^{(d-1)}}\sum_{n_1,n_2}
\int_{0}^{1}dx\int_{0}^{1}dy\int_{0}^{1}dz
\psi^{4}_{n_{1},n_{2}}\cr}\no $$
where
$$\eqalign{\psi^{4}_{n_{1},n_{2}}
=&{{[x(1-x)]}^{(d-5)/2}(1-y)y^{(3-d)/2}
\{s(y,z)\a_{1}+r(x,y)\a_{2}\}
\o{\left[s(y,z)\a_{1}+r(x,y)\a_{2}+{\omega(n_1,n_2,x,y)\o\k^2L^2}\right]}^{(6-d)}}
\cr}\no$$
For a mass independent symmetric point normalization one simply sets
$\a_{1}=1$ and $\a_{2}=0$, and for standard mass normalization at
zero symmetric point momentum one sets $\a_{1}=0$ and $\a_{2}=1$

The expressions for $A_{2(4-d)}\if$ still have a simple pole at $d=4$
arising from the second order pole in the preceding expression. The pole is
cancelled, however, by the diagram $\bub^2$ which also appears in the
two loop renormalization of $\G^{(4)}$. The expression
for the combination
$A_{2(4-d)}(\if-{1\o2}\bub^2)$ is non singular in the limit
$d\ra4$. It is clearly desireable to combine our expressions so as to obtain
manifestly finite quantities. To this end we note that
${1\o2}A_{2(4-d)}\bub^2=\bub A_{(4-d)}\bub$ has the expression
$$\bub A_{(4-d)}\bub=-{2\G(6-d)\o{(\kl)}^2{(4\pi)}^{(d-1)}}\sum_{n_{1},n_{2}}
\int_{0}^{1}dx\int_{0}^{1}dy\int_{0}^{1}dz\psi^{2}_{n_{1},n_{2}}\no $$
where
$$\psi^{2}_{n_{1},n_{2}}
={{[x(1-x)]}^{(d-5)/2}y^{(3-d)/2}{(1-y)}^{(5-d)/2}\{z(1-z)\a_{1}+r(x,y)\a_{2}\}
\o{\left[\tilde s(x,y)\a_{1}+r(x,y)\a_{2}+
{\tilde\omega(n_1,n_2,x,y)\o\k^2L^2}\right]}^{(6-d)}}\no $$
where $\tilde s(x,y)=y+z(1-z)(1-y)$ and
$\tilde\omega(n_1,n_2,x,y)={\omega(n_{2})y\o x(1-x)}+\omega(n_{1})(1-y)$.

The two expressions for $A_{2(4-d)}\if$ and $\bub A_{(4-d)}\bub$ are now
sufficently similar that we can combine
them under the Feynman parameter integrals. The resulting expression is
$$\eqalign{A_{2(4-d)}(\if-{1\o2}\bub^2)&=-{2\G(6-d)\o{{(\kl)}^2(4\pi)}^{(d-1)}}
\sum_{n_1,n_2}
\int_{0}^{1}dx\int_{0}^{1}dy\int_{0}^{1}dz\Phi^{4}_{n_{1},n_{2}}}\no $$
where
$$\eqalign{\Phi^4_{n_1,n_2}
&=\psi^4_{n_1,n_2}-\psi^2_{n_1,n_2}={[x(1-x)]}^{(d-5)/2}y^{(3-d)/2}\times\cr
&\quad\left[{(1-y)[s\a_1+r\a_2]
\o{\left[s\a_1+r\a_{2}+{\oxy\o\k^2L^2}\right]}^{(6-d)}}
-{{(1-y)}^{(5-d)/2}[z(1-z)\a_{1}+\a_{2}]
\o{\left[\tilde s\a_{1}+r\a_{2}+{\tilde\oxy\o\k^2 L^2}\right]}^{(6-d)}}
\right]\cr}\no$$
The singularities of the integrands near $y=0$
each give rise to pole contributions in the individual diagrams,
but these now cancel at the level of
the combined integrand and the expression is regular at $d=4$.

Now, by our rules for extracting the asymptotic values of these
expressions we obtain
$$\eqalign{&A_{2(4-d)}(\if-{1\o2}\bub^2)|_{\kl=0}=-{2\G(6-d)\o{(4\pi)}^{(d-1)}}
\int_{0}^{1}dx\int_0^1dy\int_0^1dz\Phi_{0,0}}\no $$
where
$$\eqalign{\Phi_{0,0}&={[x(1-x)]}^{(d-5)/2}y^{(3-d)/2}\times\cr
&\left[{(1-y)[s\a_1+r\a_2]
\o{\left[s\a_1+r\a_2\right]}^{(6-d)}}
-{{(1-y)}^{(5-d)/2}[z(1-z)\a_1+a_2]\o{\left[\tilde s\a_1+r\a_2\right]}^{(6-d)}}
\right]\cr}\no$$
and the $\kl\ra\infty$ limit is obtained by replacing $d-1$ by $d$
and multiplying by $\k^2L^2$ (one power of $\kl$ for each sum).
Note that
$$f_{1}=-\fone\no $$
is a finite expression for all values of $\kl$ and
interpolates between ${1\o3}$ and ${1\o2}$
as $\kl$ ranges from $0$ to $\infty$ for $d=4$, $\a_1=0$ and $\a_2=1$, and more
generally it
interpolates between the values for $d-1$ and $d$ as $\kl$ ranges
from $0$ to $\infty$.

Let us now consider the diagram arising due to wavefunction
renormalization.
$$\eqalign{\itr={1\o\k^2L^2}\sum_{n_{1},n_{2}}\int_{q_{1}}\int_{q_{2}}{1\o
[q_{1}^2+\tpn{1}]}{1\o[q_{2}+\tpn{2}]}
{1\o[{(k+q_{1}-q_{2})}^2+\tpn{12}]}\cr}\no $$
and again $n_{12}=n_{1}-n_{2}$. Introducing a Feynman parameter
to combine the last two propagators and performing the integral over
$q_{2}$ we get
$$\eqalign{\itr &=
{\G({5-d\o2})\o{(\kl)}^2{(4\pi)}^{(d-1)/2}}\times\cr
&\sum_{n_{1},n_{2}}\int_{q_{1}}{1\o
[q_{1}^2+\tpn{1}]}\int_{0}^{1}dx
{[{x(1-x)(k+q_{1})}^2+\tpnx]}^{(5-d)/2}\cr}\no $$
Introducing an additional Feynman parameter
and performing the integration over $q_{1}$ yields
$$\eqalign{\itr=
&{\G(4-d)\o(\kl)^2{(4\pi)}^{(d-1)}}\sum_{n_{1},n_{2}}
\int_{0}^{1}dx\int_{0}^{1}dy
{{[x(1-x)]}^{(d-5)/2}y^{(5-d)/2}\o{[y(1-y)k^2+r(x,y)\t+{\oxy\o\k^2
L^2}]}^{(4-d)}}\cr}\no $$
We next differentiate with respect to $k^2$ and evaluate at our
generic normalization point obtaining
$$\itrp={-\G(5-d)\o(\kl)^2{(4\pi)}^{(d-1)}}\sum_{n_{1},n_{2}}
\int_{0}^{1}dx\int_{0}^{1}dy
{{[x(1-x)]}^{(d-5)/2}y^{(5-d)/2}(1-y)\o{[y(1-y)\a_{1}+r(x,y)\a_{2}+{\oxy\o\k^2L^2}]}^{(5-d)}}
\no $$
Finally we have
$$\eqalign{&\ \ \ A_{2(4-d)}\itrp=\cr
&{2\G(6-d)\o{(\kl)}^{2}{(4\pi)}^{(d-1)}}\sum_{n_{1},n_{2}}
\int_{0}^{1}dx\int_{0}^{1}dy
{{[x(1-x)]}^{{(d-5)/2}}y^{(5-d)/2}(1-y)\{y(1-y)\a_{1}+r(x,y)\a_{2}\}
\o{[y(1-y)\a_{1}+r(x,y)\a_{2}+{\oxy\o\k^2L^2}]}^{(6-d)}}\cr}
\no $$
The expression for
$$f_{2}=\ftwo\no $$
is therefore finite for all values of  $\kl$.
In particular in the prescription $\a_1=0$, $\a_2=1$ with $d=4$ we have $f_{2}$
ranging from ${4\o27}$ to ${1\o4}$ as
$\kl$ goes from $0$ to $\infty$.

\beginappendix{EVALUATION OF THE DIAGRAMS
ON $R^3\times S^1$ WITH MASS RENORMALIZATION POINT}
In this appendix, working strictly on $R^3\times S^1$ we give an alternative
calculation of the diagrams $\bub$, $\if$, $\itr$, and
the resulting expressions derived from them required for the RG
equations. We calculate our
expressions taking advantage of the fact that the propagator in
three dimensions has the simple form
$$G(x,y)={1\o 4\pi |x-y|}e^{-m|x-y|}\no$$
In our four dimensional problem,
we Fourier transform on the periodic coordinate (the $S^{1}$
of $R^3\times S^{1}$) to obtain the propagator
$$G(x,y;n)={1\o 4\pi|x-y|}e^{-\sqrt{[\t+\nkol]}|x-y|}\no$$
We will evaluate the diagrams at zero external momentum and
also extract a scale $\k$ from each of our coordinates to
work with dimensionless diagrams. Finally we observe that the
corresponding three dimensional diagrams are UV finite,
hence the diagrams are convergent for fixed values of $n_{i}$
labeling the discrete momenta propagating in the loops. The
four dimensional UV divergences arise in this case from the
sum over $n_{i}$. We will therefore regulate the
diagrams by treating the sums not as being sums over all integers
but as sums ranging from $-N$ to $N$ where $N$ is some very
large but finite cut off value.
The combinations of these diagrams that enter the RG
equations are strictly finite and so no restriction on the sums in
this case is necessary.

We begin with the one loop contribution
$$\eqalign{\bub&={1\o\kl}\sum_{n}{1\o V}\int d^{3}x_{1}\int d^{3}x_{2}
G^{2}(x_{1},x_{2};n)\cr
&=\sum_{n}\int d^3x {1\o{(4\pi)}^2}
e^{-2\sqrt{[\t+\nkol]}|x-y|}\cr}\no$$
where the translation invariance of the propagator is used to cancel
the volume of $R^3$, which we formally took to be $V$.
Transforming to spherical coordinates this
integral can be evaluated. We find
$$\bub={1\o 8\pi\kl}\sum_{n}{1\o {[\t+\nkol]}^{1\o2}}\no$$
Therefore, setting $\t=1$, i.e. $\alpha_2=1$, and acting with
$A_0$ we get
$$A_{0}\bub=-{1\o 8\pi\kl}\sum_{n}{1\o{[1+\nkol]}^{3/2}}\no$$
The restriction on the sum over $n$ can be removed for this
contribution since it does not lead to any divergence.

We next need to evaluate the diagram $\if$, which is given in
position space by
$$\eqalign{\if&={1\o V}\sum_{n_1,n_2}\int dx_1\int dx_2
G(x_1,x_2;m_2+m_{12})^2 G(x_2,x_3;m_1)
G(x_1,x_3;m_1)\cr
&={1\o V(4\pi)^4}
\sum_{n_1,n_2}\int d^3x_1\int d^3x_2\int d^3x_3
{e^{-(m_2+m_{12})|x_1-x_2|}\o{|x_1-x_2|}^{2}}
{e^{-m_1|x_2-x_3|}\o
|x_2-x_3|}{e^{-m_1|x_1-x_3|}\o|x_1-x_3|}\cr}\no$$
where we have introduced the notation
$$m_{i}=\t+\nkol$$
with $n_{12}=n_{1}+n_{2}$.
Again the sums are treated as being cut off at $n=N$.

This diagram, which has the geometrical shape of a cone, can be
evaluated most simply by
making a judicious choice of coordinates, taking full advantage of
the translational invariance of the propagators. The most convenient is to
choose the origin to be at one of the vertices, not
at the apex of the cone. We choose the origin at $x_{2}=0$.
With this choice the integration over the location of this vertex
cancels the volume factor arising in the definition of the vertex.
The cone as a geometrical object in this setting is in fact a
triangle since two of the propagators coincide and as such is specified
by an angle and the length of the two sides.
Thus  if the angle between the points $x_1$ and $x_3$
is $\theta$ the graph becomes
$$\if=
\sum_{n_1,n_2}{1\over32\pi^2}\int_{0}^{\infty}
dx_1\int_0^{\infty} dx_3 x_3 e^{-(m_2+m_{12})x_1}e^{-m_1 x_3}
\int_{-1}^{1} d\mu {e^{-m_1|x_1-x_3|}
\over{|x_1-x_3|}}\no$$
where now $|x_1-x_3|=\sqrt{x_1^2+x_3^2-2x_1x_3\mu}$
The integral over $\mu=\cos\theta$ can be done using
$$\int_{-1}^1d\mu {e^{-m\sqrt{(a-b\mu)}}\o\sqrt{a-b\mu}}
={2\o b m}(e^{-m\sqrt{a-b}}-e^{-m\sqrt{a+b}})\no$$
Going to the coordinates $x={1\o2}(x_{1}+x_{3})$ and
$z={x_{1}-x_{3}\o x_{1}+x_{3}}$ our expression for the diagram becomes
$$\if= \sum_{n_{1},n_{2}}{1\o{(4\pi)}^2m_1}\int_{0}^{\infty}dx\int_{-1}^{1}dz
{e^{-[(m_{2}+m_{12})(1+z)+m_{1}(1-z)]x}\o1+z}[e^{-2m_{1}|z|x}-e^{-2m_{1}x}]\no$$
The integral over $x$ can now be performed to get
$$\if=\sum_{n_{1},n_{2}}{1\o{(4\pi)}^{2}m_{1}}
\int_{-1}^{1}{dz\o(1+z)}\left\{{1\o
M+(M-2m_{1})z+2m_{1}|z|}
-{1\o M+2m_{1}+(M-2m_{1})z}\right\}\no$$
where $M=m_{1}+m_{2}+m_{12}$.
This integral can be performed by dividing the range of integration
up and noting that
$$\int_{0}^{1}{dz\o{(1+z)}^2}={1\o2}$$
$$\int{dz\o{(1-z)(a+b z)}}
={1\o a+b}\ln\left[{a+bz\o 1-z}\right]$$
$$\int{1\o (1+z)(c+dz)}
={1\o c-d}\ln\left[{1+z\o c+d z}\right]$$
Care is needed with the limits as there are apparent logarithmic
singularities. These however cancel.
Completing the evaluation of the diagram one obtains
$$\if
={1\over 2(4\pi)^2}\sum_{n_{1},n_{2}}{1\over m_1 M}\no$$
Note that if we keep our cutoff $N$ fixed, then in the limit
$\kl\ra0$ the expression reduces reduces for $\t=1$ to
$\if=1/(96\pi^2)$.
Evaluating $A_{0}\if$ as with $A_0\bub$ by setting $\t=1$ and then acting with
$A_0$ we obtain
$$A_{0}\if=-\sum_{n_{1},n_{2}}{1\o32\pi^2\k^2L^2 m_1M}({1\o
m_{1}^2}+{1\o M}{1\o \tilde M})\no$$
where
$${1\o\tilde M}={1\o m_{1}}+{1\o m_{2}}+{1\o m_{12}}$$
Similarly
$${1\o2}A_{0}(\bub^2)=\bub A_{0}\bub
=-{1\o{(8\pi)^2\k^2L^2}}{\sum_{n_1,n_2}}{1\o m_1^3m_2}\no$$
We are now in a position to put the two diagrams together
to evaluate $A_{0}(\if-{1\o2}\bub^2)$. We first evaluate
$$A_{0}(\if-{1\o2}\bub^2)=
-{1\o32\pi^2\k^2L^2}\sum_{n_{1},n_{2}}
\left[{1\o m_{1}^3}({1\o M}-{1\o 2m_{2}})+{1\o m_{1} M^{2}\tilde M}\right]\no$$
where there is no need to truncate the sum since this combination of
diagrams is no longer UV divergent in four dimensions.

Putting the expressions together, we obtain
$$f_{1}={2\displaystyle\sum_{n_{1},n_{2}}
\left[{1\o m_{1}^3}({1\o M}-{1\o 2m_{2}})+{1\o m_{1} M^{2}\tilde M}\right]
\o {(\displaystyle\sum_{n}{1\o {m}^3})}^{2}}\no$$

The final diagram to be computed is the diagram contributing to
wavefunction renormalization.
We evaluate it in position space but with a momentum $k$, in
the layers $R^3$, flowing through the diagram. Again we Fourier
transform in the layers, i.e. the $S^1$.
The diagram is therefore given by
$$\itr=\sum_{n_{1},n_{2}}{1\o{(\kl)}^2}\int d^3 z e^{i k z}
G(z;m_{1})G(z;m_{2})G(z;m_{12})
\no$$
where translation invariance of the propagator has been used to
cancel the volume factor, and the sums are again regulated by a large
$n$ cutoff at $N$.  This still has a logarithmic divergence due to
the corresponding three dimensional divergence. The quantity of
interest, to us at any rate, is the contribution to wave function
renormalization, which is obtained from the derivative of this with
respect to $k^2$. It is therefore this quantity we will focus on.
Using the explicit representation for the propagator and choosing
the angle between the in coming momentum $k$ and the direction $z$
so that $k z =k r \mu$, where $\mu=\cos\theta$,  we have
$$\itrp={1\o 32\pi^2\k^2L^2}\sum_{n_{1},n_{2}}\int_{0}^{\infty}{dr\o
r}e^{-M r}{d \o d k^2}\int_{-1}^{1}d\mu e^{i k r \mu}\no$$
Noting that
$$\int_{-1}^{1}d\mu e^{i k r\mu} = {2\sin( k r)\o k r}\no$$
we have
$$\itrp ={1\o 32\pi^2\k^2L^2k^2} \sum_{n_{1},n_{2}}\int_{0}^{\infty}dy
e^{-{M\o k} y}{d\o dy}({\sin y \o y})\no$$
This can easily be evaluated using integration by parts yielding
$$\itrp ={1\o 32\pi^2\k^2L^2k^2} \sum_{n_{1}n_{2}}
\left[-1+{M\o k}\tan^{-1}({k\o M})\right]\no$$
Noting that as $k\ra0$
$$-1+{M\o k} \tan^{-1}({k\o M})=-{1\o3}{({k\o M})}^2+\dots$$
Evaluating this in the limit $k\ra0$ yields
$$\itrp =-\sum_{n_{1}n_{2}}{1\o 96\pi^2\k^2L^2M^2}\no$$
Finally the contribution needed for the RG equations is
$$A_{0}\itrp=\sum_{n_{1},n_{2}}{1\o48\pi^2\k^2L^2 M^{3}\tilde M}\no$$

The interpolating function $f_{2}=\ftwo$, remembering that
here $d=4$ is then given by
$$f_{2}={4\o3}{\displaystyle\sum_{n_{1},n_{2}}{1\o M^3 \tilde M}\o
{\left(\displaystyle\sum_{n}{1\o {m}^3}\right)}^2}\no$$
With $\tau=1$, in the limit $\kl\ra 0$, only the $n_{i}=0$ modes contribute,
therefore $m\ra 1$ $M\ra 3$ and $\tilde M\ra {1\o 3}$. Thus
we obtain $f_{2}(0)={4\o27}$.
In the other limit, $\kl\ra\infty$, the summations become integrals
and we find $f_{2}(\infty)={1\o4}$.

\references

\item{[1]}M.E. Fisher, Rev. Mod. Phys. {\bf 46} (1974) 597.

\item{[2]}R. Frowein, J. K\"otzler, B. Schaub and H.G. Schuster, Phys. Rev.
{\bf B}25, (1982) 4905.

\item{[3]}M.E. Fisher and D.R. Nelson, Phys. Rev. Lett. {\bf 32}, (1974) 1350.

\item{[4]}I. Rhee, F.M. Gasparini and D.J. Bishop, Phys. Rev. Lett. {\bf 63},
(1989) 410.

\item{[5]}R. B. Stinchcombe J. Phys. {\bf C6}, (1973), 2459.

\item{[6]} K. Binder,
Phase Transitions and Critical Phenomena Vol. 5B, ed.s Domb and
Green (1976).

\item{[7]}M. N. Barber, in Phase Transitions and Critical Phenomena,
vol.8, eds.  C. Domb and J. L. Lebowitz
(Academic Press, London 1983); CPSC vol.2,
ed. J. L. Cardy (North Holland, 1988).

\item{[8]}M. Suzuki, Prog. Theor. Phys. {\bf 56}, (1976), 1454;
Stat. Phys. {\bf 18} ed S. Hess, North Holland
(1993), 432.

\item{[9]}E.K. Riedel and F. Wegner, Z. Phys. {\bf 225}, (1969) 195.

\item{[10]}P. Pfeuty, D. Jasnow and M.E. Fisher, Phys. Rev. {\bf B10} (1974)
2088; S. Singh and D. Jasnow, Phys. Rev. {\bf B11} (1975) 3445.

\item{[11]}T.W. Capehart and M.E. Fisher, Phys. Rev. {\bf B13}, (1976) 5021.

\item{[12]}A.E. Ferdinand and M.E. Fisher, Phys. Rev. {\bf 185} (1969) 832.

\item{[13]} M. Barber and M.E. Fisher, Ann. Phys., {\bf 77} (1973) 1.

\item{[14]}E. K. Riedel and F. Wegner, Phys. Rev. {\bf B9}, (1974), 294.

\item{[15]}D.J. Amit and Y.Y. Goldschmidt, Ann. Phys., {\bf 114} (1978) 356.

\item{[16]} D.R. Nelson  and E. Domany, Phys. Rev. {\bf B13} (1976) 236.

\item{[17]}P. Seglar and M.E. Fisher, J. Phys. {\bf C13} (1980) 6613.

\item{[18]} Denjoe O'Connor and C.R. Stephens, Proc. Roy. Soc. {\bf A}
(1993) to be published.

\item{[19]}E. Frey and F. Schwabl, Phys. Rev. B, {\bf 42} (1990) 8261.

\item{[20]} C.R. Stephens, J. of Magnetism and Magnetic Materials, 104-107
(1992) 297.

\item{[21]}D. Schmeltzer, Phys. Rev. {\bf B32}, 7512 (1985).

\item{[22]}I. D. Lawrie, J. Phys. {\bf C 11}, (1978), 3857.

\item{[23]}A.M. Nemirovsky and K.F. Freed, Nucl. Phys. {\bf B270}[FS16]
(1986) 423.

\item{[24]} E. Br\'ezin and J. Zinn-Justin,
Nucl. Phys. {\bf B257}\ [FS14](1985) 867;
J. Rudnick, H. Guo and D. Jasnow, Jour. Stat. Phys. {\bf 41} (1985) 353.

\item {[25]} H. Matsumoto, Y. Nakano and H. Umezawa, Phys. Rev. D29 (1984)
1116.

\item{[26]} Denjoe O'Connor, C.R. Stephens and F. Freire,
Class. Quan. Grav. {\bf 23} (1993) S243; F. Freire and C.R. Stephens, Zeit.
Phys. {\bf C} (1993) to be published; Denjoe O'Connor, C.R. Stephens and F.
Freire, Mod. Phys. Lett. {\bf A25} (1993); M. van Eijck, C.R. Stephens and C.W.
van Weert, ``Temperature Dependence of the QCD Coupling'' Utrecht/Amsterdam
preprint ITFA-93-11, THU-93/08.

\item{[27]}Denjoe O'Connor and C. R. Stephens,
Nucl. Phys. {\bf B360} (1991) 297; J. Phys. {\bf A25} (1992) 101; J. of
Magnetism and Magnetic Materials, 104-107 (1992) 300.

\item{[28]}Denjoe O'Connor and C.R. Stephens,
``Finite Size Scaling and the Renormalization Group'', preprint DIAS-STP-90-26,
Imperial TP/89/90/36 (1990).

\item{[29]}F. Freire, Denjoe O'Connor and C.R. Stephens, Dimensional Crossover
and Finite Size Scaling Below $T_c$, Univ. Utrecht preprint THU 92/36, to be
published in J. Stat. Phys..

\item{[30]}F. Wegner, Phase Transitions and Critical Phenomena Vol. 6, ed.s
Domb and Green (1976).

\item{[31]}E. Stueckelberg and A. Peterman, Helv. Phys. Acta. {\bf 26} (1953)
499; M. Gell-Mann and F. Low, Phys. Rev. {\bf 95} (1954) 1300;
N.N. Bogoliubov and D.V. Shirkov, Doklady AN SSSR {\bf 103} (1955) 203.

\item{[32]} D.V. Shirkov,  RG'91, (World Scientific 1992).

\item{[33]}L. Kadanoff, Physica {\bf 2} (1966) 263; K. Wilson, Phys. Rev. {\bf
179} (1969) 1499.

\item{[34]}G. Jona-Lasinio, Phase Transitions and Critical Phenomena Vol. 6,
ed.s  Domb and Green (1976).

\item{[35]}K. Wilson and J. Kogut Phys. Rep. {\bf C12} (1974) 75.

\item{[36]}A.C.D. van Enter, R. Fernandez and A. Sokal, preprint.

\item{[37]}A. Polyakov, Phys. Lett. {\bf B59} (1975) 79.

\item{[38]}J.F. Nicoll, T.S. Chang and H.E. Stanley,
Phys. Rev. {\bf B12} (1975) 458.

\item{[39]}Denjoe O'Connor, C.R. Stephens and B.L. Hu, Annals of Physics {\bf
190} (1989);
Denjoe O'Connor and C.R. Stephens, Phys. Rev. {\bf B43} (1991) 3652.

\item{[40]}G. Parisi, J. Stat. Phys. {\bf 23} (1980) 49.

\item{[41]}J. Zinn-Justin, Quantum Field Theory and Critical Phenomena,
Clarendon Press, (Oxford 1989).

\item{[42]}A. Aharony and M. E. Fisher, Phys. Rev. Lett.
{\bf 45} (1980) 679; A. Aharony and G. Ahlers, Phys. Rev. Lett. {\bf 44} (1980)
782.

\item{[43]}J. A. Hertz, Phys. Rev. {\bf B14} (1976) 1165.

\item{[44]}P. Pfeuty, J. Phys. C.: Solid St. Phys. {\bf 9} (1976)
3993.

\item{[45]}L. Dolan and R. Jackiw. Phys. Rev. {\bf D9} (1974) 3320.

\item{[46]}E. Br\'ezin, D.J. Wallace and K.G. Wilson, Phys. Rev. Lett.
{\bf 29} (1972) 591.

\item{[47]}Denjoe O'Connor and C.R. Stephens, ``Dimensional Crossover in the
Non-linear $\sigma$ Model'', Utrecht/DIAS preprint.

\item{[48]}G.A. Baker, B.G. Nickel and D.I. Meiron, Phys. Rev. {\bf B17}
(1978) 1365.

\item{[49]}Denjoe O'Connor, C.R. Stephens and A. Bray, ``Dimensional
Crossover in the $N\ra\i$ limit'', Utrecht/DIAS preprint.

\item{[50]}R.D. Kamien and D.R. Nelson, Jou. Stat. Phys. {\bf 71} (1993) 23.

\item{[51]}I.D. Lawrie and M.E. Fisher, J. Appl. Phys. {\bf 49} (1978) 1353.

\item{[52]}Y. Kubyshin, Denjoe O'Connor and C.R. Stephens, Class. Quan. Grav.
(1993) to be published.

\item{[53]}M. Benhamou, Int. Jou. Mod. Phys. {\bf A8} (1993) 2581.

\item{[54]}L. Leibler, Macromolecules {\bf 13} (1980) 1602.

\item{[55]}P.G. de Gennes, J. Phys. Lett. {\bf 40} (1979) 69.

\vfill\eject
\centerline{\bf FIGURE CAPTIONS}
Figure 1: Graph of $\e$, $f_1$ and $f_2$ for $d=4$ and periodic boundary
conditions.

Figure 2: Two loop [2,1] Pad\'e resummed floating coupling, h,
for $d=4$ and periodic
boundary conditions. $N=0$ (polymers), $N=1$ (Ising model), $N=2$ (XY-model),
$N=3$ (Heisenberg model), $N=\i$ (Spherical model) and $N=-2$ are shown.

Figure 3: Graph of two loop [2,1] Pad\'e resummed $\nu\ef$ for $d=4$
and periodic boundary conditions,
$N=0,1,2,3,\i$. For $N=-2$ we have
$\nu(z)=1/2$ identically.

Figure 4: Graph of two loop [2,1] Pad\'e resummed $\eta\ef$ for $d=4$ and
periodic boundary conditions, $N=0,1,2,3$. For
$N=-2$ and $N=\i$, $\eta(z)=0$ identically.

Figure 5: Graph of two loop [2,1] Pad\'e resummed $\gamma\ef$ for
$d=4$ and periodic boundary conditions,
$N=0,1,2,3,\i$. For $N=-2$, $\gamma(z)=1$ identically.

Figure 6: Graph of two loop [2,1] Pad\'e resummed $\beta\ef$ for $d=4$ and
periodic boundary conditions, $N=-2,0,1,2,3,\i$.

Figure 7: Graph of two loop [2,1] Pad\'e resummed $\delta\ef$ for $d=4$ and
periodic boundary conditions, $N=-2,0,1,2,3,\i$.

Figure 8: Graph of two loop [2,1] Pad\'e resummed $\alpha\ef$ for $d=4$ and
periodic boundary conditions, $N=-2,0,1,2,3,\i$.

Figure 9: Graph of two loop [2,1] Pad\'e resummed $\gl(=4-\de)$ for $d=4$
and periodic boundary conditions, $N=-2,0,1,2,3,\i$.

Figure 10: Graph comparing different approximations for $\nu\ef$ for a four
dimensional layered Heisenberg model and periodic boundary conditions. The
upper
graphs correspond to two loop [2,1] Pad\'e resummed results.
In the topmost curve
$\gft$ was resummed as well as $\beta$ and gives $\nu(\i)=0.709$, whereas in
the lower only $\beta$ was resummed and $\nu$ expanded to second order in $h$
yielding $\nu(\i)=0.706$. The bottom two graphs are the corresponding one loop
results.

Figure 11: Graph comparing two loop [2,1] Pad\'e resummed values for
the floating coupling and the floating fixed point.

Figure 12: Graph of $\nu\ef$ at one loop for periodic, Dirichlet and
antiperiodic boundary conditions. $d=3$ and $N=1$.

Figure 13: Graph of $\alpha\ef$ at one loop for periodic, Dirichlet and
antiperiodic boundary conditions. $d=3$ and $N=1$.

Figure 14: Graph of $\beta\ef$ at one loop for periodic, Dirichlet and
antiperiodic boundary conditions. $d=3$ and $N=1$.

Figure 15: Graph of $\de$ at one loop for periodic, Dirichlet and
antiperiodic boundary conditions. $d=3$ and $N=1$.

Figure 16: Graph of two loop [2,1] Pad\'e resummed $\gft$ and $\nu\ef$ for
three dimensional quantal Ising model.

Figure 17: Graph of two loop [2,1] Pad\'e resummed $\alpha\ef$ and $\eta\ef$
for
three dimensional quantal Ising model.

Figure 18: Graph of two loop [2,1] Pad\'e resummed $\gamma\ef$ and $\gl$ for
three dimensional quantal Ising model.

Figure 19: Graph of $\nu\ef$, $\gl$ and $\gft$ at one loop for two
dimensional quantum Ising model.

Figure 20: Graph of two loop [2,1] Pad\'e resummed $\gamma\ef$ for $d=4$ and
periodic boundary conditions showing the effect of different initial
conditions.

Figure 21: Comparison of $\nu\ef$ at one loop for uniaxial dipolar ferromagnet,
three dimensional quantum Ising model and four dimensional layered Ising model
with periodic boundary conditions.

Figure 22: The floating coupling $h$ for three dimensional layered Ising model
including crossover to mean field theory.

\vfill\eject
\end